\documentclass[%
 aip,
 amsmath,amssymb,
 reprint,%
]{revtex4-1}

\usepackage[pdftex]{graphicx}
\usepackage{hyperref}
\usepackage{graphicx}
\usepackage[none]{hyphenat}
\usepackage{dcolumn}
\usepackage{bm}
\usepackage{amsmath}
\usepackage{xcolor}
\usepackage{float}
\graphicspath{{Figures/}}

\newcommand{\RomanNumeralCaps}[1]
    {\MakeUppercase{\romannumeral #1}}

\begin{document}


\preprint{AIP/123-QED}

\title{Ion-driven destabilization of a toroidal electron plasma - A 3D3VPIC Simulation}

\author{S. Khamaru}
\affiliation{ 
Institute for Plasma Research, Bhat, Gandhinagar 382428, India}
\email[]{swapnali.khamaru@ipr.res.in}
\affiliation{ 
Homi Bhabha National Institute, Training School Complex, Anushaktinagar, Mumbai 400094, India}
\author{R. Ganesh}
\affiliation{ 
Institute for Plasma Research, Bhat, Gandhinagar 382428, India}
\affiliation{ 
Homi Bhabha National Institute, Training School Complex, Anushaktinagar, Mumbai
400094, India}
\author{M. Sengupta}%
\affiliation{
Lawrence Livermore National Laboratory, Livermore, California, 94551, USA}

\date{\today}


\begin{abstract}
Ion resonance instability of toroidal electron plasmas in a tight aspect ratio axisymmetric toroidal device is reported for ${Ar}^+$ ions of different initial density values using a high fidelity 3D3V PIC solver. Stability of a recently discovered quiescent quasi-steady state (QQS) of a toroidal electron plasma obtained from ``seed" solution as a result of entropy extremization at zero inertia, is addressed to the presence of a small ion population. An ion fraction ($f$) and corresponding number of secondary electrons are preloaded into the system after the electron plasma attains a QQS state. Driven by the ions, the electron plasma exhibits destabilized ``center of charge motion" ($m$ $=$ $1$) along with increased poloidal mode coupling ($m$ $=$ $1$ to $9$) with dominant $m$ $=$ $2$ mode. The growth in wall probe current is algebraic in nature and increases for $f$ $\geq$ $0.005$, showing saturation at later time. Higher values of ion temperatures than the electron temperatures indicate a resonant energy transfer from electron plasma to ions via ion-resonance and concomitant ion heating. The volume averaged temperature value of the electron plasma rises with simulation time, attaining a quasi-steady nature near the end of the simulation time. As can be expected from conservation of adiabatic invariants, the flux tube averaged electron temperatures along parallel and perpendicular directions are found to scale as $1/R^{2}$ and $1/R$ respectively, where $R$ is the major radial variable, though the plasma is nearly collision-less.  

\end{abstract}



\maketitle

\section{Introduction}
\label{sec:Introduction}
Instability induced confinement issues in charged particle traps are ubiquitous and effective mitigation of such issues requires knowledge of the underlying fundamental processes, which in turn often result in further technical improvement of the device. Besides providing useful insight into several fundamental physical processes, these particle trapping devices become very useful in various applications, as discussed next.\

External electric field and homogeneous magnetic field aided linear traps such as Penning-Malmberg\cite{Malmberg1975} (PM) trap, Paul trap\cite{Paul} and some modified PM traps\cite{Hurst2014,Yamada2016,Lane2019} are used extensively to trap pure electron plasma\cite{Davidson1998}, pure ion plasma\cite{ONeil1998,Landa2019} and unconventional charge species such as positron plasma\cite{Greaves2000}, anti-protron plasma.\cite{Ahmadi2016,Ahmadi2018} In straight cylindrical PM trap, radial and axial confinement are provided to the pure electron (or ion) plasma, with the aid of an axial homogeneous magnetic field and electrostatic end plugs, respectively. The result achieved in such an arrangement is so robust that the electron plasma plasma can be confined for a long time (over several days) in near absolute equilibrium state\cite{ONeil1979,Malmberg1980,Driscoll1988,Dubin1999}, allowing numerous experiments to be performed. For example, pure electron plasmas share fundamental properties with inviscid 2D Euler fluids at low density limit ($m \rightarrow 0$) and hence are often investigated experimentally\cite{Fine1995} to understand complex nonlinear dynamics of such fluids. These findings often corroborate with the corresponding theoretical\cite{Briggs1970} and numerical\cite{Ganesh2006,Rome2000} studies. Extensive numerical studies\cite{Sengupta2014} using two dimensional Particle-in-Cell (PIC)\cite{Birdsall} simulation also provide essential insight into two dimensional electron plasma dynamics in the cylindrical cross-section of the PM trap. Experimental applications of pure electron plasmas are found in high-precision measurement techniques\cite{Niemann2019,Schuh2019}, mass spectrometry\cite{Guti2019,Peurrung1996} etc. Positron, antiprotons are confined in separate PM traps with modified electric fields and axial magnetic field, to synthesize antihydrogen used in anti-matter investigations\cite{Ahmadi2016,Ahmadi2018,Fajans2020}. Linear Paul traps\cite{Paul,Landa2019} are mainly motivated to trap low density ions using time varying RF fields and axial magnetic fields, whereas stabilization of the ions is performed with laser cooling\cite{Morigi1999}. Such stable ions are used in quantum information processing\cite{Morigi2001}, mass spectrometry\cite{Douglas2005} etc.\ 
Stable pair plasmas such as positron-electron, found in astrophysical environment, are studied experimentally using combined Penning-Paul trap\cite{Greaves2002}, inhomogeneous magnetic field in magnetic mirror configuration\cite{Higaki2017} and recently reported numerical experiment\cite{Hicks2019} with RF multipole electric field. In a recent experiment, stable electron-ion coexistence is achieved in a PM like\cite{nakajima2021} trap under certain conditions.\ 

In contrast to the above discussed devices, toroidal traps\cite{Janes1966,Daugherty1969,Clark1976,Zaveri1992,Khirwadkar1993,Lachhvani2017,Stoneking2004} of different aspect ratios ($a/b$, $a$ is the major radius, $b$ is the minor radius) apply an inhomogeneous external magnetic field with inherent curvature to confine an electron plasma. The electron plasma goes through rotational cross-field $\bf{E} \times \bf{B}$ drift where $\bf{E}$ field is the self electric field of the plasma and $\bf{B}$ is the external toroidal field. Because of the strong rotational $\bf{E} \times \bf{B}$ drift, electron plasma avoids the vertical loss of particles due to ${\bf{\nabla}} B$ and curvature drift. Additionally, the electron plasma undergoes toroidal $m =1$ Diocotron\cite{Davidson} rotation or center of charge dynamics, which is similar to the Diocotron motion of cylindrical electron plasma but due to magnetic field curvature are strongly coupled to elliptical (i.e. $m=2$) mode and often to other higher order poloidal modes (i.e. $m = 3, 4, 5$ etc)\cite{Khamaru2019} depending on the aspect ratio of the device and the amplitude of the $m=1$ mode. Apart from toroidal magnetic field, electron plasma is also confined using magnetic surfaces in stellarator\cite{Marksteiner2008} and dipole magnetic field\cite{Saitoh2010} based devices.\ 

Earliest experiments of trapped electron plasmas in toroidal geometry were performed by Janes et al. ($a/b$ $\sim$ 5)\cite{Janes1966}, Daugherty et al. ($a/b$ $\sim$ 4.6)\cite{Daugherty1969}, where the axisymmetric toroidal device was proposed to be used as heavy ion plasma accelerator (HIPAC) producing highly stripped heavy ions. The device was intended to utilize the potential well of the electron plasma to trap and accelerate heavy ions. Subsequent experiments were performed to confine electron plasma in axisymmetric toroidal geometry of different aspect ratios ($a/b$ $\sim$ 6.2\cite{Clark1976} and $\sim$ 1.5\cite{Zaveri1992,Khirwadkar1993}), followed by a number of experiments in partial toroidal geometry: small aspect ratio device SMARTEX-C ($a/b$ $\sim$ 1.6)\cite{Lachhvani2017}, large aspect ratio device LNT \RomanNumeralCaps{2} ($a/b$ $\sim$ 13.7)\cite{Stoneking2004}.\  

Such an electron plasma in inhomogeneous magnetic field is useful to understand nonlinear coupling\cite{Khamaru2019,Khamaru2021,Khamaru2021Er} between different toroidal Diocotron modes, collisional/non-collisional transport mechanisms\cite{Crooks1996} of electron plasma energy and also crucial to initiate an ongoing construction of electron-positron plasma experiment using dipole magnetic field\cite{Stoneking2020,Singer2021}. These systems can often serve as test-beds to understand plasma dynamics in curved magnetic field which are often found in astrophysical phenomena\cite{Brambilla2018}.\

As indicated before, near-absolute equilibria of the electron plasma in PM traps with uniform magnetic fields allows various experiments to be performed in the device. Such near-absolute equilibrium in toroidal trap was recently addressed and a quiescent quasi-steady (QQS)\cite{Khamaru2021,Khamaru2021Er} toroidal electron plasma has been found in a numerical experiment with similar parametric space of that of an existing small aspect ratio torus\cite{Lachhvani2017}. This finding also suggests the possibility of realizing such a quiescent pure ion plasma in toroidal equilibrium, which might provide excellent testbed/new directions for quantum information processing\cite{Landa2019,Morigi2001} in future.\

Coming back to electron plasmas, trapped pure electron plasmas are prone to loss of electrons due to various instabilities associated with the electron plasma dynamics. These instabilities are mainly resistive wall instability\cite{White1982}, electrostatic instability induced by electron-neutral collisions\cite{Davidson1996} and ion resonance instability\cite{Levy1969,Davidson1977,Fajans1993,Peurrung1993}. Most of these instabilities are subdued in devices with confining walls made of low resistivity materials and better vacuum conditions (even at low working pressure $\sim$ $10^{-9}$ torr\cite{Lachhvani2016} or better). Amongst these instabilities, ion resonance instability in cylindrical traps as well as in toroidal traps, is more frequent and predominant over the rest. Even in the low background pressure $\sim$ $10^{-9}$ torr, residual gases (mainly $N_2$ and $H_2$) remain in the system. Through the process of electron-impact ionization of these background natural gas, ions and secondary electrons are generated. It is believed that the ions are trapped in the device by the potential well of the electron plasma leading to ion resonance instability, inciting the destabilization of the electron plasma by the onset of an $m=1$ like toroidal Diocotron instability. To illuminate the underlying processes, various analytical models of this instability that were constructed in the past should be addressed first.\

 Initial analytical model of the cylindrical electron plasma in the presence of few ``trapped" ions was proposed by Levy et al.\cite{Levy1969} suggesting an exponential growth (i.e. mode amplitude is $\propto$ $exp(\gamma t)$, with $\gamma > 0$) of the $m=1$ Diocotron mode of the electron plasma. It was shown that the exact resonance condition arises if the oscillation frequency of the ion trapped in the potential well of the electron plasma matches the Diocotron frequency of the off centered electron plasma. In another simplified linear model developed by Davidson and Uhm\cite{Davidson1977} for infinitely long cylindrical electron plasma under weak neutralization by ``trapped" ions, collisionless rotating two stream instability between electron and ion plasma approach was adopted. The results of this model are found to be in good agreement with those proposed by Levy el al. showing the $m=1$ Diocotron mode of the electron plasma as fastest growing mode among other exponentially growing modes. Later, for a 3D straight cylindrical trap (i.e. PM trap) with electrostatic end plugs, ion resonance instability of the cylindrical electron plasma in the presence of axially drifting ions or ``transient" ions (due to rapid loss of ions in the axial direction as compared to an infinitely long cylinder) was investigated analytically by Fajans\cite{Fajans1993}. This theory suggested algebraic growth of the Diocotron instability, which was also reported to be confirmed in the experiment performed by Peurrung et al.\cite{Peurrung1993} under similar experimental conditions. Earlier, simulation based studies using 2D Particle-in-Cell(PIC) code PEC2PIC\cite{Sengupta2015,Sengupta2016,Sengupta2017}, have been performed addressing the ion resonance instability of cylindrically confined partially neutralized electron plasma, under different initial conditions. Through collisional and collisionless processes, the growth rate of different Diocotron modes were addressed where the results were found to be in good agreement with the Davidson and Uhm model. It was also shown that the non-ionizing collisions do not initiate instabilities in contrary to then existing theory\cite{Sengupta2016}.\ 

As of now, numerous experiments of cylindrical electron plasmas\cite{Peurrung1993,Eckhouse1981,Bettega2006,Kabantsev2007} have exhibited ion resonance instability under different configurations. These experiments show exponential\cite{Eckhouse1981,Kabantsev2007} or algebraic growth\cite{Peurrung1993,Bettega2006} depending on the nature of the ion entrapment. A toroidal electron plasma, is also greatly affected by the presence of ions in the system\cite{Lachhvani2016,Stoneking2002,Marksteiner2008}. Extensive investigation of the dependency of the growth on ion species, magnetic field strength and pressure has been performed in stellarator\cite{Marksteiner2008} trap under magnetic surface trapping. In SMARTEX-C\cite{Lachhvani2016} trap, it is reported that the improved vacuum condition of the device resulted in the damping of the growth of the instability.\

Recently, collisionless dynamics of toroidal electron plasma has been addressed in tight aspect ratio device (in a similar parametric space as SMARTEX-C) using the 3D3V Particle-in-Cell code PEC3PIC\cite{Khamaru2019}, for both axisymmetric\cite{Khamaru2019,Khamaru2021,Khamaru2021Er} and partial\cite{Sengupta2021} or nonaxisymmetric configurations with electrostatic end-plugs along the toroidal direction. Of late, the existence of a quiescent quasi-steady state (QQS)\cite{Khamaru2021,Khamaru2021Er} has been discovered, for the toroidal electron plasma in axisymmetric configuration using a maximum-entropy based zero-inertia solution as an initial condition to the PEC3PIC, which then solves the full inertia-based equation of motion of electrons. This slowly evolving state is shown to exhibit nearly free of center of charge motion of the naturally shaped toroidal electron plasma with finite parallel and perpendicular temperatures. We believe that the discovery of such a quiescent quasi-steady state of the toroidal electron plasma may become crucial to the future of quantum information processing\cite{Landa2019,Morigi2001} and quantum computing. As a subsequent step, collisionless ion-driven destabilization of the toroidal electron in QQS state has been investigated in this paper. In an ongoing study, impact ionization and various collisional mechanisms between electron-neutral, ion-neutral collisions have been considered and shall be reported in future.\  

In the present study, we have addressed the effect of ion resonance instability on the toroidal electron plasma in QQS state, via preloaded $Ar^+$ ion plasma for different fractional neutralization\cite{Davidson} values $f$ $=$ $n_i/n_e$ i.e. the ratio of ion density to electron density, in an axisymmetric tight aspect ratio device using PEC3PIC\cite{Khamaru2019,Khamaru2021,Khamaru2021Er,Sengupta2021}. It is shown that the electron plasma is destabilized and the wall probe current shows algebraic growth. It is found that the growth rate increases with increasing values of $f$. Along with ``center of charge motion" ($m$ $=$ $1$) of the electron plasma, mode coupling between toroidal Diocotron modes ($m$ $=$ $1$ to $9$) has been observed with $m$ $=$ $2$ as the dominant mode. The growth of the wall probe current later saturates as ion (and electron) losses increase with simulation time. It is shown that the calculated value of ion plasma rotational frequency of the $Ar^+$ ion plasma in cylindrical approximation is at $ \sim 31 \%$ deviation from the toroidal $m=1$ Diocotron\cite{Davidson} frequency of the electron plasma. Destabilization of the electron plasma and algebraic growth of the wall probe current is brought out unambiguously. It has been observed that the ion resonance instability of the electron plasma is possible for a certain range of the ion plasma rotational frequency (via adding $N^+$ and $H^+$ ions into the system and observing the electron plasma dynamics, though the results are not shown here). Collision-less heating of the ions has been observed resulting in high ion temperature values, which is interpreted as due to a resonant transfer of energy of the electron plasma to ions via toroidal ion-resonance instability.\ 

In Sec. \ref{sec:Initial condition and simulation parameters}, the toroidal device configuration is given and the initial conditions including the initialization steps of electron/ion plasma loading in the toroidal device are explained. Sec. \ref{sec:Diagnostics} various diagnostics implemented for this study are shown. In Sec. \ref{sec:Conclusions and Discussions} the conclusion of the present study is described.\  

\section{Initial condition and simulation parameters} 
\label{sec:Initial condition and simulation parameters} 
As discussed previously, 3D3V particle-in-cell\cite{Birdsall} code PEC3PIC\cite{Khamaru2019,Khamaru2021} has been used in the present study, addressing the dynamics of toroidal electron plasma in tight aspect ratio axisymmetric toroidal device (in a similar parametric space as SMARTEX-C\cite{Lachhvani2017}). Correct loading of the electron and ion plasma in this device is crucial for the desired experiments and needs careful study along with corresponding simulation parameters used in the study. In the next Subsec. \ref{subsec:Device parameters and initialization of the electron and ion plasma}, the device parameters and initialization process of the plasmas are explained in details.

\subsection{Device parameters and initialization of the electron and ion plasma}
\label{subsec:Device parameters and initialization of the electron and ion plasma}
 \begin{figure}[htp]
 \includegraphics[scale=0.35]{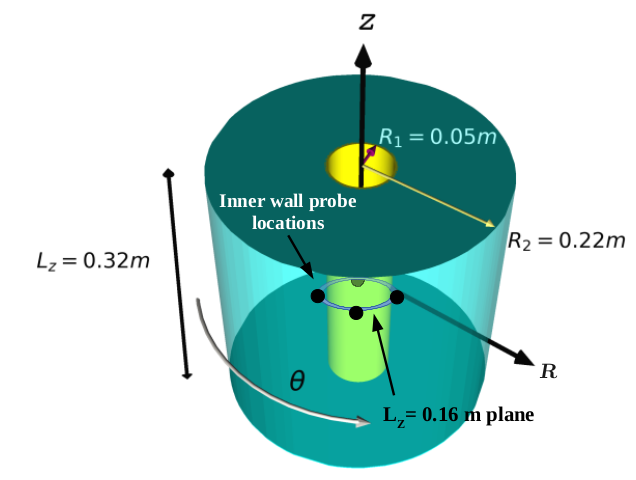}
 \caption{A schematic diagram of the axisymmetric toroidal vessel with conducting boundary walls. The radius of the inner wall boundary is $R_1$ $=$ $0.05~m$, outer wall is $R_2$ $=$ $0.22~m$ (aspect ratio $\sim$ 1.6) and the height is $L_z$ $=$ $0.32~m$. The vessel is closed by upper and lower boundary walls. Wall probes located near the inner boundary wall, separated by $\theta$ $=$ $\pi /2$ from each other are shown at $L_z$ $=$ $0.16~m$ mid-plane. The device sizes described are identical to SMARTEX-C\cite{Lachhvani2017} device. However unlike the experimental device, here electrostatic endplugs in the toroidal direction are absent, though the particle dynamics is fully 3D3V.}  
\label{fig:Fig1_3D_vessel_RGanesh}
 \end{figure}
The schematic diagram of the axisymmetric toroidal vessel used in this study is presented in Fig. \ref{fig:Fig1_3D_vessel_RGanesh} in cylindrical geometry. Inner wall radius of the torus is $R_1$ $=$ $0.05~m$ and the outer wall radius is $R_2$ $=$ $0.22~m$, making the device aspect ratio $\sim$ $1.6$. Height of the torus is $L_z$ $=$ $0.32~m$. The vessel is closed by upper and lower boundary walls. The boundary walls of the torus are conducting. Wall probes are located near the inner boundary wall, separated by $\theta$ $=$ $\pi /2$ locations at $L_z$ $=$ $0.16~m$ mid-plane are shown. The device size is identical to that of SMARTEX-C\cite{Lachhvani2017}.\

At background pressure of $10^{-9}$ to $10^{-7}$ torr\cite{Marksteiner2008,Lachhvani2016}, ions and secondary electrons are generated by the energetic primary electrons due to ionization of the background neutral atoms. In typical tight aspect ratio toroidal experiments\cite{Lachhvani2016} with pressure $10^{-8}$ torr, the ion build up is upto $\sim$ $2.5\%$ within $5 \times 10^{-4}$ sec time span. To investigate the effects of low density ion population on the primary electron plasma in our study, the primary electron plasma, $Ar^{+}$ ion plasma along with secondary electron plasma are loaded in the system in two consecutive steps, described next in detail.\

\begin{figure}[htp]
 \includegraphics[scale=0.22]{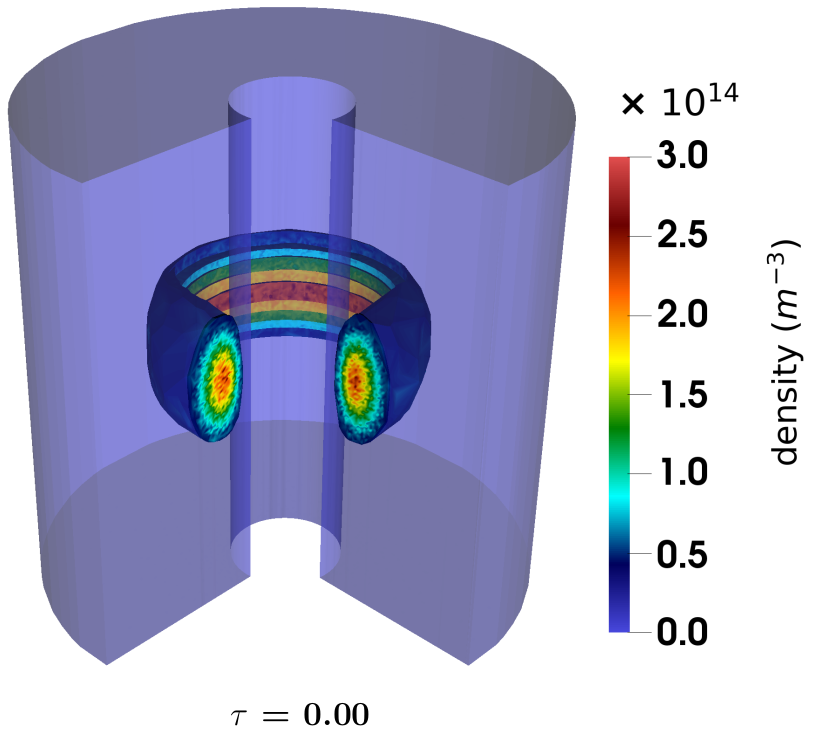}
 \caption{Initial density distribution of the primary electrons in the 3D toroidal vessel is shown at $\tau$ $=$ $0.00$, where $\tau$ $=$ $t/\overline{t}_D$ and $\overline{t}_D$ is toroidal Diocotron period\cite{Khamaru2019} ($\sim$ $1.8 \times 10^{-6} s$) for QQS state\cite{Khamaru2021,Khamaru2021Er}. Only $3/4 th$ of the toroidal direction is shown to present the poloidal density contours along with the toroidal density contours. The initial density of the electron plasma has the radial variation of two orders of magnitude, viz $ \sim 7.1 \times 10^{12} m^{-3} - 2.485 \times 10^{14} m^{-3}$ i.e. $f_b \sim 0.002 - 0.07$, where $f_b \sim 0.07$ corresponds to the mode (the locally maximum value of the density) of the total density distribution at $z$ $=$ $L_z/2$ plane ($0.16~m$) of the torus. Few electrons are present outside the contour region making the density negligible in that area. In the absence of ions, this QQS state is shown to suffer very little radial transport\cite{Khamaru2021,Khamaru2021Er}.}  
\label{fig:Fig2_3D_electron_RGanesh}
\end{figure}

 \begin{figure*}[htp]
 \includegraphics[scale=0.25]{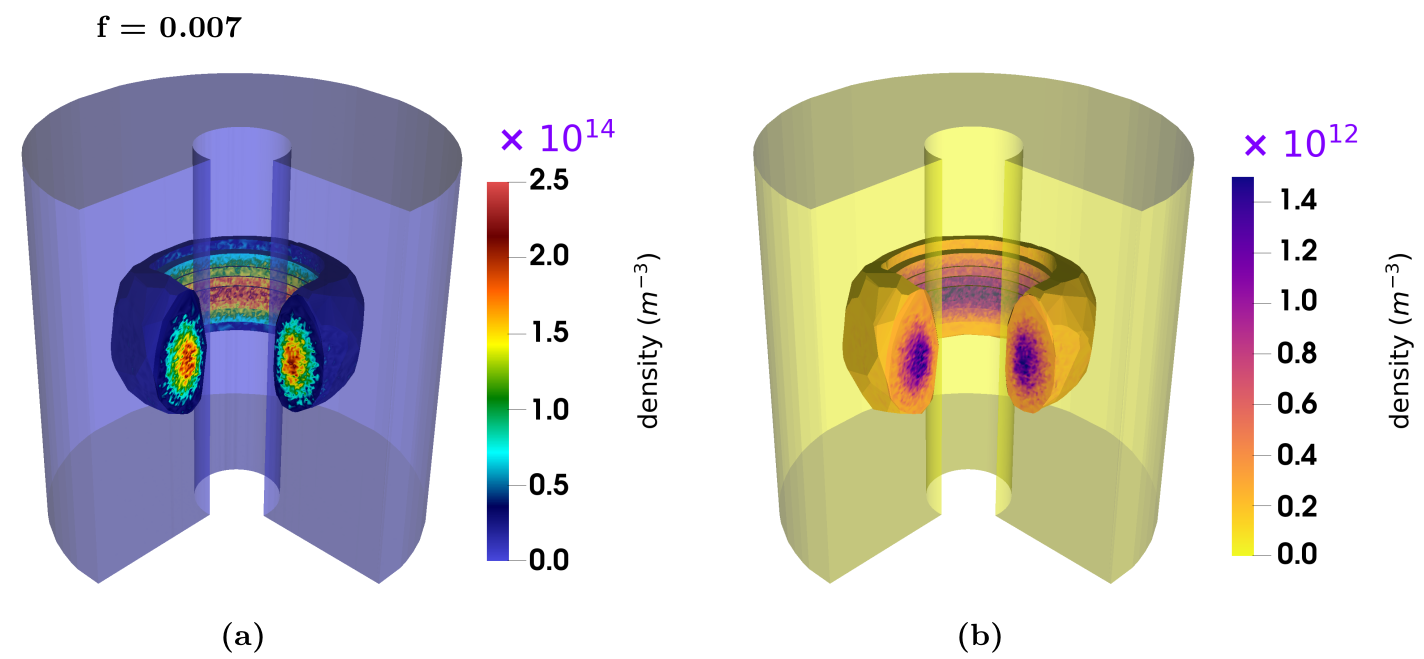}
 \caption{(a)The density distributions of primary electron plasma at $\tau$ $=$ $11.11$, where $\tau$ $=$ $t/\overline{t}_D$ and $\overline{t}_D$ is toroidal Diocotron period for the QQS state. The density variation of the electron plasma is in the range $\sim 6.83 \times 10^{12} m^{-3} - 2.39 \times 10^{14} m^{-3}$ i.e. $f_b \sim 0.0019 - 0.067$. Here $f_b \sim 0.067$ corresponds to the mode of the total electron density distribution at $z$ $=$ $L_z/2$ plane of the torus. (b) The density distribution of $Ar^{+}$ ion plasma at $\tau$ $=$ $11.11$ for fractional neutralization factor $f$ $=$ $0.007$ which corresponds to the ratio of ion density to the electron density i.e. $n_i/n_e$. The density of the ion plasma is in the range $ \sim 4.78 \times 10^{10} m^{-3} - 1.67 \times 10^{12} m^{-3}$. Here $1.67 \times 10^{12} m^{-3}$ corresponds to the mode of the total ion density distribution at $z$ $=$ $L_z/2$ plane of the torus.}  
\label{fig:Fig3_3D_RGanesh}
 \end{figure*}

As described in Ref.45\cite{Khamaru2021}, in the first step, zero-inertia maximum entropy density distribution function\cite{Khamaru2021} is used as the ``seed" particle density profile of the primary electron plasma in the system at the beginning of the simulation which has been previously shown to result in a novel quiescent quasi-steady state (QQS)\cite{Khamaru2021} of the electron plasma. Thus the initial spatial density distribution of the electron plasma, following this initial ``seed" solution, is shown to evolve into an axisymmetric ($\partial / \partial \theta = 0$) nonuniform toroidal plasma plasma which now is accurate to all orders in ${\rho _{\strut{Le}}} /L$, where ${\rho _{\strut{Le}}}$ is the average electron Larmor radius and $L$ is a typical magnetic field gradient length scale. Corresponding density distribution of the electrons in the 3D toroidal vessel is shown in the Fig. \ref{fig:Fig2_3D_electron_RGanesh} at $\tau$ $=$ $0.00$, where $\tau$ $=$ $t/\overline{t}_D$ and $\overline{t}_D$ is toroidal Diocotron period\cite{Khamaru2019} $\sim$ $1.8 \times 10^{-6} s$ for the QQS\cite{Khamaru2021,Khamaru2021Er} state of the electron plasma. The method to obtain $\overline{t}_D$ was elaborated in a previous study\cite{Khamaru2019}. The quantity toroidal Diocotron period is explained later in Sec. \ref{sec:Diagnostics}.\ 

The electrons are cold loaded with zero velocities. This method of initial loading is in general different than the conventional electron injection technique used in typical toroidal pure electron plasma experiments\cite{Lachhvani2017,Stoneking2004}, where a negatively biased injector grid filament emit the electrons in toroidal direction, filling the toroidal vessel via another positive biased grid placed ``nearby", followed by axial confinement of electrons by negative end plugs. This procedure results in a toroidal or parallel velocity such that toroidal or parallel bounce time is much smaller than the typical $\bf{E} \times \bf{B}$ time scale. The parallel kinetic energy of electrons often ionize the weak background neutrals, giving rise to ions. In the present study, zero-inertia maximum entropy distribution results in nonuniform density distribution of the electrons featuring contoured structures with peaked centers in toroidal and poloidal planes, as shown in the Fig. \ref{fig:Fig2_3D_electron_RGanesh}. Though the electrons fill up the whole torus, only $3/4 th$ of the whole torus is shown in this figure (Fig. \ref{fig:Fig2_3D_electron_RGanesh}) to present the poloidal density contours along with the toroidal density contours. This configuration of the density distribution will be used throughout this paper. The initial density of the electron plasma is in the range $ \sim 7.1 \times 10^{12} m^{-3} - 2.485 \times 10^{14} m^{-3}$ i.e. $f_b \sim 0.002 - 0.07$. Here $f_b$ is the ``cylindrical" Brillouin fraction\cite{Davidson} ratio, $f_b=2{\omega_p}^2/{\omega_c}^2$, ($\omega_p$ is the plasma frequency and $\omega_c$ is the cyclotron frequency for magnetic field $B_0$ ($0.03 T$) at inner wall radius). (Inferred range of $f_b$ is in the range $f_b \sim 10^{-3} - 10^{-4}$ in the experiments\cite{Lachhvani2016}, which is typically an order of magnitude lower than our simulation parameters.) As there are very few electrons in the outer regions of the plasma, the density contours are seen to sharply fall to zero in this region. The electron plasma attains QQS state shortly after the initialization and is left to evolve upto $\tau$ $=$ $11.11$.\

In the second step, the $Ar^{+}$ ions are introduced in the system at $\tau$ $=$ $11.11$ (which corresponds to $t$ $=$ $2 \times 10^{-5}$ s) according to same spatial distribution of QQS as the primary electrons have evolved into. Though at $\tau$ $=$ $11.11$ the electrons acquire certain finite width of velocity distribution function, ions are loaded cold at this instant. In cylindrical experiments\cite{Peurrung1993,Bettega2006,Kabantsev2007} in PM traps with ion resonance instability, the values of the fractional neutralization factor\cite{Davidson} $f$, which corresponds to the fraction of ion population to the primary electron population i.e. $n_i/n_e$, is reported as typically $\sim$ $10^{-5}$ to $10^{-4}$. For partial toroidal experiments\cite{Marksteiner2008,Lachhvani2016}, $f$ $\sim$ $10^{-2}$ to $10^{-1}$. In our axisymmetric toroidal simulation, $f$ values are chosen between $0.001$ and $0.01$, which is in the range of experimental estimates. The 3D density distributions of primary electron plasma and $Ar^{+}$ ion plasma at $\tau$ $=$ $11.11$, is shown in the Fig. \ref{fig:Fig3_3D_RGanesh} (a) and (b) respectively for $f$ $=$ $0.007$. In Fig. \ref{fig:Fig3_3D_RGanesh} (a), the density of the electron plasma is in the range $\sim 6.83 \times 10^{12} m^{-3} - 2.39 \times 10^{14} m^{-3}$ i.e. $f_b \sim 0.0019 - 0.067$. Fig. \ref{fig:Fig3_3D_RGanesh} (b), the density of the ion plasma is in the range $ \sim 4.78 \times 10^{10} m^{-3} - 1.67 \times 10^{12} m^{-3}$. The ion densities used are two orders of magnitude smaller than the electron density at any given spatial location.\

\begin{figure*}[htp]
 \includegraphics[scale=0.4]{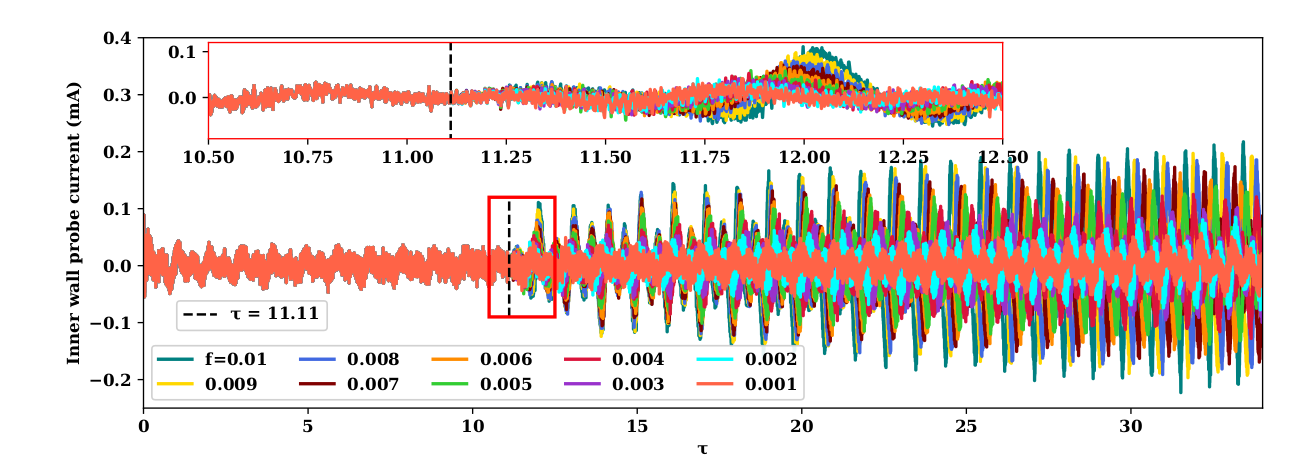}
 \caption{Inner wall probe current for fractional neutralization factor $f$ (i.e. $n_i/n_e$) ranging from $0.001$ to $0.01$ for $Ar^{+}$ upto $\tau$ $=$ $34$, where $\tau$ $=$ $t/\overline{t}_D$ and $\overline{t}_D$ is toroidal Diocotron period for QQS state. The time of introduction of ions is at $\tau$ $=$ $11.11$, as indicated by black dotted line. The inset plot shows onset of the growth of wall probe current amplitude.}  
\label{fig:Fig5_probe_current_RGanesh}
 \end{figure*}

Along with the ions, cold (zero velocity) secondary electrons are also loaded at $\tau$ $=$ $11.11$ with density value $f$ times the density of the primary electrons i.e. the density of secondary electron plasma is same as the ion plasma loaded. This method mimics the ionization process in the experiments and conserves the total charge of the system at $\tau$ $=$ $11.11$. Total energy of the system is not effected by addition of the secondary electron population/ion population because of the low density $\sim$ $10^{-3}$ to $10^{-2}$. After the loading processes, this nonneutral plasma consisting of the primary and secondary electrons along with ions is left to evolve and the dynamics is monitored through the simulation time period upto $34$ $\overline{t}_D$. In the next Subsec. \ref{subsec:Simulation Parameters}, the simulation parameters are described.\\
\subsection{Simulation Parameters}
\label{subsec:Simulation Parameters}

In this study, 3D3V OPEN-MP parallelized particle-in-cell\cite{Birdsall} (PIC) code PEC3PIC\cite{Khamaru2019,Khamaru2021} is used. More details on the code and the algorithm can be found in the recent QQS study\cite{Khamaru2021} of electron plasmas where same PIC code has been used. Here initial super particle number for the primary electron plasma is 2560640 in a grid size of $192\times 192\times 192$ in Cartesian coordinates. At $\tau$ $=$ $11.11$, super particle number for the primary electron plasma becomes 2464401 due to initial loss of the electrons. Also initial super particle numbers for the ion plasma and secondary electron plasma are 2464401 each at $\tau$ $=$ $11.11$. The density values of the ion plasma and secondary electron plasma are made $f$ times lesser than the primary electron plasma using the fractional neutralization factor $f$ $=$ $0.007$ in the code. At $\tau$ $=$ $11.11$, most of the electron and ion particles are confined in toroidal tube with high density value which has lesser volume than the whole torus, as shown in the Fig. \ref{fig:Fig3_3D_RGanesh}. Thus the effective number of particles (electrons and ions) per 3D cell of this toroidal tube is $\sim 10$ and the number of particles per 3D cell of the whole toroidal simulation device is $\sim 3.4$. Numerical conservation tests for the current PEC3PIC solver has been performed in a previous study\cite{Khamaru2019}. $Ar^{+}$ ion used in the study have mass $72820.77$ $m_e$ where $m_e$ is the electron mass.\
 
The toroidal magnetic field used in this device is spatially inhomogeneous. At inner wall radius $R_1$, the magnitude of the magnetic field is highest ($B_0 = 0.03 T$) and gradually falls with cylindrical radial coordinate $R$ as $B = {B_0R_1} / R$. At $R = R_1$ the estimated electron cyclotron time period $T_{ce} = 1.19 \times 10^{-9} s$ and the ion cyclotron time period $T_{ci} = 8.68 \times 10^{-5} s$. As is obvious, the cyclotron periods are smallest at the largest magnetic field in the system, which is at the inner wall at $R = R_1$. The simulation time step chosen is $10^{-10} s$ which is about 10 times smaller than the smallest cyclotron time in the system.\

The parameters used here are typical of recent experiments \cite{Lachhvani2016}, though any collisional processes are not included in the present study. As neutrals are not present in our simulation, collisional processes with neutrals are exempted as well. Ion-ion collision are also exempted because of their mass and extremely low density. Other possible processes are the electron-electron and electron-ion collisions. Here we have estimated the respective collisional parameters for electron-electron collision and electron-ion collision (for $Ar^{+}$), $f=0.007$ case as follows: Large angle electron-electron Coulomb collisions time scale\cite{Huba2013} ${\tau}_{col}^{e}$ $\simeq$ $\frac{{{T_e}^{3/2}}}{2.91 \times 10^{-6} ~n_e ln \Gamma }$ sec $\simeq$ $0.099$ sec i.e. $\overline{\tau}_{col}^{e}$ $\simeq$ ${{\tau}_{col}^{e}}/{\overline{t}_D}$ $\simeq$ $5.5$ $\times$ $10^{4}$. Electron-ion Coulomb collisions time scale ${\tau}_{col}^{i}$ $\simeq$ $\frac{{{T_i}^{3/2}}}{4.8 \times 10^{-8} ~n_i \mu Z_a^4 ln \Gamma }$ sec $\simeq$ $141.44$ sec i.e. $\overline{\tau}_{col}^{i}$ $\simeq$ $7.86$ $\times$ $10^{7}$. (In the formulae for collision times, temperature is in~eV units, whereas all the other quantities are in CGS units). Consequently, ${\overline{t}_D} \sim 1.8 \times 10^{-6} s$ is the toroidal Diocotron time period, $ln \Gamma$ $\sim$ 10, $T_{max}^{e}$ $\sim$ $80.0~eV$, $n_e$ $\sim$ $2.485 \times 10^{8} {cm}^{-3}$, $n_i$ $\sim$ $1.74 \times 10^{6} {cm}^{-3}$, $T_{max}^{i}$ $\sim$ $280.0~eV$, atomic number $Z_a$ $=$ $1$, ratio of ion mass to proton mass $\mu$ $=$ $\frac{m_i}{m_p}$ $=$ 39.66. In the present study, total simulation time is $\tau$ $=$ $t/\overline{t}_D$ $=$ $34$, where $t$ is time. Both electron-electron Coulomb collisions time scale and electron-ion collisions time scale are larger than the present simulation time scale. Thus the plasma is considered as collision-less plasma.\

 \begin{figure*}[htp]
 \includegraphics[scale=0.45]{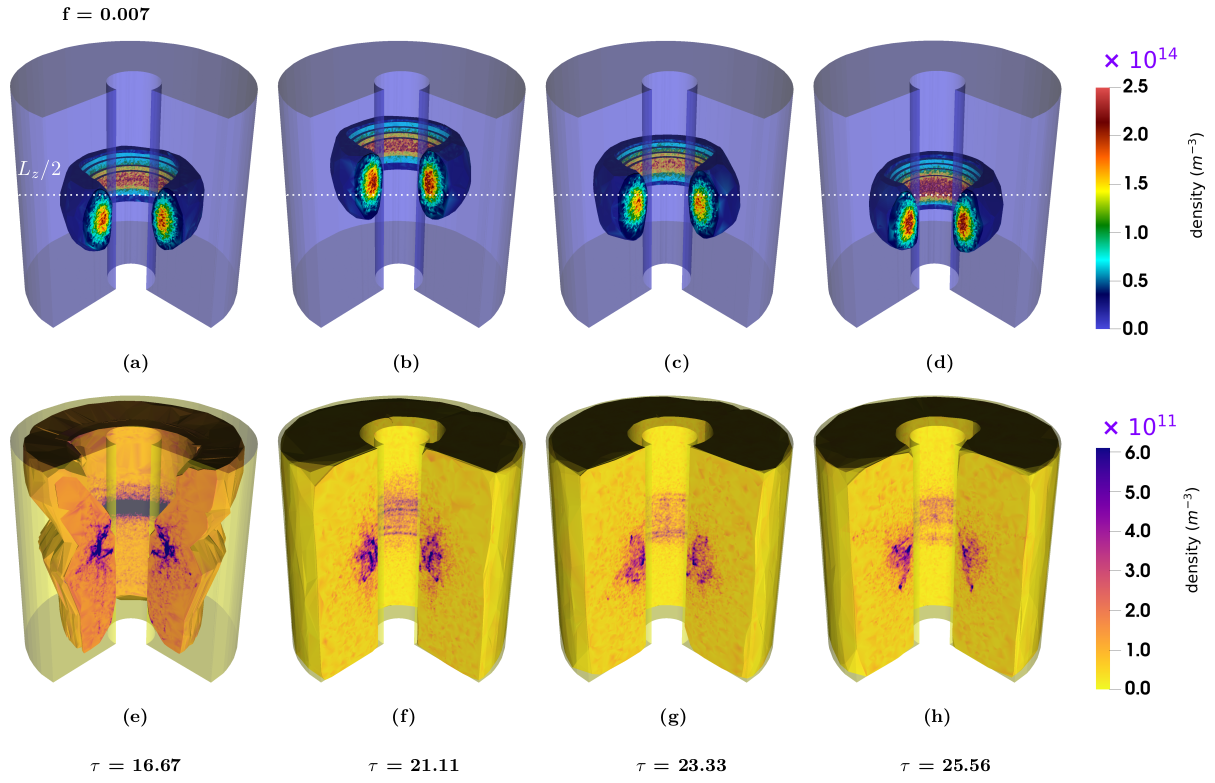}
 \caption{Time evolution of primary electron plasma and ion plasma with respective density values for $f$ $=$ $0.007$ at different simulation time periods $\tau$ $=$ $16.67$, $21.11$, $23.33$ and $25.56$, where $\tau$ $=$ $t/\overline{t}_D$ and $\overline{t}_D$ is toroidal Diocotron period for QQS state. The density contours of electron plasma along with the spatial position of the plasma is shown in (a)-(d). These figures show displaced electron plasma with ``center of charge motion" ($m$ $=$ $1$ mode) in the vertical and horizontal direction, with distinguishable compressed and expanded forms of the plasma along with periodic peak density value evolution. The time evolution of the ion plasma and density is shown in (e)-(h). (e) at $\tau$ $=$ $16.67$ shows that the initially cold loaded ion plasma starts to gain energy from the system and initial shape of the plasma (Fig. \ref{fig:Fig3_3D_RGanesh} (b) and (d)) at $\tau$ $=$ $11.11$) is distorted. In (f)-(h), at $\tau$ $=$ $21.11$, $23.33$, $25.56$ respectively, the ion plasma spreads out into the torus representing rapid loss form the torus boundaries. This loss of ions is also reflected in decrement of the peak poloidal and toroidal density values of the ion plasma through the simulation time.}  
\label{fig:Fig4_3D_time_evolution_RGanesh}
 \end{figure*}
 
 \begin{figure*}[htp]
 \includegraphics[scale=0.35]{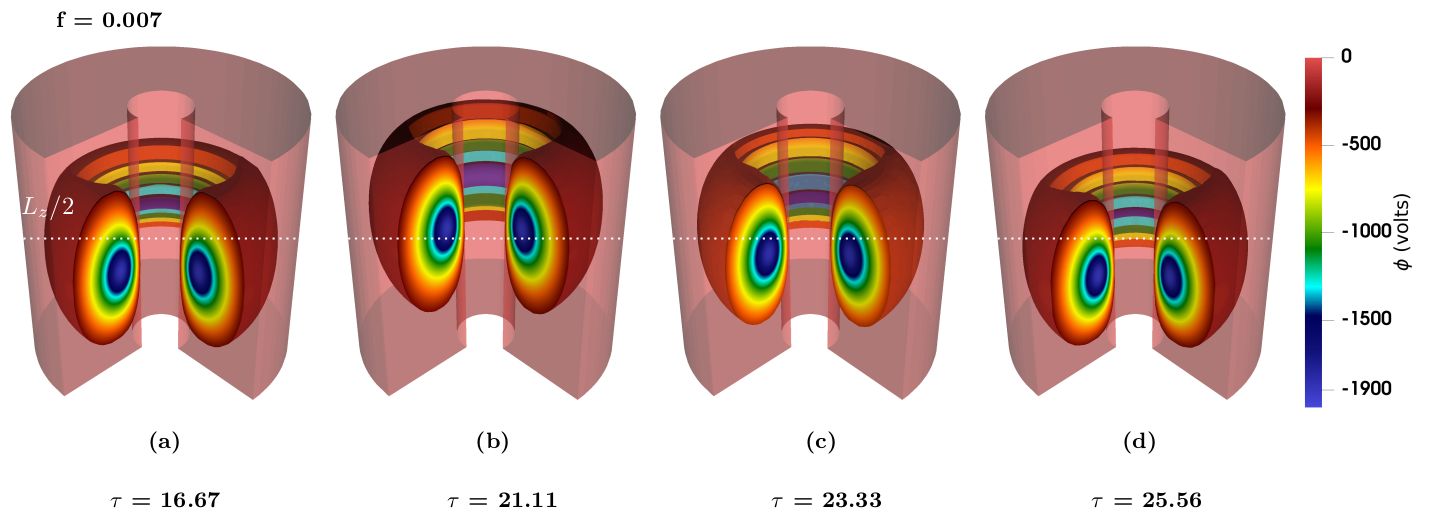}
 \caption{(a)-(d) shows electrostatic potential surface plots of the system for $f$ $=$ $0.007$ at different simulation time periods $\tau$ $=$ $16.67$, $21.11$, $23.33$ and $25.56$, where $\tau$ $=$ $t/\overline{t}_D$ and $\overline{t}_D$ is toroidal Diocotron period for QQS state. The potential surface plots display closed axisymmetric toroidal potential well of the system, with spatial location of the well minima around the same spatial location of the electron plasma density maxima (Fig. \ref{fig:Fig4_3D_time_evolution_RGanesh} (a)-(d)).}  
\label{fig:potential}
 \end{figure*}

Wall probes similar to experimental devices are placed at four different azimuthal locations (Fig. \ref{fig:Fig1_3D_vessel_RGanesh}) (separated by $\theta$ $=$ $\pi /2$) near the inner wall at $z$ $=$ $L_z/2$ plane. Wall probe current obtained from the inner wall probe is used as a diagnostic method to analyse the electron plasma dynamics in the vessel. This ``wall probe current" depends mainly on the charge induced by center of charge motion of the electron plasma and expressed as $I_{wall} =  \frac{d(CV)}{dt}$. Here $V (t)$ and $C (t)$ are the instantaneous probe voltage and capacitance respectively. Thus the charge induced on the probe is $C(t)V(t)$ at that instance. In this work, a constant capacitance value is chosen to be $C =0.08 pF$ so that the obtained order of magnitude of probe current is similar to typical experimental values \cite{Ganesh2006}.\

In the next Sec. \ref{sec:Diagnostics}, the effect of the ion resonance instability on electron dynamics is investigated via different diagnostics methods: (i) Growth in wall probe current and dependency on $f$, electron/ion plasma spatial evolution in the torus and corresponding density evolution, spectrogram analysis of the wall probe current and electrostatic potential of the system, cylindrically approximated ion plasma rotational frequency and corresponding toroidal $m=1$ Diocotron frequency of the electron plasma (ii) electron/ion particle loss in the system (iii) energy of the electron/ion plasma (iv) growth rate of the wall probe current at different periods of the simulation, and (v) temperature analysis of the electron and ion plasmas.\\

\section{Diagnostics}
\label{sec:Diagnostics}
As indicated earlier, to investigate the electron and ion plasma dynamics, we have implemented several diagnostics suitable for this study. The diagnostics results covers the simulation time period $\tau$ $=$ $0.00$ to $\tau$ $=$ $34$. 
\subsection{Growth in wall probe current and dependency on $f$}
\label{subsec:Growth in wall probe current and dependency on $f$}

To study the electron dynamics in the presence of ions in the system, wall probe signal (wall probe voltage in Volts) is obtained as a function of time from one of the inner wall probe and the corresponding wall probe current is shown in Fig. \ref{fig:Fig5_probe_current_RGanesh}. Wall probe current from three other probes at three remaining toroidal locations provide identical information which suggests absence of any toroidal mode in the system (not shown here). In Fig. \ref{fig:Fig5_probe_current_RGanesh}, wall probe current growth is observed for $f$ values ranging from $0.001$ to $0.01$ upto $\tau$ $=$ $34$. As discussed earlier, at $\tau$ $=$ $0.00$, a cold electron plasma density distribution obtained by extremizing inertia-less entropy is initialized\cite{Khamaru2019,Khamaru2021}, which evolves self consistently into a quiescent quasi-steady (QQS) state, as reflected in the wall probe data shown in Fig. \ref{fig:Fig5_probe_current_RGanesh}. As discussed earlier, ions are then introduced ``by hand" or ``pre-loaded" at $\tau$ $=$ $11.11$ (along with equal number of secondary electrons, as it would happen in an ionizing process), soon after which, the amplitudes of the probe currents start to grow with time and the growth is more for higher values of $f$ for any instant of time. This is clearly demonstrated in the inset plot displayed at the onset of the growth. Also an initial phase difference between the probe currents for different values of $f$ can be observed. The growth of the wall probe current is algebraic in nature for all values of $f$ and saturates at later time of the simulation, mainly for higher $f$ values. More analysis on the growth rate is performed in Subsec. \ref{subsec:Growth rate of the wall probe current at different periods of the simulation}.\

To aid the above analysis, spatial positions of the electron plasma inside the torus along with density contours are shown in Fig. \ref{fig:Fig4_3D_time_evolution_RGanesh} (a)-(d), at simulation times $\tau$ $=$ $16.67$, $21.11$, $23.33$, $25.56$ respectively with $f$ $=$ $0.007$ case. Before $\tau$ $=$ $11.11$, the electron plasma has attained a quiescent state with negligible displacement, shown in previous study of QQS state\cite{Khamaru2021,Khamaru2021Er}. After ions were introduced at $\tau$ $=$ $11.11$, the ``center of charge motion" ($m$ $=$ $1$) of the electron plasma in the poloidal plane gains amplitude as time progresses. The electron plasma is displaced along the vertical and horizontal directions, with a distinguishable compression-expansion cycle due to toroidicity. The peak density values at the toroidal/poloidal plane evolve periodically along with the ``center of charge motion" of the plasma. In these figures, the poloidal shape of the plasma shows elliptic nature representing $m$ $=$ $2$ mode. Referring to Fig. \ref{fig:Fig5_probe_current_RGanesh}, the probe currents attain ``double peak" nature which actually corresponds to the ``center of charge motion" ($m$ $=$ $1$) of the electron plasma in the poloidal plane and onset of strong compression-expansion cycle. Due to this center of charge motion, electron plasma makes a finite amplitude elliptical trajectory in the poloidal plane, which we dub here as nonlinear toroidal Diocotron motion and effectively similar to the cylindrical $m$ $=$ $1$ Diocotron motion of the electron plasma with a remarkable difference that at such small aspect ratios as studies here, the modes are coupled in a poloidal plane due to strong toroidicity or $1/R$ dependence of the toroidal magnetic field. Among these double peaks, the higher amplitude peak corresponds to the electron plasma position near the inner wall of the torus with most compressed form of the plasma and the lower amplitude peak appears while the plasma is away from the inner wall in most expanded form\cite{Khamaru2019}. This can be seen from the Fig. \ref{fig:Fig5_probe_current_RGanesh}, (i) the growth rate of the amplitudes of the peaks increases with time for every $f$ values (for higher $f$ values growth with time is more) and (ii) at any instant of time, the amplitude values of the peaks depend on $f$ values i.e. with high value of $f$ the peak amplitude is higher. This brings out clearly the fact that with higher ion density the ``center of charge" motion becomes stronger, which means an extended elliptical trajectory of the electron plasma from the QQS state, in the poloidal plane of the torus. This displacement of the electron plasma increases rapidly with time for higher values of the ion density. These signatures point out clearly that the physics of destabilization of QQS state is strongly connected to the presence of ions and its amplitude to the initial fraction of ions $f$ present in the system. Following the above said arguments, we believe that the wall probe growth or amplitude increase is indeed due to a toroidal ion resonance instability.\

\begin{figure*}[htp]
 \includegraphics[scale=0.35]{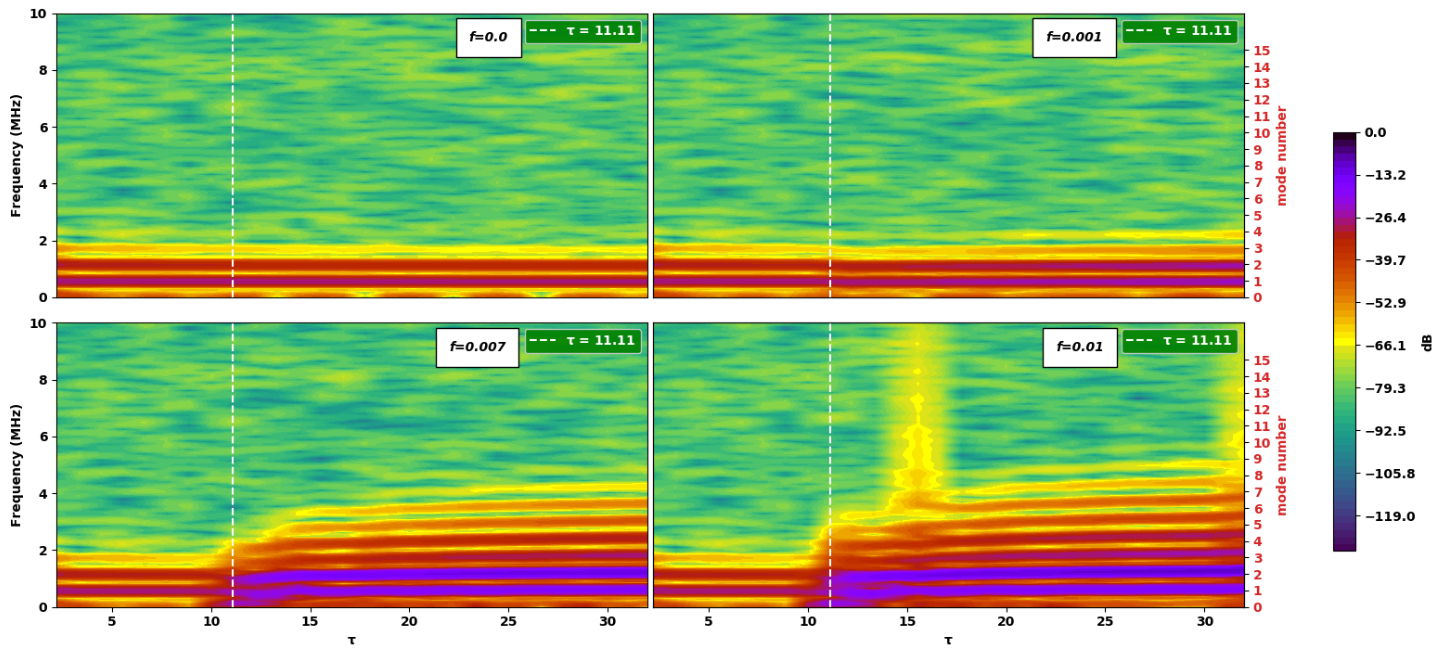}
 \caption{Spectrogram of the wall probe current for $f$ $=$ $0.0$ (QQS state), $0.001$, $0.005$, $0.01$ upto $\tau$ $=$ $34$, where $\tau$ $=$ $t/\overline{t}_D$ and $\overline{t}_D$ is toroidal Diocotron period for QQS state. White dotted line represents the introduction time of the ion plasma at $\tau$ $=$ $11.11$. Left y axis presents the frequency of the data where as in the right y axis the frequency is normalized by the nonlinear toroidal Diocotron frequency ($\sim$ $0.58$ MHz) of the QQS state. For all values of $f$, without the ion plasma before $\tau$ $=$ $11.11$, $m$ $=$ $1$ and $m$ $=$ $2$ Diocotron modes are present in the system. For $f$ $=$ $0.001$, $0.005$ and $0.01$ cases higher modes (upto $m$ $=$ $9$ for $f$ $=$ $0.01$) start to appear with time and $m$ $=$ $2$ mode becomes dominant. The temporal evolution of the spectrogram plots for $f$ $=$ $0.001$, $0.005$ and $0.01$ cases reveal the ion density dependency of the dominant modes ($m$ $=$ $1$ and $m$ $=$ $2$) and dynamical chirping infused coupling of low poloidal modes ($m$ $=$ $1$ to $m$ $=$ $9$). For $f$ $=$ $0.01$ case, a high chirping region between $\tau$ $\sim$ $15$ and $16$ is noted.}  
\label{fig:spectrogram_for_paper}
 \end{figure*}
 
The time evolution of the ion plasma and density is shown in Fig. \ref{fig:Fig4_3D_time_evolution_RGanesh}. At $\tau$ $=$ $16.67$ (see Fig. \ref{fig:Fig4_3D_time_evolution_RGanesh} (e)) the initially cold loaded ion plasma starts to gain energy from the system and initial shape of the plasma (Fig. \ref{fig:Fig3_3D_RGanesh} (b) at $\tau$ $=$ $11.11$) is distorted. In Fig. (f)-(h), at $\tau$ $=$ $21.11$, $23.33$, $25.56$ respectively, the plasma spreads out into the torus representing rapid loss of ions from the torus boundaries. Initial loss of the ions is higher at the top boundary wall and at later simulation times from other boundary walls also. This loss of ions is also reflected in decrement of the peak poloidal and toroidal density values of the ion plasma through the simulation time. Also rapid ion loss is discussed in the next Subsec. \ref{subsec:Loss of electron and ion}. The cold loaded ion plasma gains enough kinetic energy from the electron plasma as a result of the ion resonance instability. As our simulation is collisionless, we believe that the underlying exchange of energy between ions and electrons is a collisionless resonant process. The thermal energy gain of the ions results in increase of ion Larmour radii (for example, at perpendicular ion temperature $50~eV$, average Larmour radius of an ion is ${\rho _{\strut{Li}}}$ $\sim$ $0.214~m$ near the inner wall location and ${\rho _{\strut{Li}}}$ $\sim$ $1.073~m$ near outer wall location). Also various inertia driven drifts due to the toroidal inhomogeneous magnetic fields ($\bm{\nabla}B$ and curvature drift) aid the loss of ions. Thus the ions are rapidly lost from the system. The energy transfer process is investigated further in the Subsec. \ref{subsec:Energy variation with simulation time}.\

Corresponding to the presented spatial locations and density plots of the electron and ion plasma, electrostatic potential surface plots of the system are shown in Fig. \ref{fig:potential} (a)-(d). The potential surface plots display closed axisymmetric toroidal potential well of the system, with spatial location of the well minima around the same spatial location of the electron plasma density maxima (Fig. \ref{fig:Fig4_3D_time_evolution_RGanesh} (a)-(d)). This potential well has parabolic nature in the poloidal plane and the ions are initially trapped in the potential well. Though the location of the maxima of the ion plasma density is near to that of the electron plasma/potential minima at $\tau$ $=$ $16.67$, after that the ion plasma density maxima do not exhibit the same. This observation suggests the decoupling of the ion plasma from the electron plasma at later simulation time period after $\tau$ $=$ $16.67$. More analysis on this is presented in Subsec. \ref{subsec:Energy variation with simulation time}.\

To further understand the dynamics of the dynamic toroidal Diocotron modes of the system and possible mode coupling associated with the electron and ion plasma dynamics, spectrogram analysis of the wall probe current has been performed for $f$ $=$ $0.0$ (QQS state), $0.001$, $0.005$, $0.01$ upto $\tau$ $=$ $34$ and shown in Fig. \ref{fig:spectrogram_for_paper}. Spectrogram frequency is normalized by the nonlinear toroidal Diocotron frequency ($\sim$ $0.58$ MHz) of the QQS state and shown in the right y axis. For all values of $f$ considered, without the ion plasma before $\tau$ $=$ $11.11$, $m$ $=$ $1$ and $m$ $=$ $2$ Diocotron modes are the only dominant ones that are present in the system. For $f$ $=$ $0.001$, $0.005$ and $0.01$ cases higher modes (upto $m$ $=$ $9$ for $f$ $=$ $0.01$) start to appear with time with introduction of ions at $\tau$ $=$ $11.11$ and $m$ $=$ $2$ mode becomes dominant. The temporal evolution of the spectrogram plots for $f$ $=$ $0.001$, $0.005$ and $0.01$ cases reveal the ion density dependency of the dominant modes ($m$ $=$ $1$ and $m$ $=$ $2$) and dynamical chirping infused coupling of low poloidal modes ($m$ $=$ $1$ to $m$ $=$ $9$). Similar feature of the poloidal modes with $m$ $=$ $2$ as dominant mode, was also seen in our previous study\cite{Khamaru2019} with an initial loading of the electron plasma while the electron plasma had stronger ``center of charge" motion than that of the QQS state. Also for $f$ $=$ $0.01$ case, a high chirping region at very low power has been observed between $\tau$ $\sim$ $15$ and $16$.\

\subsection{Ion resonance instability and ion plasma rotational frequency}
\label{subsec:Ion resonance instability and ion plasma rotational frequency}
In straight cylinder geometry\cite{Davidson}, ion plasma rotational frequency in a partially neutralized electron plasma is given by

{\centering
\begin{equation}
    \omega_{ri}^{-} = - \frac{\omega_{ci}}{2}[1- (1+\frac{2 \omega_{pe}^2}{\omega_{ce}^2}(1-f)\frac{m_i}{m_e})^{1/2}] 
\end{equation}
}where $\omega_{ci}$ and $\omega_{ce}$ are the cyclotron frequency of ion and electron respectively at uniform $B$ field, $\omega_{pe}$ is the electron plasma frequency, $m_i$ and $m_e$ are the ion mass and electron mass respectively. The condition for ion resonance instability\cite{Davidson} of the electron plasma in cylindrical geometry is that the ion plasma rotational frequency should be nearly equal to the cylindrical $m=1$ Diocotron frequency of the electron plasma. In the present study with radially varying toroidal magnetic field, the an approximated ion plasma rotational frequency has been considered to compare with the toroidal $m=1$ Diocotron rotation of the electron plasma, as discussed in several ion resonance instability studies\cite{Marksteiner2008,Lachhvani2016}. With the present parameters for $Ar^+$ ions, the ion plasma rotational frequency (magnetic field at the minor axis location of the torus $B = 0.011T$) $\sim$ 0.4 MHz under cylindrical approximation where ``measured" toroidal Diocotron frequency of the electron plasma $\sim$ 0.58 MHz in our system. To the lowest order in inverse aspect ratio, thus the ion rotational frequency is at $ \sim 31 \%$ deviation from toroidal Diocotron frequency of the electron plasma. As the electron plasma and ion plasma dynamics are in toroidal geometry with inhomogeneous magnetic field, the presented deviation should be acceptable, if a permissible range of the deviation can be obtained. To estimate the permissible range of the deviation, $N^+$ ($m_i = 25532.65 m_e$) and $H^+$ ($m_i = 1836 m_e$) ions are also introduced in the device replacing $Ar^+$ ions to observe the electron plasma dynamics. For $N^+$ and $H^+$ ions, corresponding ion plasma rotational frequencies are $\sim$ 0.67 MHz ($ \sim 9 \%$ deviation) and $\sim$ 2.4 MHz ($ \sim 182 \%$ deviation) respectively. Algebraic growth of the wall probe current similar to ${Ar}^+$ ions has been observed for $N^+$ ions where as for $H^+$ ions, no growth is observed (the results are not shown here). These findings indicate the dependency of the ion resonance instability on the ion mass. Thus we can conclude that the permissible value of the ion plasma rotational frequencies to initiate an ion resonance instability of the electron plasma, for the current simulation parameters, lies between a certain range. Further analysis in the present study mainly focused on the ${Ar}^+$ ions and detailed investigation of the ion plasma rotational frequency range can be performed in future with different ion species.\

To investigate the ion resonance induced instability in this system in more detail, loss of electron and ion particles and temporal evolution of the ion/electron plasma's energy with simulation time, are described next.

\subsection{Loss of electron and ion}
\label{subsec:Loss of electron and ion}

 \begin{figure}[htp]
 \includegraphics[scale=0.28]{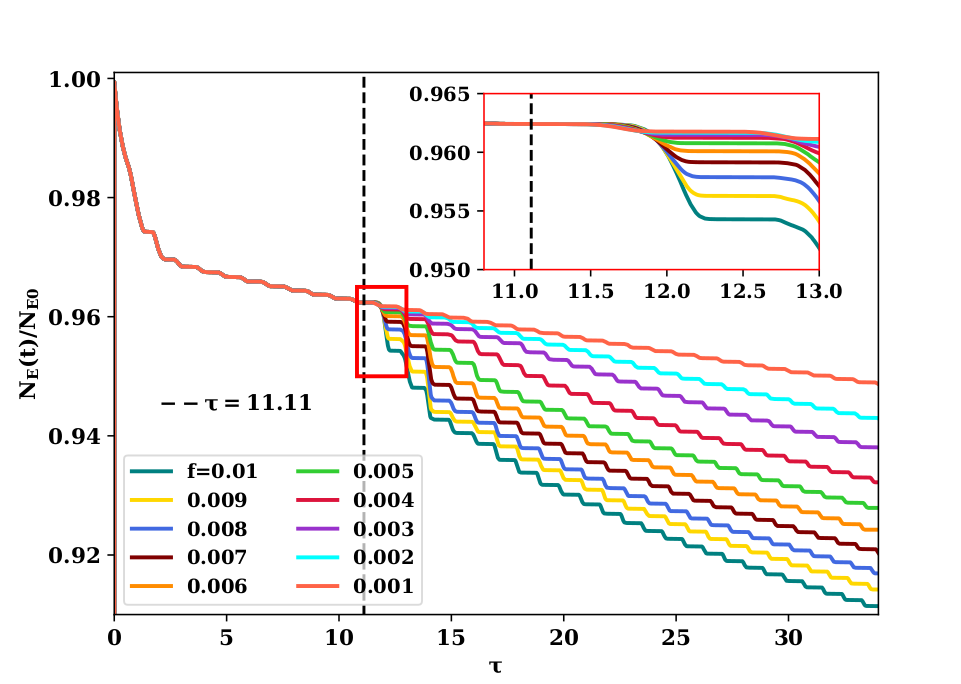}
 \caption{Ratio of the instantaneous electron particle number ($N_E(t)$) to the initial value of the electron particle number ($N_{E0}$) i.e. ${N_E(t)}/N_{E0}$ is shown for different values of $f$ upto $\tau$ $=$ $34$, where $\tau$ $=$ $t/\overline{t}_D$ and $\overline{t}_D$ is toroidal Diocotron period for QQS state. At $\tau$ $=$ $11.11$ (indicated by black dotted line) introduction of ions results into a dip in ${N_E(t)}/N_{E0}$ values which is more for higher values of $f$. The inset plot shows the effective electron loss, around $\tau$ $=$ $11.11$.}  
\label{fig:Fig_NE_RGanesh}
 \end{figure}
 
 \begin{figure}[htp]
 \includegraphics[scale=0.28]{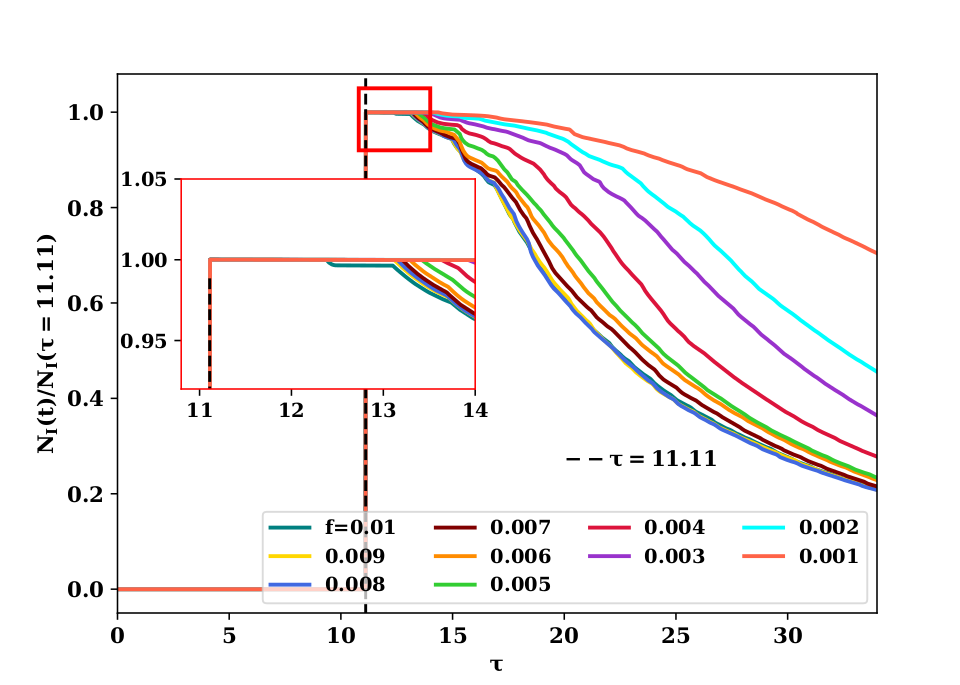}
 \caption{Ratio of the instantaneous ion particle number (${N_I(t)}$) to the ion particle number ($N_{I} (\tau = 11.11)$) at $\tau = 11.11$, i.e. ${N_I(t)}/N_{I} (\tau = 11.11)$ is shown for different values of $f$ upto $\tau$ $=$ $34$, where $\tau$ $=$ $t/\overline{t}_D$ and $\overline{t}_D$ is toroidal Diocotron period for QQS state. After $\tau$ $=$ $11.11$ (indicated by black dotted line), ${N_I(t)}/N_{I} (\tau = 11.11)$ show finite value (ion particles are introduced in the system at $\tau$ $=$ $11.11$) and its value decreases with simulation time. The decrement is more for higher values of $f$. The inset plot shows the starting of ion loss, around $\tau$ $=$ $12.5$.}  
\label{fig:Fig_NI_RGanesh}
 \end{figure}
 
Though electron plasma in a quiescent quasi-steady state has better stability properties, inertia effects i.e. various drifts due to the toroidal inhomogeneous magnetic fields: $\bm{\nabla}B$ and curvature drift (and $\overline{\rho}\strut_{Le}/{L}$ to all orders\cite{Khamaru2021,Khamaru2021Er}) are present in the system leading to some electron losses ($\sim$ $2 \%$)\cite{Khamaru2021} in the system. As ions are introduced in the system, destabilization of electron plasma due to toroidal ion resonance instability, initiates more rapid loss of electron particles from the system. Such loss results in confinement issues of electron plasma in the experimental studies. For the present study, primary electron particle loss is shown in Fig. \ref{fig:Fig_NE_RGanesh} where ratio of the instantaneous electron particle number $N_E(t)$ to the initial value of the electron particle number $N_{E0}$ i.e. ${N_E(t)}/N_{E0}$ for different values of $f$ upto $\tau$ $=$ $34$ is shown. Here $\tau$ $=$ $t/\overline{t}_D$ and $\overline{t}_D$ is toroidal Diocotron period for QQS state. The time of introduction of ions is at $\tau$ $=$ $11.11$, as indicated by black dotted line. Before $\tau$ $=$ $11.11$, the electron plasma evolves into QQS state and the the plasma losses $\sim$ $3.5 \%$ of initial total number of electrons. At $\tau$ $=$ $11.11$ introduction of ions results into a dip in the quantity ${N_E(t)}/N_{E0}$ and the loss rate increases with increasing value of $f$ or ion density. At $\tau$ $=$ $34$, the quantity ${N_E(t)}/N_{E0}$ losses $\sim$ $5 \%$ of the initial value for $f$ $=$ $0.001$ case, and $\sim$ $8 \%$ for $f$ $=$ $0.01$ case. The inset plot around $\tau$ $=$ $11.11$ shows the the effective electron loss, around the introduction period of ions in the system.\
 
\begin{figure*}[htp]
 \includegraphics[scale=0.35]{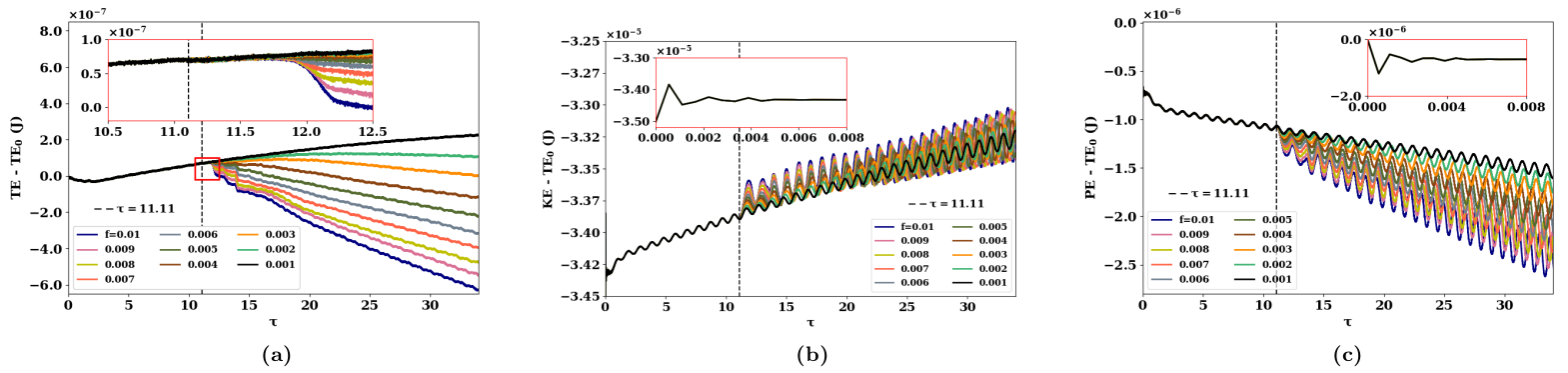}
 \caption{Energy variations of the system upto $\tau$ $=$ $34$, where $\tau$ $=$ $t/\overline{t}_D$ and $\overline{t}_D$ is toroidal Diocotron period for QQS state. The initial value of the total energy ${TE}_0$ ($3.503 \times {10}^{-5}$ J) of the system at $\tau$ $=$ $0.00$ is subtracted from the total energy ($TE$), kinetic energy ($KE$) and potential energy ($PE$) of the system and shown for the values of $f$ ranging from $0.001$ to $0.01$ in (a), (b) and (c) respectively. The time of introduction of ions is at $\tau$ $=$ $11.11$, as indicated by black dotted line in all three plots. The inset plot in (a) around $\tau$ $=$ $11.11$ shows that total energy of the system starts to decrease before $\sim$ one toroidal Diocotron time period after $\tau$ $=$ $11.11$. The inset plots in (b) and (c) shows the initial values of ${KE}$ $-$ ${TE}_0$ and ${PE}$ $-$ ${TE}_0$ respectively.}  
\label{fig:Fig6_energy_RGanesh}
 \end{figure*} 

In Fig. \ref{fig:Fig_NI_RGanesh}, ratio of the instantaneous ion particle number (${N_I(t)}$) to the ion particle number ($N_{I} (\tau = 11.11)$) at $\tau = 11.11$, i.e. ${N_I(t)}/N_{I} (\tau = 11.11)$ is shown for different values of $f$ upto $\tau$ $=$ $34$, where $\tau$ $=$ $t/\overline{t}_D$ and $\overline{t}_D$ is toroidal Diocotron period for QQS state. The time of introduction of ions is at $\tau$ $=$ $11.11$, as indicated by black dotted line. Upto $\tau$ $=$ $11.11$, ${N_I(t)}/N_{I} (\tau = 11.11)$ shows zero value as the system contained no ions at all, whereas after $\tau$ $=$ $11.11$, ${N_I(t)}/N_{I} (\tau = 11.11)$ has finite value (ion particles are introduced in the system at $\tau$ $=$ $11.11$). At $\tau$ $=$ $34$, ion plasma losses $\sim$ $32 \%$ of its initial value for $f$ $=$ $0.001$ case and $\sim$ $80 \%$ for $f$ $=$ $0.01$ case. The inset plot around $\tau$ $=$ $11.11$ shows the introduction period of ions and the starting point of the loss of ions from the system. After $\tau$ $=$ $11.11$, the loss of ion particles starts after $\sim$ one toroidal Diocotron cycle ($\sim$ $\tau$ $=$ $12.5$) and the starting time varies with different values of $f$. In the case of electron loss, the effective electron loss (after $\tau$ $=$ $11.11$) starts just before $\sim$ one toroidal Diocotron cycle $\sim$ $\tau$ $=$ $12.5$). This observation indicates that though the loss of ion at later stage of the simulation is rapid and significantly high when compared to the electron loss within the same simulation time period, initially the ion particles were trapped and gained energy while the system endured electron particle loss. Also, in spite of the ion loss at later stage of the simulation, the electrons plasma remains destabilized throughout that simulation time period showing saturation of the amplitude of the wall probe current (Fig. \ref{fig:Fig5_probe_current_RGanesh}). This suggests, the initial presence of the ion plasma at $\tau$ $=$ $11.11$ started the process of electron plasma destabilization which became independent of the presence of ion plasma in the system at a later time. It is also to be noted that, as the ion loss rate is more than the electron loss rate along with the simulation time, the dynamic $f$ values become lower than the initial values. To understand the plasma dynamics in detail, temporal evolution of the kinetic and potential energy of the total system as well as for electron/ion plasma are described.\\ 

\subsection{Energy variation with simulation time}
\label{subsec:Energy variation with simulation time}
 
 
In Fig. \ref{fig:Fig6_energy_RGanesh} energy evolution of the system are shown upto $\tau$ $=$ $34$. The initial value of the total energy ${TE}_0$ ($3.503 \times {10}^{-5}$ J) of the system at $\tau$ $=$ $0.00$ is subtracted from the total energy ($TE$), kinetic energy ($KE$) and potential energy ($PE$) of the system and shown for the values of $f$ ranging from $0.001$ to $0.01$ in Fig. \ref{fig:Fig6_energy_RGanesh} (a), (b) and (c) respectively. The time of introduction of ions is at $\tau$ $=$ $11.11$, as indicated by black dotted line in all three plots. Fig. \ref{fig:Fig6_energy_RGanesh} (a) shows ${TE}$ $-$ ${TE}_0$ $\sim$ ${10}^{-7}$, indicating that the total energy of the system is conserved for all $f$ values (for lowest $f$ $=$ $0.001$, the increment is $<$ $1 \%$). The inset plot in Fig. \ref{fig:Fig6_energy_RGanesh} (a) around $\tau$ $=$ $11.11$ shows that total energy (${TE}$ $-$ ${TE}_0$) of the system starts to decrease before one toroidal Diocotron time period after $\tau$ $=$ $11.11$. The inset plots in Fig. \ref{fig:Fig6_energy_RGanesh} (b) and (c) shows the initial values of ${KE}$ $-$ ${TE}_0$ and ${PE}$ $-$ ${TE}_0$ respectively. From $\tau$ $=$ $0.00$ to $\tau$ $=$ $11.11$, while the system remains in QQS state without ion plasma, kinetic energy of the system/the electron plasma increases, along with an increase oscillation amplitude, at the cost of electrostatic potential energy of the system (Fig. \ref{fig:Fig6_energy_RGanesh} (c): $\tau$ $=$ $0.00$ to $\tau$ $=$ $11.11$). After the introduction of cold ion plasma in the system at $\tau$ $=$ $11.11$, average kinetic energy of the system (electrons and ions) rises with time while the oscillation amplitude starts to grow with time for all values of $f$. For higher values of $f$, the growth in the oscillation amplitude are higher, where as average kinetic energy of the system reaches to close values for all $f$ at $\tau$ $=$ $34$. In the same time interval, average potential energy of the system decreases for all values of $f$ (Fig. \ref{fig:Fig6_energy_RGanesh} (c)), along with growth of the oscillation amplitudes similar as the kinetic energy. However, the average potential energy values drops with higher $f$ values at $\tau$ $=$ $34$, mainly because of its high transfer to the kinetic energy of the system (electrons and ions) with higher $f$ values. Also loss of particles from the system with time (Fig. \ref{fig:Fig_NE_RGanesh} and \ref{fig:Fig_NI_RGanesh}) aids to small order decrement of the potential energy of the system. Clearly, rise in the oscillation amplitude of kinetic energy as well as potential energy with higher values of $f$ needs to be addressed separately. In the ion resonance instability process, energy is transferred from electron plasma to ion plasma. Thus the energy profile variations of the electron and ion plasma are investigated separately and explained in the next paragraph.\

 \begin{figure*}[htp]
 \centering
 \includegraphics[scale=0.35]{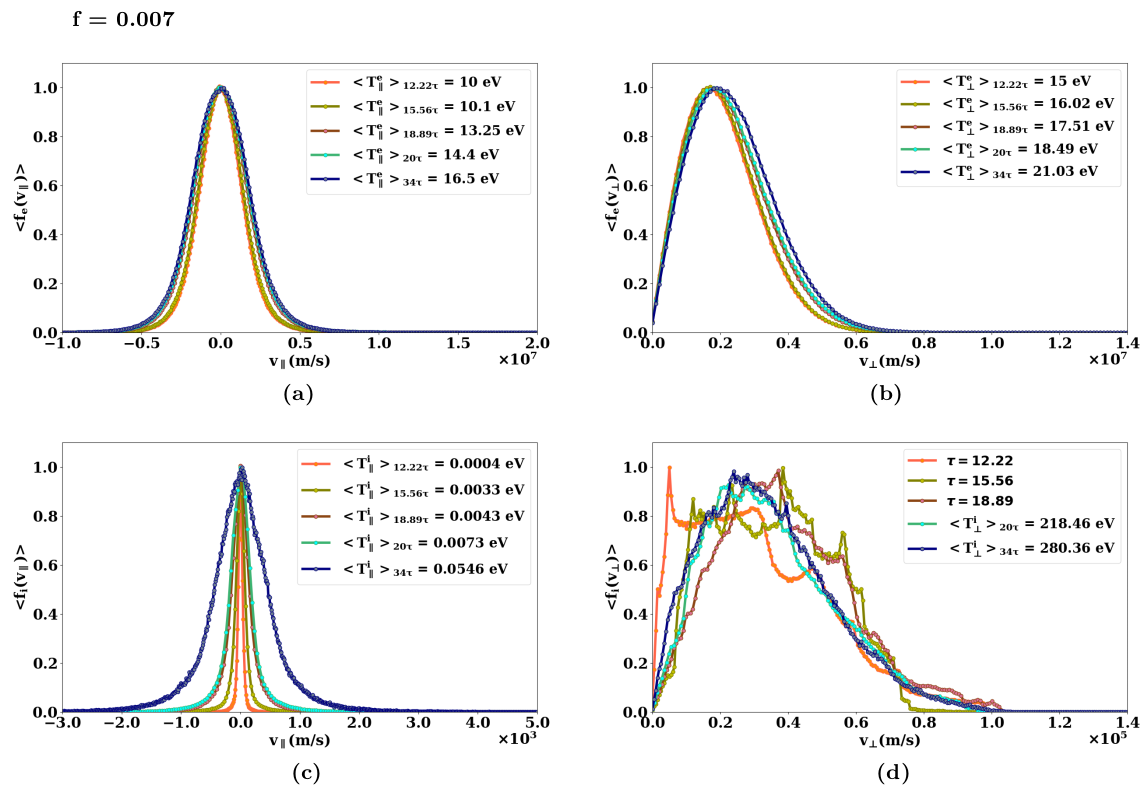}
 \caption{(a) and (b) represent volume averaged parallel and perpendicular velocity distribution functions of electron plasma, $\langle f_e(v_\parallel) \rangle$ and $\langle f_e(v_\bot) \rangle$ respectively, at different simulation time periods $\tau$ $=$ $12.22$, $15.56$, $18.89$, $20.00$ and $34.00$ along with corresponding parallel and perpendicular temperatures $\langle T_{\parallel}^e \rangle$ and $\langle T_{\bot}^e \rangle$, for the case of $f$ $=$ $0.007$. In (c) and (d), volume averaged parallel and perpendicular velocity distribution functions of ion plasma, $\langle f_i(v_\parallel) \rangle$ and $\langle f_i(v_\bot) \rangle$, at same simulation time periods, are shown respectively along with corresponding parallel and perpendicular temperatures, $\langle T_{\parallel}^i \rangle$ and $\langle T_{\bot}^i \rangle$, for same $f$ value. In (a) and (b), the temperatures increase from ${\langle T_{\parallel}^e \rangle}_{12.22 \tau}$ $=$ $10~eV$ to ${\langle T_{\parallel}^e \rangle}_{34 \tau}$ $=$ $16.5~eV$ and from ${\langle T_{\bot}^e \rangle}_{12.22 \tau}$ $=$ $15~eV$ to ${\langle T_{\bot}^e \rangle}_{34 \tau}$ $=$ $21.03~eV$ respectively. In (c), parallel distribution function of the ion plasma at $\tau$ $=$ $12.22$, results into ${\langle T_{\bot}^i \rangle}_{12.22 \tau}$ $=$ $0.0004~eV$. The parallel temperature attains ${\langle T_{\bot}^i \rangle}_{34 \tau}$ $=$ $0.0546~eV$. In (d), for $\tau$ $=$ $12.22$, $15.56$ and $18.89$, the distribution functions do not form near-Maxwellian shape, though $v_{\perp}^i$ $\sim$ $10^{5}$ m/s, $10^{2}$ order $>$ $v_{\parallel}^i$. For $\tau$ $=$ $20.00$ and $34.00$, the distribution function starts to form near-Maxwellian shape resulting ${\langle T_{\bot}^i \rangle}_{20 \tau}$ $=$ $218.46~eV$ and ${\langle T_{\bot}^i \rangle}_{34 \tau}$ $=$ $280.36~eV$.}  
\label{fig:distribution}
 \end{figure*}

\begin{figure*}[htp]
 \centering
 \includegraphics[scale=0.35]{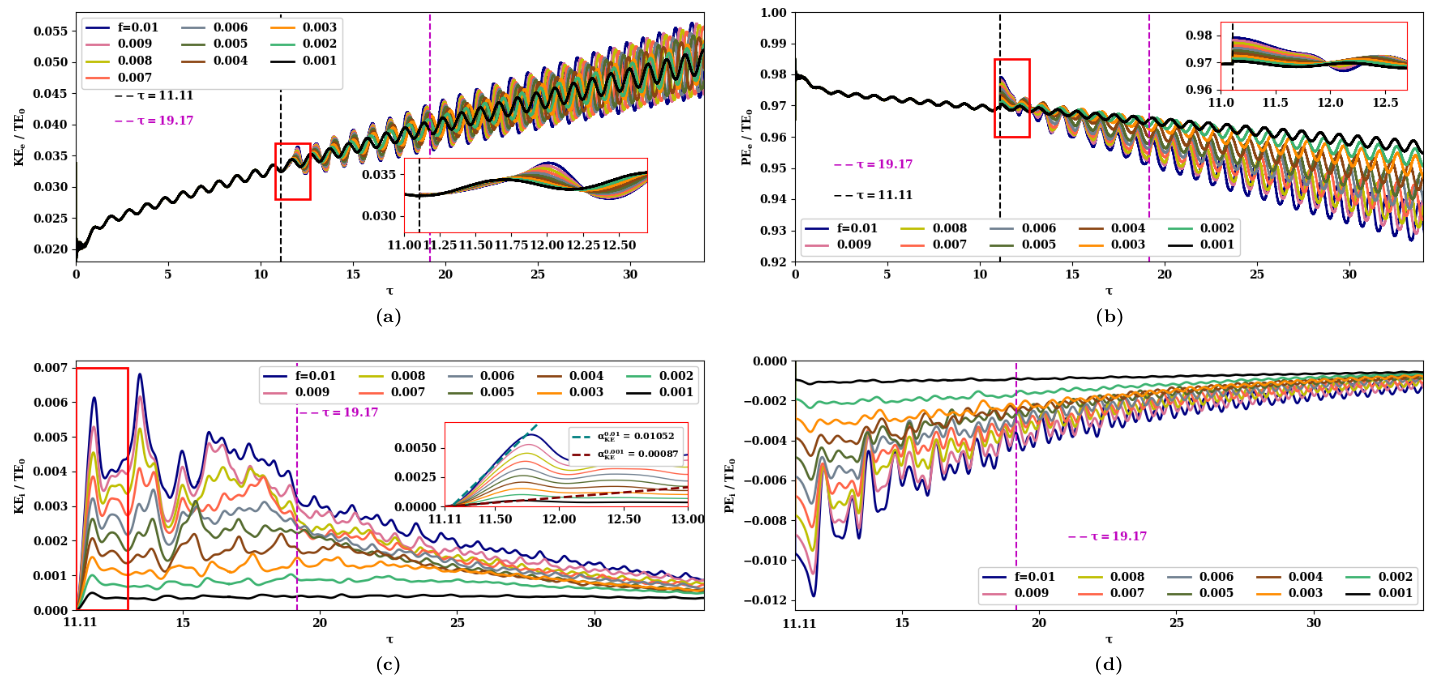}
 \caption{Normalized kinetic energy and potential energy variations of the electron plasma and ion plasma are shown upto $\tau$ $=$ $34$ for $f$ $=$ $0.001$ to $0.01$. The energy values are normalized by initial total energy ${TE}_0$ ($3.503 \times {10}^{-5}$ J) of the system at $\tau$ $=$ $0.00$. The time of introduction of ions is at $\tau$ $=$ $11.11$, as indicated by black dotted line in all plots. (a) and (b) represents the variation of kinetic energy ${KE}_e$/${TE}_0$ and potential energy ${PE}_e$/${TE}_0$ of the electron plasma respectively. The inset plot in (a) around $\tau$ $=$ $11.11$ shows that oscillation amplitude in the kinetic energy of the electron plasma starts to increase at $\sim$ $\tau$ $=$ $11.11$. The inset plots in (b), around $\tau$ $=$ $11.11$ shows $\sim$ ${10}^{-3}$ to ${10}^{-2}$ order increase in the amplitude of the ${PE}_e$/${TE}_0$ values at $\tau$ $=$ $11.11$ because of the addition of secondary electrons. (c) and (d) shows kinetic energy ${KE}_i$/${TE}_0$ and potential energy ${PE}_i$/${TE}_0$ of the ion plasma respectively from $\tau$ $=$ $11.11$ to $34$. ${KE}_i$/${TE}_0$ linearly increases after $\tau$ $=$ $11.11$ for all $f$ values. The linear fits yield: growth rates ${\alpha}_{KE}^{0.001}$ $=$ $0.00087$ and ${\alpha}_{KE}^{0.01}$ $=$ $0.01052$. After the time period $\tau$ $=$ $11.11$ to $\tau$ $\sim$ $19.17$, the fall of the quantity ${KE}_i$/${TE}_0$ is not followed by rise of the order $>$ ${10}^{-3}$ for any values of $f$. In the period form $\tau$ $=$ $11.11$ to $\tau$ $\sim$ $19.17$ the ion plasma gains energy from the potential of the electron plasma. In (d), the potential energy of the ion plasma, increases after $\tau$ $=$ $11.11$ (at the cost of the potential energy of the electron plasma) with similar fluctuating trend as the kinetic energy. Around $\tau$ $\sim$ $19.17$, the fluctuation reduces and the potential energy of the plasma slowly decreases with simulation time.}  
\label{fig:Fig_ion_energy_RGanesh}
 \end{figure*}

In the above discussion, it is found that the kinetic energy of the system (i.e. the electrons and ions) increases with simulation time. This suggests that the average temperature also increases. To verify, volume averaged parallel and perpendicular velocity distribution functions of the ion plasma as well as electron plasma are shown in Fig. \ref{fig:distribution} to check if the electron or ion plasma reach a certain equilibrium with a Maxwellian-like distribution function. Here parallel velocity, $v_\parallel$ is defined as the 1D velocity vector with velocity component along toroidal direction ($\theta$). Perpendicular velocity, $v_\bot$ is defined as a 2D vector having a minor-radial component and a poloidal component, or equivalently through a basis transformation, a major-radial component, $v_R$ together with a cylindrical-axial component, $v_Z$. Choosing the later basis for convenience, the perpendicular speed is $|v_{\perp}|=\sqrt{{v_{R}}^2 + {v_{Z}}^2}$. To calculate corresponding volume averaged parallel and perpendicular temperatures, volume averaged distribution functions, $\langle f(v_\parallel) \rangle$ and $\langle f(v_\bot) \rangle$, which are near-Maxwellian, are shown and fitted with respective perfect Maxwellian distribution function (1D/2D) for a certain temperature value. Then $\langle T_\parallel \rangle$ and $\langle T_\bot \rangle$ are calculated from the respective fitted temperature values. Using the expressions given, for case of $f$ $=$ $0.007$, volume averaged parallel and perpendicular velocity distribution functions of electron plasma, $\langle f_e(v_\parallel) \rangle$ and $\langle f_e(v_\bot) \rangle$, at different simulation time periods (at $\tau$ $=$ $12.22$, $15.56$, $18.89$, $20.00$ and $34.00$) are shown in Fig. \ref{fig:distribution} (a) and (b) respectively along with corresponding parallel and perpendicular temperatures, $\langle T_{\parallel}^e \rangle$ and $\langle T_{\bot}^e \rangle$. In Fig. \ref{fig:distribution} (c) and (d), volume averaged parallel and perpendicular velocity distribution functions of ion plasma, $\langle f_i(v_\parallel) \rangle$ and $\langle f_i(v_\bot) \rangle$, at same simulation time periods, are shown respectively along with corresponding parallel and perpendicular temperatures, $\langle T_{\parallel}^i \rangle$ and $\langle T_{\bot}^i \rangle$, for same $f$ value. In Fig. \ref{fig:distribution} (a) and (b), the widths of the electron distribution functions spread with simulation time and the temperatures also increase as ${\langle T_{\parallel}^e \rangle}_{12.22 \tau}$ $=$ $10~eV$ to ${\langle T_{\parallel}^e \rangle}_{34 \tau}$ $=$ $16.5~eV$ (Fig. \ref{fig:distribution} (a)) and ${\langle T_{\bot}^e \rangle}_{12.22 \tau}$ $=$ $15~eV$ to ${\langle T_{\bot}^e \rangle}_{34 \tau}$ $=$ $21.03~eV$ (Fig. \ref{fig:distribution} (b)). The rise in temperatures are also reflecting the rise in the kinetic energy of the electron plasma. However, the ion distribution functions are more informative about the energy transfer process and need to be discussed. In Fig. \ref{fig:distribution} (c), parallel distribution function of the ion plasma at $\tau$ $=$ $12.22$, with narrow width, corresponds to ${\langle T_{\parallel}^i \rangle}_{12.22 \tau}$ $=$ $0.0004~eV$. The parallel temperature does not increase much and attains ${\langle T_{\parallel}^i \rangle}_{34 \tau}$ $=$ $0.0546~eV$. This result suggests that cold loaded ion plasma at $\tau$ $=$ $11.11$ does not acquire enough parallel energy ($v_{\perp}^i$ $\sim$ $10^{3}$ m/s) obstructing the ion particle motion in the toroidal plane. In Fig. \ref{fig:distribution} (d), for $\tau$ $=$ $12.22$, $15.56$ and $18.89$, the distribution functions do not form near-Maxwellian shape, thus it is inaccurate to attribute any temperature information to the plasma in these time periods. However, the spread in these distribution functions ($v_{\perp}^i$ $\sim$ $10^{5}$ m/s, $10^{2}$ order higher than $v_{\parallel}^i$) suggest the ion particle motion mainly in the poloidal plane. For $\tau$ $=$ $20.00$ and $34.00$, the distribution function starts to form near-Maxwellian shape which suggests that the ion plasma approaches equilibrium state. By fitting the curves we obtain ${\langle T_{\bot}^i \rangle}_{20 \tau}$ $=$ $218.46~eV$ and ${\langle T_{\bot}^i \rangle}_{34 \tau}$ $=$ $280.36~eV$. As seen in this analysis, the ion plasma forms near-Maxwellian distribution function in the region between $\tau$ $=$ $18.89$ to $\tau$ $=$ $20.00$, indicating that the ion plasma attains equilibrium after $\tau$ $=$ $18.89$. The temperature values ${\langle T_{\bot}^i \rangle}_{20 \tau}$ $=$ $218.46~eV$ and ${\langle T_{\bot}^i \rangle}_{34 \tau}$ $=$ $280.36~eV$ are higher than that of the corresponding electron temperatures. Such high temperature of ion plasma indicates anomalous heating of ions via toroidal ion resonance instability process and to the best of our knowledge has never been reported prior to this work.\ 

To get a better idea of the energy transfer processes, kinetic and potential energies of the ion and electron plasmas have been obtained. Normalized kinetic energy and potential energy variations of the electron plasma and ion plasma with simulation time are shown in Fig. \ref{fig:Fig_ion_energy_RGanesh} upto $\tau$ $=$ $34$ for $f$ $=$ $0.001$ to $0.01$. The energy values are normalized by initial total energy ${TE}_0$ of the system at $\tau$ $=$ $0.00$. The time of introduction of ions is at $\tau$ $=$ $11.11$, as indicated by black dotted line in all plots. Fig. \ref{fig:Fig_ion_energy_RGanesh} (a) and (b) represents the variation of kinetic energy ${KE}_e$/${TE}_0$ and potential energy ${PE}_e$/${TE}_0$ of the electron plasma respectively, upto the simulation time $\tau$ $=$ $34$. The inset plot in Fig. \ref{fig:Fig_ion_energy_RGanesh} (a) around $\tau$ $=$ $11.11$ shows that oscillation amplitude in the kinetic energy of the electron plasma starts to increase at $\sim$ $\tau$ $=$ $11.11$. In Fig. \ref{fig:Fig_ion_energy_RGanesh} (b), the inset plots around $\tau$ $=$ $11.11$ shows $\sim$ ${10}^{-3}$ to ${10}^{-2}$ order increase in the amplitude of the ${PE}_e$/${TE}_0$ values at $\tau$ $=$ $11.11$. This is because of the addition of secondary electrons of density $\sim$ ${10}^{-3}$ to ${10}^{-2}$ times the density of the primary electrons. After $\tau$ $=$ $11.11$, the kinetic energy and potential energy of the electron plasma show similar trend as seen in the total kinetic energy and potential energy variation of the system (Fig. \ref{fig:Fig6_energy_RGanesh} (b) and (c)) through the simulation time with the values of $f$. Following initial drop after $\tau$ $=$ $11.11$, potential energy of the electron plasma is drained out as the kinetic energy of the plasma for all values of $f$. The drop is more for higher $f$ values as seen in Fig. \ref{fig:Fig6_energy_RGanesh} (c). Loss of electron particles from the system with time (Fig. \ref{fig:Fig_NE_RGanesh}) also aids to small order decrement of the potential energy of the plasma. The orders of the values of ${KE}_e$/${TE}_0$ are that of the total kinetic energy of the system throughout the simulation, as the ions and secondary electrons are loaded cold contributing zero kinetic energies to the system. Similarly, the orders of the values of ${PE}_e$/${TE}_0$ are that of the total potential energy of the system as the ions and secondary electrons are loaded with lower density than that of the primary electron plasma ($f$ $=$ $0.001$ to $0.01$). In Fig. \ref{fig:Fig_ion_energy_RGanesh} (a) and (b), $\tau$ $=$ $19.17$ is marked by vertical dotted line in magenta and explained in next paragraph.\  

To understand the energy variation of ion plasma through the simulation time, normalized kinetic energy ${KE}_i$/${TE}_0$ and potential energy ${PE}_i$/${TE}_0$ of the ion plasma are shown in Fig. \ref{fig:Fig_ion_energy_RGanesh} (c) and (d) respectively for the values of $f$ ranging from $0.001$ to $0.01$ from simulation time $\tau$ $=$ $11.11$ to $\tau$ $=$ $34$. As the ion plasma was cold loaded at $\tau$ $=$ $11.11$, kinetic energy is zero at $\tau$ $=$ $11.11$ in Fig. \ref{fig:Fig_ion_energy_RGanesh} (c). The parameter ${PE}_i$/${TE}_0$ $\sim$ ${10}^{-3}$ to ${10}^{-2}$ times the total energy of the system at $\tau$ $=$ $11.11$ in Fig. \ref{fig:Fig_ion_energy_RGanesh} (d). Demonstrating typical outcome of the ion resonance instability process, the ion particles gain energy (kinetic/potential) from the potential energy of the electron plasma. As our simulation is collisionless, we believe that there exists an underlying collisionless process, resulting in collisionless exchange of energy between ions and electrons. In Fig. \ref{fig:Fig_ion_energy_RGanesh} (c), the parameter ${KE}_i$/${TE}_0$ linearly increases after $\tau$ $=$ $11.11$ for all $f$ values. To check the growth rate with respect to $f$ values, the parameter ${KE}_i$/${TE}_0$ is linearly fitted upto $\tau$ $=$ $11.8$ for $f$ $=$ $0.001$ and $0.01$ cases and shown in the inset plot. The linear fits yield slope of ${KE}_i$/${TE}_0$ growth as ${\alpha}_{KE}^{0.001}$ $=$ $0.00087$ for $f$ $=$ $0.001$ and ${\alpha}_{KE}^{0.01}$ $=$ $0.01052$ for $f$ $=$ $0.01$, suggesting $\sim$ ${10}$ time higher growth rate for one order of magnitude increase in $f$ value ($\frac{{\alpha}_{KE}^{0.01}}{{\alpha}_{KE}^{0.001}}$ $\sim$ ${12.09}$). After $\tau$ $=$ $11.11$, the quantity ${KE}_i$/${TE}_0$ displays fluctuating rise/fall profile with occasional gain/loss of ion kinetic energy, for all values of $f$. We have identified a region, $\tau$ $=$ $11.11$ to $\tau$ $\sim$ $19.17$ ($\tau$ $=$ $19.17$ marked by vertical dotted line in magenta), after when the fall of the quantity ${KE}_i$/${TE}_0$ is not followed by rise of the order $>$ ${10}^{-3}$ for any values of $f$. After $\tau$ $\sim$ $19.17$, ${KE}_i$/${TE}_0$ slowly decreases with simulation time with small amount of fluctuations (though for $f$ $=$ $0.001$ to $0.005$, the gradual fall of the kinetic energy starts at later simulation time). In the period form $\tau$ $=$ $11.11$ to $\tau$ $\sim$ $19.17$, the ion plasma gains energy from the potential of the electron plasma. After $\tau$ $\sim$ $19.17$ the ion plasma starts to equilibrate. This transition time is within the region between $\tau$ $=$ $18.89$ to $\tau$ $=$ $20.00$, as seen in the velocity distribution function analysis of Fig. \ref{fig:distribution} (d). Further loss of kinetic energy after $\tau$ $\sim$ $19.17$ can also be related to the loss of ion particles from the system though the particle loss starts sooner than $\tau$ $\sim$ $19.17$ (Fig. \ref{fig:Fig_NI_RGanesh}). The potential energy of the ion plasma, shown in Fig. \ref{fig:Fig_ion_energy_RGanesh} (d), also increases after $\tau$ $=$ $11.11$ (at the cost of the potential energy of the electron plasma) with similar fluctuating trend as the kinetic energy. Around $\tau$ $\sim$ $19.17$, the fluctuation reduces and the potential energy of the plasma slowly decreases with simulation time, though the transition of the fluctuation amount around $\tau$ $\sim$ $19.17$ is not strikingly distinguishable as the kinetic energy case. The region around $\tau$ $\sim$ $19.17$ is also identified in Fig. \ref{fig:Fig_ion_energy_RGanesh} (a) and (b) (vertical dotted line in magenta) for the electron plasma, but there is no noticeable transition of kinetic/potential energy of the electron plasma around that time period. The reason might be the low order magnitude of the ion plasma energies compared to the electron plasma energies.\



\subsection{Growth rate of the wall probe current at different periods of the simulation}
\label{subsec:Growth rate of the wall probe current at different periods of the simulation}

\begin{figure*}[htp]
 \includegraphics[scale=0.35]{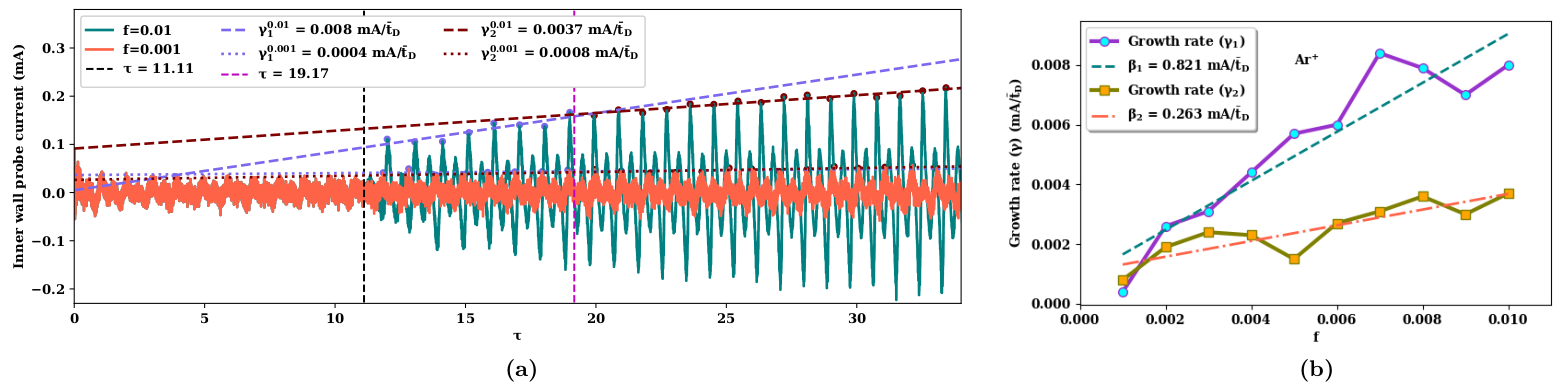}
 \caption{(a) Growth rate analysis of the inner wall probe current for $f$ $=$ $0.001$ and $0.01$ upto $\tau$ $=$ $34$. $\tau$ $=$ $t/\overline{t}_D$ and $\overline{t}_D$ is toroidal Diocotron period for QQS state. The time of introduction of ions is at $\tau$ $=$ $11.11$, as indicated by black dotted line. $\tau$ $=$ $19.17$ is indicated by vertical dotted line in magenta. Algebraic growth rate of the wall probe current is calculated by linear fitting of the maximum peak among the double peaks of the probe current, in two regions: (1) $\tau$ $=$ $11.11$ to $19.17$ and (2) $\tau$ $=$ $19.17$ to $34$, for both of $f$ $=$ $0.001$ and $0.01$. The fits yield ${\gamma}_{1}^{0.001}$ $=$ $0.0004$ $mA/\overline{t}_D$, ${\gamma}_{1}^{0.01}$ $=$ $0.008$ $mA/\overline{t}_D$ and ${\gamma}_{1}^{0.001}$ $=$ $0.0008$ $mA/\overline{t}_D$, ${\gamma}_{1}^{0.01}$ $=$ $0.0037$ $mA/\overline{t}_D$. (b) Growth rates for the two regions (1) $\tau$ $=$ $11.11$ to $19.17$ and (2) $\tau$ $=$ $19.17$ to $34$ are shown as a function of $f$, as ${\gamma}_1$ and ${\gamma}_2$ respectively. Linear fit of ${\gamma}_1$ and ${\gamma}_2$ reveals growth rate as ${\beta}_1$ $=$ $0.821$ $mA/\overline{t}_D$ and ${\beta}_1$ $=$ $0.263$ $mA/\overline{t}_D$.}  
\label{fig:Fig_probe_current_RGanesh}
 \end{figure*}

Fig. \ref{fig:Fig_probe_current_RGanesh} (a) shows growth rate analysis of the inner wall probe current for $f$ $=$ $0.001$ and $0.01$ upto $\tau$ $=$ $34$. Wall probe currents for extrema values of $f$ are chosen for comparison purpose i.e. to reveal the dependency of growth on lower and higher $f$ values. The time of introduction of ions is at $\tau$ $=$ $11.11$, as indicated by black dotted line. $\tau$ $=$ $19.17$, simulation time close to the transition time of the ion plasma to equilibrium state, is indicated by vertical dotted line in magenta. Algebraic growth rate of the wall probe current is calculated by linear fitting of the maximum peak among the double peaks of the probe current, in two regions: (1) $\tau$ $=$ $11.11$ to $19.17$ and (2) $\tau$ $=$ $19.17$ to $34$, for both of $f$ $=$ $0.001$ and $0.01$. In the initial ($\tau$ $=$ $11.11$ to $19.17$) time period, the linear fit reveals growth rates ${\gamma}_{1}^{0.001}$ $=$ $0.0004$ $mA/\overline{t}_D$ and ${\gamma}_{1}^{0.01}$ $=$ $0.008$ $mA/\overline{t}_D$. For later time period ($\tau$ $=$ $19.17$ to $34$), growth rates are ${\gamma}_{1}^{0.001}$ $=$ $0.0008$ $mA/\overline{t}_D$ and ${\gamma}_{1}^{0.01}$ $=$ $0.0037$ $mA/\overline{t}_D$. As expected, the growth rate is high in the initial region and attains lower value in the later region, indicating saturation of the wall probe current for $f$ $=$ $0.01$ case. This result echos the same explanation of energy transfer given in the ion plasma energy analysis of Subsec. \ref{subsec:Energy variation with simulation time}. For $f$ $=$ $0.001$ case though, the growth rate is higher at later part (lower values than $f$ $=$ $0.01$ case). The discrepancy suggests that for lower $f$ values, the ion plasma gains energy for longer time (low energy gain than the high $f$ value) and equilibrates at a later time than $\tau$ $=$ $19.17$. This feature is also seen in Fig. \ref{fig:Fig_ion_energy_RGanesh} (a), where for $f$ $=$ $0.001$ to $0.005$, the fall of the kinetic energy started at later simulation time ($\sim$ $\tau$ $=$ $25$). To get an idea about the variation of the growth rate for every $f$ values in the present study, growth rate for $f$ $=$ $0.001$ to $0.01$ are shown as a function of $f$ in Fig. \ref{fig:Fig_probe_current_RGanesh} (b), for the two regions (1) $\tau$ $=$ $11.11$ to $19.17$ and (2) $\tau$ $=$ $19.17$ to $34$ as ${\gamma}_1$ and ${\gamma}_2$ respectively. Linear fit of ${\gamma}_1$ and ${\gamma}_2$ values with respect to $f$ values shows that, ${\gamma}_1$ (${\beta}_1$ $=$ $0.821$ $mA/\overline{t}_D$) has higher slope value than ${\gamma}_2$ (${\beta}_1$ $=$ $0.263$ $mA/\overline{t}_D$) i.e. the average growth rate of the wall probe current is higher at the initial time period $\tau$ $=$ $11.11$ to $19.17$. Interestingly, ${\gamma}_1$ and ${\gamma}_2$ values decreases with low values of $f$ and approaches to a common value ($\sim$ $f$ $=$ $0.0015$).\

As discussed in the Sec. \ref{sec:Introduction} (Introduction), algebraic growth\cite{Fajans1993,Peurrung1993,Bettega2006} is seen in the ion resonance instability with transient nature of the ion in the cylindrical trap. In tight aspect ratio toroidal trap experiment\cite{Lachhvani2016} with toroidal end plugs, the ions are produced and lost along toroidal direction and thus are effectively transient. Though the growth rate is high but slower\cite{Lachhvani2016} than theoretically proposed rates by Levy et al.\cite{Levy1969}. Here growth of the wall probe current has been found out to be algebraic in nature. In our case there is no continues source of ions but ion loss occurs throughout the simulation, which is a unique situation never studied before. But it is evident that the nature of the instability is similar to that of the transient ions.\

In the next Subsec. \ref{subsec:Temperature of the electron and ion plasma}, the global temperature profiles of electron and ion plasma are described as function time. Also the radial temperature profiles of the electron plasma and their evolution is addressed. 
 
 \subsection{Temperature of the electron and ion plasma}
\label{subsec:Temperature of the electron and ion plasma}
To understand the temperature evolution of electron and ion plasma, volume averaged parallel and perpendicular temperatures are evaluated through out the simulation time and explained next. Average temperatures of electron flux tubes at different radial locations are also found out and corresponding radial variations are also discussed. 

\subsubsection{Global temperature of the electron plasma}
\label{subsec:Global temperature of the electron plasma}

\begin{figure}[htp]
 \includegraphics[scale=0.3]{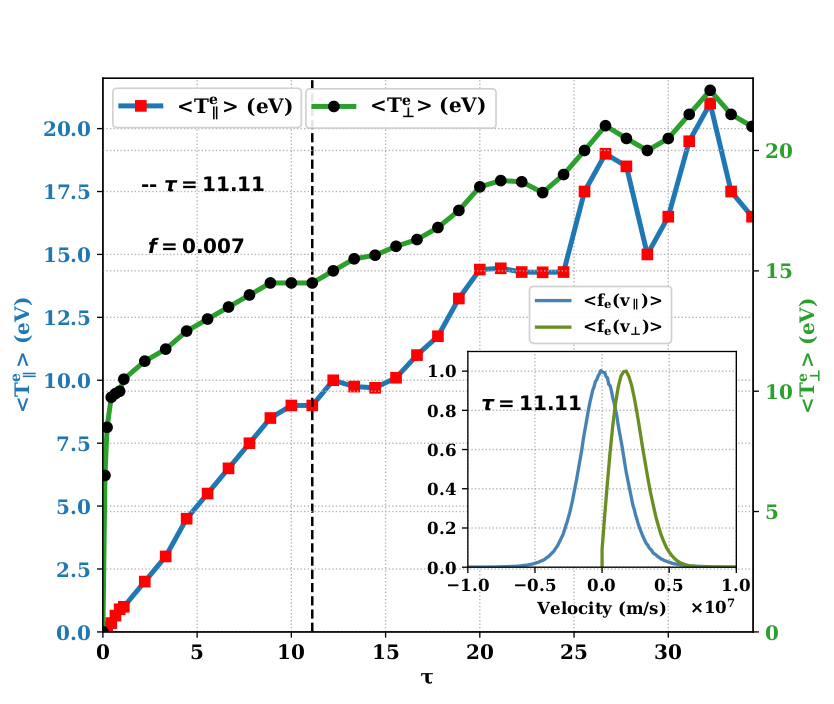}
 \caption{Volume averaged parallel and perpendicular temperatures of the electron plasma, $\langle T_{\parallel}^e \rangle$ and $\langle T_{\bot}^e \rangle$ are shown for case of $f$ $=$ $0.007$ upto $\tau$ $=$ $34$. $\langle T_{\parallel}^e \rangle$ $\sim$ $13.8$~eV and $\langle T_{\bot}^e \rangle$ $\sim$ $8.8$~eV at $\tau$ $=$ $11.11$ (indicated by black dotted line), when the ion plasma was introduced. At $\tau$ $=$ $34$, $\langle T_{\parallel}^e \rangle$ $\sim$ $21$~eV and $\langle T_{\bot}^e \rangle$ $\sim$ $17$~eV. Though the rise in the temperature values was almost steady initially, after $\tau$ $\sim$ $25$ the fluctuations in $\langle T_{\parallel}^e \rangle$ values were $\sim$ $5$ev and for $\langle T_{\bot}^e \rangle$ $\sim$ $2.5$ev. Volume averaged parallel and perpendicular distribution functions, $\langle f_e(v_\parallel) \rangle$ and $\langle f_e(v_\bot) \rangle$, at $\tau$ $=$ $11.11$, are shown in the inset plot. }  
\label{fig:electron_temp_RGanesh}
 \end{figure}
 
The volume averaged parallel and perpendicular temperatures of the electron plasma, $\langle T_{\parallel}^e \rangle$ and $\langle T_{\bot}^e \rangle$ are shown in Fig. \ref{fig:electron_temp_RGanesh} for case of $f$ $=$ $0.007$ upto $\tau$ $=$ $34$. As the electrons were loaded cold, the temperature values start from zero $\tau$ $=$ $0.00$ and reaches to finite temperature values ($\langle T_{\parallel}^e \rangle$ $\sim$ $13.8$~eV and $\langle T_{\bot}^e \rangle$ $\sim$ $8.8$~eV), at $\tau$ $=$ $11.11$, at the instant the ion plasma was introduced. Volume averaged parallel and perpendicular distribution functions, $\langle f_e(v_\parallel) \rangle$ and $\langle f_e(v_\bot) \rangle$, at $\tau$ $=$ $11.11$, are shown in the inset plot. After $\tau$ $=$ $11.11$, the temperature values rise further and attains values as $\langle T_{\parallel}^e \rangle$ $\sim$ $21$~eV and $\langle T_{\bot}^e \rangle$ $\sim$ $17$~eV. Though the rise was almost steady initially, after $\tau$ $\sim$ $25$ the fluctuations in $\langle T_{\parallel}^e \rangle$ values were $\sim$ $5$ev and for $\langle T_{\bot}^e \rangle$ $\sim$ $2.5$ev. The orders of these global temperature values are in the same range ($\simeq 10 - 20~eV$) of the values for pure electron plasma, if the ion plasma was absent from the system, though the oscillating features at later simulation time represents the high amplitude oscillation of the electron plasma after $\tau$ $\sim$ $25$ seen in Fig. \ref{fig:Fig_ion_energy_RGanesh} (a). Volume averaged parallel and perpendicular temperatures of the ion plasma are also calculated and discussed next.

\subsubsection{Global temperature of the ion plasma}
\label{subsec:Global temperature of the ion plasma}
 \begin{figure}[htp]
 \includegraphics[scale=0.25]{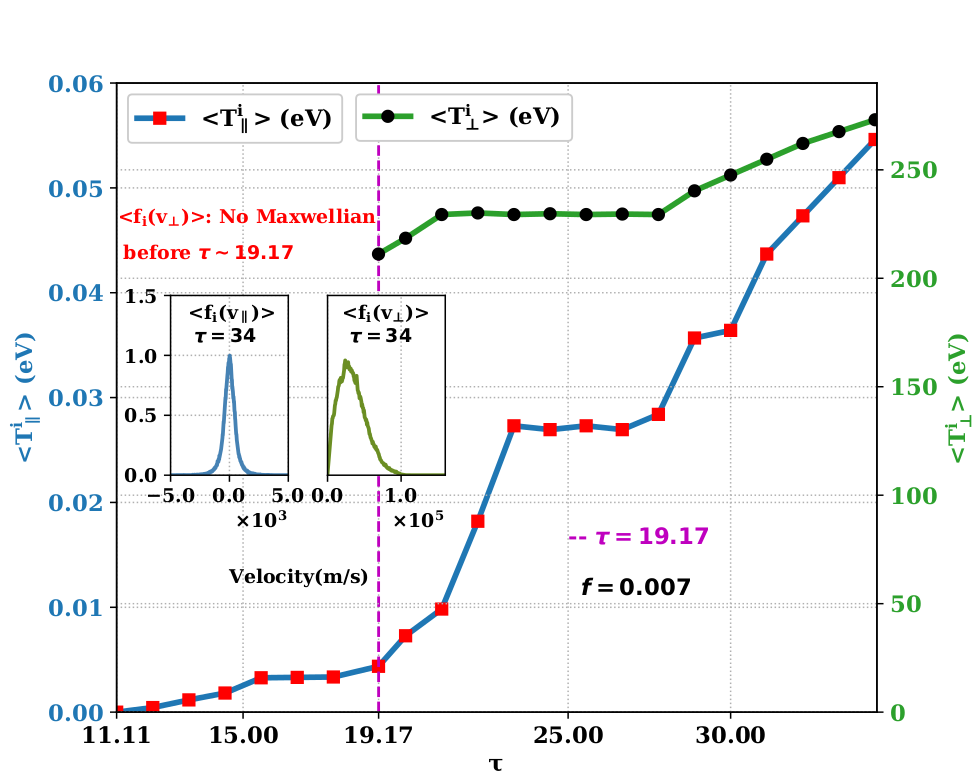}
 \caption{Volume averaged parallel and perpendicular temperatures of the ion plasma, $\langle T_{\parallel}^i \rangle$ and $\langle T_{\bot}^i \rangle$ are shown for case of $f$ $=$ $0.007$ upto $\tau$ $=$ $34$. As seen in Fig. \ref{fig:distribution}, the perpendicular distribution functions of ion plasma form near-Maxwellian shape around $\tau$ $=$ $19.17$ (indicated by vertical dotted line in magenta) with $\langle T_{\bot}^i \rangle$ $\sim$ $211$~eV at $\tau$ $=$ $19.17$. $\langle T_{\bot}^i \rangle$ $\sim$ $280$~eV at $\tau$ $=$ $34$. As the ions were loaded cold, the $\langle T_{\parallel}^i \rangle$ values start from zero at $\tau$ $=$ $11.11$ and reaches to low order finite temperature values ($\langle T_{\parallel}^i \rangle$ $\sim$ $0.055$~eV ) at $\tau$ $=$ $34$. Steep rise in $\langle T_{\parallel}^i \rangle$ values is seen after $\tau$ $=$ $19.17$. Volume averaged parallel and perpendicular distribution functions, $\langle f_i(v_\parallel) \rangle$ and $\langle f_i(v_\bot) \rangle$, at $\tau$ $=$ $34$, are shown in the inset plot.}  
\label{fig:ion_temp_RGanesh}
 \end{figure}
 
 As seen in Fig. \ref{fig:distribution}, volume averaged temperature evolution of the ion plasma also helps to clarify the equilibration process of the ion plasma. Thus, the volume averaged parallel and perpendicular temperatures of the ion plasma, $\langle T_{\parallel}^i \rangle$ and $\langle T_{\bot}^i \rangle$ are shown for case of $f$ $=$ $0.007$ upto $\tau$ $=$ $34$. $\tau$ $=$ $19.17$ is indicated by vertical dotted line in magenta. As seen in Fig. \ref{fig:distribution}, the perpendicular distribution functions of ion plasma form near-Maxwellian shape around $\tau$ $=$ $19.17$ with $\langle T_{\bot}^i \rangle$ $\sim$ $211$~eV at $\tau$ $=$ $19.17$. After $\tau$ $=$ $19.17$, $\langle T_{\bot}^i \rangle$ values increases and attains $\sim$ $280$~eV at $\tau$ $=$ $34$. As the ions were loaded cold, the $\langle T_{\parallel}^i \rangle$ values start from zero at $\tau$ $=$ $11.11$ and reaches to low order finite temperature values ($\langle T_{\parallel}^i \rangle$ $\sim$ $0.055$~eV) at $\tau$ $=$ $34$. However, steep rise in $\langle T_{\parallel}^i \rangle$ values is seen after $\tau$ $=$ $19.17$ suggesting energy gain of the ion plasma, though very small in the parallel direction. Volume averaged parallel and perpendicular distribution functions, $\langle f_i(v_\parallel) \rangle$ and $\langle f_i(v_\bot) \rangle$, at $\tau$ $=$ $34$, are shown in the inset plot.
 
\subsubsection{Radial variation of the electron plasma temperature}
\label{subsec:Radial variation of the electron plasma temperature}

\begin{figure*}[htp]
 \includegraphics[scale=0.35]{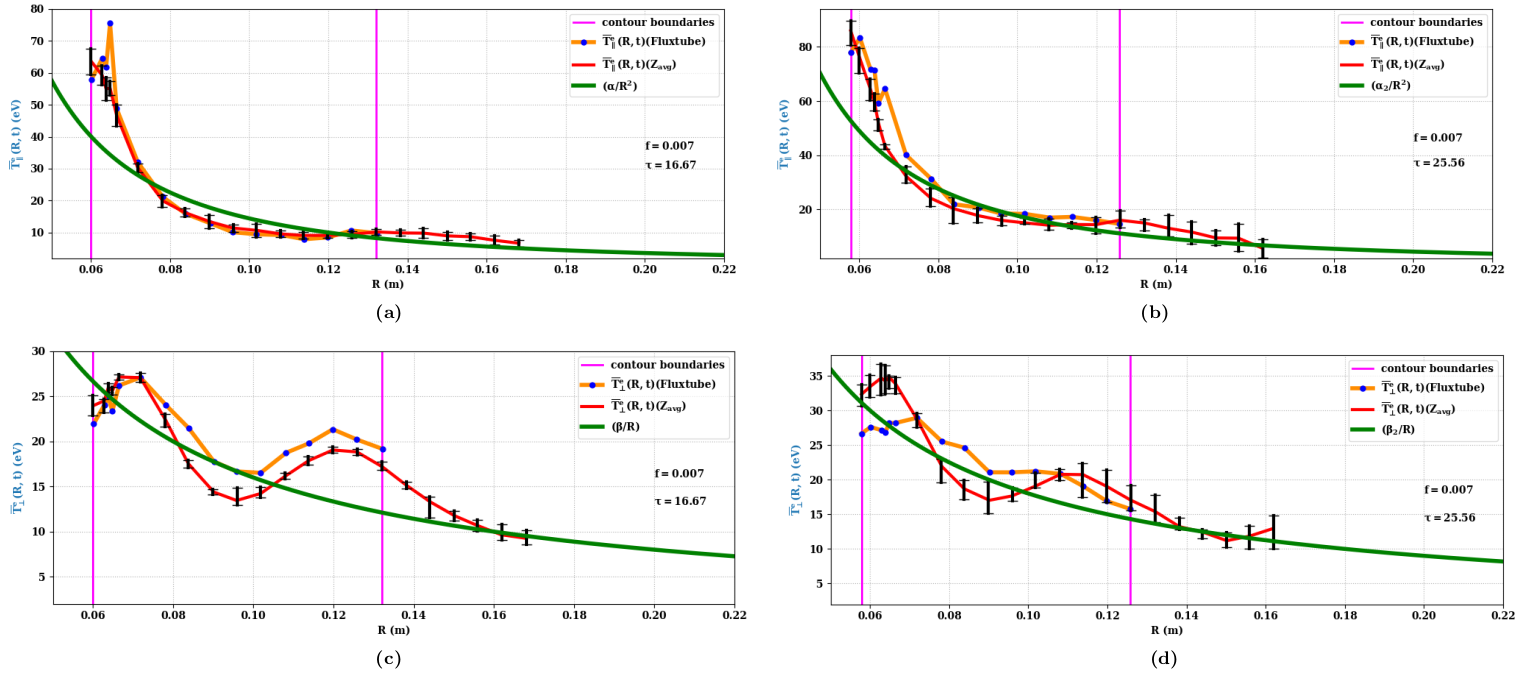}
 \caption{Averaged temperatures ${\Bar{T}}_{\parallel}^e (R,t)$ and ${\Bar{T}}_{\bot}^e (R,t)$ are shown as functions of $R$, at $\tau$ $=$ $16.67$ and $\tau$ $=$ $25.56$ for $f$ $=$ $0.007$ case. Z-averaged quantities are expressed as ${\Bar{T}}_{\parallel}^e (R,t)(Z_{avg})$, ${\Bar{T}}_{\bot}^e (R,t)(Z_{avg})$ and flux tube averaged quantities are expressed as ${\Bar{T}}_{\parallel}^e (R,t)(Fluxtube)$, ${\Bar{T}}_{\bot}^e (R,t)(Fluxtube)$ in all of these Figures. The magenta colored vertical lines represents $1/e$ times value of the peak electron density. For high accuracy the analysis is performed over different sized bins along $R$ and the respective range of the temperatures are shown by the error bars in these Figures. In (a) and (b), ${\Bar{T}}_{\parallel}^e (R,t)(Z_{avg})$ and ${\Bar{T}}_{\parallel}^e (R,t)(Fluxtube)$ values are higher at the inboard side and gradually fall with $R$. The temperature values are fitted with $\alpha /R^{2}$ and $\alpha_2 /R^{2}$ ($\alpha$ and $\alpha_2$ are an arbitrary constants) showing $1/R^{2}$ scaling with $R$. In (c) and (d), ${\Bar{T}}_{\bot}^e (R,t)(Z_{avg})$ and ${\Bar{T}}_{\bot}^e (R,t)(Fluxtube)$ values are also higher at the inboard side but showing non-monotonic variation with $R$. The average temperature values are fitted with $\beta /R$ and $\beta_2 /R$ ($\beta$ and $\beta_2$ are arbitrary constants) showing $1/R$ scaling with $R$.}  
\label{fig:temperature_time}
 \end{figure*}
 
The electron plasma performs a center of charge motion in the poloidal plane, the toroidal electron flux tubes passes across the $1/R$ varying toroidal magnetic field resulting in periodic compression–expansion cycle of the tubes (which is also known as magnetic pumping). Under this cross-field dynamics, mechanisms like electron-electron large angle collisions, results into asymmetric exchange of energies between parallel and perpendicular directions. As a result, the parallel and perpendicular temperatures of the electron flux tubes undergo radial variations which, in general, scales inversely with $R$. As our simulation is collisionless, we believe that there exists an underlying collisionless process, which introduces asymmetry in the compression–expansion cycle (i.e. MP), resulting in collisionless exchange of energy between parallel and perpendicular directions. More details of this mechanism remain to be explored in future. Extensive analysis of the radial variation of parallel and perpendicular temperature of pure toroidal electron plasma had been performed in our previous work with QQS state. It was found out that the parallel temperature of toroidal electron plasma flux tubes scales as $1/R{^2}$ and the perpendicular temperature  scales as $1/R$, which is consistent with the angular momentum conservation and mean magnetic moment conservation principles. In the present study, we have observed similar scaling laws despite the presence of the ion plasmas in the system. As performed in the previous study, parallel and perpendicular temperatures are averaged in two different ways: 1. In toroidal and vertical directions (Z-averaged). 2. In toroidal direction only, around mid-Z plane (averaged over toroidal flux tubes). In the first method, particles from different vertical coordinates contribute in the calculations. In the second method, particles within different toroidal flux tubes along with radial coordinates, around mid-z plane have been considered. In the second method, the particles within a flux tube at certain R may be expected to follow similar dynamics to each other. In Fig. \ref{fig:temperature_time} (a)-(d), these averaged temperatures ${\Bar{T}}_{\parallel}^e (R,t)$ (Fig. \ref{fig:temperature_time} (a),(b)) and ${\Bar{T}}_{\bot}^e (R,t)$ (Fig. \ref{fig:temperature_time} (c),(d)) are shown as functions of $R$, at two different time periods $\tau$ $=$ $16.67$ and $\tau$ $=$ $25.56$ for $f$ $=$ $0.007$ case. Z-averaged quantities are expressed as ${\Bar{T}}_{\parallel}^e (R,t)(Z_{avg})$, ${\Bar{T}}_{\bot}^e (R,t)(Z_{avg})$ and flux tube averaged quantities are expressed as ${\Bar{T}}_{\parallel}^e (R,t)(Fluxtube)$, ${\Bar{T}}_{\bot}^e (R,t)(Fluxtube)$ in all of these Figures. The magenta colored vertical lines represents $1/e$ times value of the peak electron density. For high accuracy the analysis is performed over different sized bins along $R$ and the respective range of the temperatures are shown by the error bars in these figures. In Fig. \ref{fig:temperature_time} (a) and (b), ${\Bar{T}}_{\parallel}^e (R,t)(Z_{avg})$ and ${\Bar{T}}_{\parallel}^e (R,t)(Fluxtube)$ values are higher at the inboard side and gradually fall with $R$. The temperature values are fitted with $\alpha /R^{2}$ and $\alpha_2 /R^{2}$ ($\alpha$ and $\alpha_2$ are an arbitrary constants) showing $1/R^{2}$ scaling with $R$. In Fig. \ref{fig:temperature_time} (c) and (d), ${\Bar{T}}_{\bot}^e (R,t)(Z_{avg})$ and ${\Bar{T}}_{\bot}^e (R,t)(Fluxtube)$ values are also higher at the inboard side but showing non-monotonic variation with $R$. The average temperature values are fitted with $\beta /R$ and $\beta_2 /R$ ($\beta$ and $\beta_2$ are arbitrary constants) showing $1/R$ scaling with $R$. In these plots, the scaling laws are preserved at later simulation times representing the conservation of angular momentum and mean magnetic moment, thus suggesting the conserved quantities of the system are not affected by the introduction of ion plasma in the system. The reason for this weak non-monotonicity in the perpendicular temperature profile is unknown and also seen in the analysis of pure electron plasma in QQS\cite{Khamaru2021Er} state.\

\section{Conclusions and Discussions}   
\label{sec:Conclusions and Discussions} 

In this paper, toroidal electron plasmas in a tight aspect ratio axisymmetric toroidal device to the presence of small fraction of ions, has been investigated with a three dimensional collisionless Particle-in-Cell simulation study with $Ar^+$ ions as the ion species. The electron plasma is loaded with a ``seed" solution obtained by zero-inertia entropy maximization and initialized in the PIC code to evolve to attain a quiescent quasi-steady state (QQS). After a QQS state is attained, $Ar^+$ ions and secondary electrons are introduced cold into the system. $Ar^+$ ions of different number density values i.e. $f$ values (0.001 to 0.01) are used and the dependency of the instability on the ion density is addressed via different diagnostics methods.\

We have demonstrated that the electron plasma dynamics and the ion plasma dynamics are affected by presence of small number of ions, for mainly $f$ values $>$ 0.001 i.e. for such ion population in the system results in growth of the wall probe current after the ions are introduced. Thus the results are mainly interpreted on the basis of high $f$ value datum ($f$ $\geq$ 0.005). The growth of the wall probe current is algebraic in nature and saturates at later simulation time, when a substantial amount of ions are lost. In our preloaded ion case there is no continuous source of ions but ions are being lost with simulation time. This is a unique situation not belonging to strictly trapped ions or transient ions in the system. Spatial evolution of the electron plasma and corresponding spectrogram analysis demonstrate the destabilized ``center of charge motion" ($m$ $=$ $1$) of the electron plasma in poloidal plane along with chirped coupling between toroidal Diocotron modes ($m$ $=$ $1$ to $9$ for $f$ $=$ $0.01$) with $m$ $=$ $2$ as dominant mode. Investigation of the evolution of the three dimensional potential well of the system, reveals interesting features. Based on the above, we have argued that the ion fraction destabilizes the electron plasma in QQS state via a toroidal ion resonance instability mechanism.\

The variations in the ion and electron plasma energies show collisionless transfer of electron potential energy to the kinetic energy of the ion plasma. Though ions start to exhibit a dynamic nature in the poloidal direction, toroidal motion of the ions are negligible. This is understandable as the $E_\parallel = - \nabla _\parallel \phi \sim 0$, there is no parallel acceleration or energizing dynamics for ions in the parallel direction. After initial energy gain, the ion plasmas start to equilibrate and attains near-Maxwellian distribution in the poloidal plane. We have identified a certain time period after which the ion energy gain is trivial. Correlating  the wall probe growth in the same time period, it has been found out that the wall probe current also starts to saturate after that time period. Also a linear scaling of the growth rate of the instability with respect to the ion density ($f$ values) has been obtained for the above mentioned two time windows.\

In this study, volume averaged parallel and perpendicular temperatures are defined and used to calculate the temperatures of the ion and electron plasmas. The volume averaged perpendicular temperature of the ion plasma is obtained when ions start to equilibrate, shows high temperature value, indicating ion heating due to gain of energy via the toroidal ion-resonance instability. However, the exact process of equilibration of ions resulting Maxwellian distributions is not clear. The volume averaged temperature values of the electron plasma rises with simulation time though attaining a fluctuating nature near the end of the simulation time.\ 

For two different simulation time periods, the averaged temperatures ${\Bar{T}}_{\parallel}^e$ and ${\Bar{T}}_{\bot}^e$ show $1/R^{2}$ and $1/R$ scaling with $R$ respectively, as can be expected from conservation of angular momentum and conservation of $\mu$.\ 

We have demonstrated toroidal ion resonance instability of toroidal electron plasma and the underlying processes through extensive analysis and showed that under controlled ion population, the growth of the instability can be addressed via a scaling law of the growth rate with the ion density value. In experimental conditions, ions are produced continuously via electron-impact ionization and other collisional processes play a crucial role in the energy exchange processes. An investigation considering the impact ionization, electron-neutral and ion-neutral collision operations is required. In an ongoing study we have implemented these collisional processes and the results will be reported soon.\

\begin{acknowledgments}
All numerical results reported in this study were obtained using the ANTYA cluster at IPR. The authors would like to thank the staff of the Computer Center of IPR for facilities of the ANTYA HPC. The authors would like to thank Dr. Nirmal Kumar Bisai for careful reading of the manuscript and for his comments.
\end{acknowledgments}
\section{Data availability}
The data that support the findings of this study are available from the corresponding author upon reasonable request.


\begin{thebibliography}{38}%
\makeatletter
\providecommand \@ifxundefined [1]{%
 \@ifx{#1\undefined}
}%
\providecommand \@ifnum [1]{%
 \ifnum #1\expandafter \@firstoftwo
 \else \expandafter \@secondoftwo
 \fi
}%
\providecommand \@ifx [1]{%
 \ifx #1\expandafter \@firstoftwo
 \else \expandafter \@secondoftwo
 \fi
}%
\providecommand \natexlab [1]{#1}%
\providecommand \enquote  [1]{``#1''}%
\providecommand \bibnamefont  [1]{#1}%
\providecommand \bibfnamefont [1]{#1}%
\providecommand \citenamefont [1]{#1}%
\providecommand \href@noop [0]{\@secondoftwo}%
\providecommand \href [0]{\begingroup \@sanitize@url \@href}%
\providecommand \@href[1]{\@@startlink{#1}\@@href}%
\providecommand \@@href[1]{\endgroup#1\@@endlink}%
\providecommand \@sanitize@url [0]{\catcode `\\12\catcode `\$12\catcode
  `\&12\catcode `\#12\catcode `\^12\catcode `\_12\catcode `\%12\relax}%
\providecommand \@@startlink[1]{}%
\providecommand \@@endlink[0]{}%
\providecommand \url  [0]{\begingroup\@sanitize@url \@url }%
\providecommand \@url [1]{\endgroup\@href {#1}{\urlprefix }}%
\providecommand \urlprefix  [0]{URL }%
\providecommand \Eprint [0]{\href }%
\providecommand \doibase [0]{http://dx.doi.org/}%
\providecommand \selectlanguage [0]{\@gobble}%
\providecommand \bibinfo  [0]{\@secondoftwo}%
\providecommand \bibfield  [0]{\@secondoftwo}%
\providecommand \translation [1]{[#1]}%
\providecommand \BibitemOpen [0]{}%
\providecommand \bibitemStop [0]{}%
\providecommand \bibitemNoStop [0]{.\EOS\space}%
\providecommand \EOS [0]{\spacefactor3000\relax}%
\providecommand \BibitemShut  [1]{\csname bibitem#1\endcsname}%
\let\auto@bib@innerbib\@empty
\bibitem [{\citenamefont {Malmberg}\ and\ \citenamefont
  {deGrassie}(1975)}]{Malmberg1975}%
  \BibitemOpen
  \bibfield  {author} {\bibinfo {author} {\bibfnamefont {J.~H.}\ \bibnamefont
  {Malmberg}}\ and\ \bibinfo {author} {\bibfnamefont {J.~S.}\ \bibnamefont
  {deGrassie}},\ }\bibfield  {title} {\enquote {\bibinfo {title} {Properties of
  nonneutral plasma},}\ }\href {\doibase 10.1103/PhysRevLett.35.577} {\bibfield
   {journal} {\bibinfo  {journal} {Phys. Rev. Lett.}\ }\textbf {\bibinfo
  {volume} {35}},\ \bibinfo {pages} {577--580} (\bibinfo {year}
  {1975})}\BibitemShut {NoStop}%
  \bibitem [{\citenamefont {Paul}(1990)}]{Paul}%
  \BibitemOpen
  \bibfield  {author} {\bibinfo {author} {\bibfnamefont {W.}~\bibnamefont
  {Paul}},\ }\bibfield  {title} {\enquote {\bibinfo {title} {Electromagnetic
  traps for charged and neutral particles (nobel lecture)},}\ }\href {\doibase
  https://doi.org/10.1002/anie.199007391} {\bibfield  {journal} {\bibinfo
  {journal} {Angewandte Chemie International Edition in English}\ }\textbf
  {\bibinfo {volume} {29}},\ \bibinfo {pages} {739--748} (\bibinfo {year}
  {1990})},\ \Eprint
  {http://arxiv.org/abs/https://onlinelibrary.wiley.com/doi/pdf/10.1002/anie.199007391}
  {https://onlinelibrary.wiley.com/doi/pdf/10.1002/anie.199007391} \BibitemShut
  {NoStop}%
  \bibitem [{\citenamefont {Yamada}\ and\ \citenamefont
  {Himura}(2016)}]{Yamada2016}%
  \BibitemOpen
  \bibfield  {author} {\bibinfo {author} {\bibfnamefont {S.}~\bibnamefont
  {Yamada}}\ and\ \bibinfo {author} {\bibfnamefont {H.}~\bibnamefont
  {Himura}},\ }\bibfield  {title} {\enquote {\bibinfo {title} {Note:
  Consecutive capture of images of ions and electrons using high-voltage vacuum
  relay},}\ }\href {\doibase 10.1063/1.4944861} {\bibfield  {journal} {\bibinfo
   {journal} {Review of Scientific Instruments}\ }\textbf {\bibinfo {volume}
  {87}},\ \bibinfo {pages} {036109} (\bibinfo {year} {2016})},\ \Eprint
  {http://arxiv.org/abs/https://aip.scitation.org/doi/pdf/10.1063/1.4944861}
  {https://aip.scitation.org/doi/pdf/10.1063/1.4944861} \BibitemShut {NoStop}%
  \bibitem [{\citenamefont {Hurst}\ \emph {et~al.}(2014)\citenamefont {Hurst},
  \citenamefont {Danielson}, \citenamefont {Baker},\ and\ \citenamefont
  {Surko}}]{Hurst2014}%
  \BibitemOpen
  \bibfield  {author} {\bibinfo {author} {\bibfnamefont {N.~C.}\ \bibnamefont
  {Hurst}}, \bibinfo {author} {\bibfnamefont {J.~R.}\ \bibnamefont
  {Danielson}}, \bibinfo {author} {\bibfnamefont {C.~J.}\ \bibnamefont
  {Baker}}, \ and\ \bibinfo {author} {\bibfnamefont {C.~M.}\ \bibnamefont
  {Surko}},\ }\bibfield  {title} {\enquote {\bibinfo {title} {Electron plasma
  orbits from competing Diocotron drifts},}\ }\href {\doibase
  10.1103/PhysRevLett.113.025004} {\bibfield  {journal} {\bibinfo  {journal}
  {Phys. Rev. Lett.}\ }\textbf {\bibinfo {volume} {113}},\ \bibinfo {pages}
  {025004} (\bibinfo {year} {2014})}\BibitemShut {NoStop}%
  \bibitem [{\citenamefont {Lane}\ and\ \citenamefont
  {Orperformedz}(2019)}]{Lane2019}%
  \BibitemOpen
  \bibfield  {author} {\bibinfo {author} {\bibfnamefont {R.~A.}\ \bibnamefont
  {Lane}}\ and\ \bibinfo {author} {\bibfnamefont {C.~A.}\ \bibnamefont
  {Orperformedz}},\ }\bibfield  {title} {\enquote {\bibinfo {title} {Electrostatic
  equilibria of non-neutral plasmas confined in a penning trap with axially
  varying magnetic field},}\ }\href {\doibase 10.1063/1.5092136} {\bibfield
  {journal} {\bibinfo  {journal} {Physics of Plasmas}\ }\textbf {\bibinfo
  {volume} {26}},\ \bibinfo {pages} {052511} (\bibinfo {year} {2019})},\
  \Eprint {http://arxiv.org/abs/https://doi.org/10.1063/1.5092136}
  {https://doi.org/10.1063/1.5092136} \BibitemShut {NoStop}%
  \bibitem [{\citenamefont {Davidson}\ and\ \citenamefont
  {Felice}(1998)}]{Davidson1998}%
  \BibitemOpen
  \bibfield  {author} {\bibinfo {author} {\bibfnamefont {R.~C.}\ \bibnamefont
  {Davidson}}\ and\ \bibinfo {author} {\bibfnamefont {G.~M.}\ \bibnamefont
  {Felice}},\ }\bibfield  {title} {\enquote {\bibinfo {title} {Influence of
  profile shape on the Diocotron instability in a non-neutral plasma column},}\
  }\href {\doibase 10.1063/1.873067} {\bibfield  {journal} {\bibinfo  {journal}
  {Physics of Plasmas}\ }\textbf {\bibinfo {volume} {5}},\ \bibinfo {pages}
  {3497--3511} (\bibinfo {year} {1998})},\ \Eprint
  {http://arxiv.org/abs/https://doi.org/10.1063/1.873067}
  {https://doi.org/10.1063/1.873067} \BibitemShut {NoStop}%
\bibitem [{\citenamefont {O’Neil}\ and\ \citenamefont
  {Dubin}(1998)}]{ONeil1998}%
  \BibitemOpen
  \bibfield  {author} {\bibinfo {author} {\bibfnamefont {T.~M.}\ \bibnamefont
  {O’Neil}}\ and\ \bibinfo {author} {\bibfnamefont {D.~H.~E.}\ \bibnamefont
  {Dubin}},\ }\bibfield  {title} {\enquote {\bibinfo {title} {Thermal
  equilibria and thermodynamics of trapped plasmas with a single sign of
  charge},}\ }\href {\doibase 10.1063/1.872925} {\bibfield  {journal} {\bibinfo
   {journal} {Physics of Plasmas}\ }\textbf {\bibinfo {volume} {5}},\ \bibinfo
  {pages} {2163--2193} (\bibinfo {year} {1998})},\ \Eprint
  {http://arxiv.org/abs/https://doi.org/10.1063/1.872925}
  {https://doi.org/10.1063/1.872925} \BibitemShut {NoStop}%
\bibitem [{\citenamefont {Landa}(2019)}]{Landa2019}%
  \BibitemOpen
  \bibfield  {author} {\bibinfo {author} {\bibfnamefont {H.}~\bibnamefont
  {Landa}},\ }\bibfield  {title} {\enquote {\bibinfo {title} {Tuning nonthermal
  distributions to thermal ones in time-dependent paul traps},}\ }\href
  {\doibase 10.1103/PhysRevA.100.013413} {\bibfield  {journal} {\bibinfo
  {journal} {Phys. Rev. A}\ }\textbf {\bibinfo {volume} {100}},\ \bibinfo
  {pages} {013413} (\bibinfo {year} {2019})}\BibitemShut {NoStop}%
  \bibitem [{\citenamefont {Greaves}\ and\ \citenamefont
  {Surko}(2000)}]{Greaves2000}%
  \BibitemOpen
  \bibfield  {author} {\bibinfo {author} {\bibfnamefont {R.~G.}\ \bibnamefont
  {Greaves}}\ and\ \bibinfo {author} {\bibfnamefont {C.~M.}\ \bibnamefont
  {Surko}},\ }\bibfield  {title} {\enquote {\bibinfo {title} {Inward transport
  and compression of a positron plasma by a rotating electric field},}\ }\href
  {\doibase 10.1103/PhysRevLett.85.1883} {\bibfield  {journal} {\bibinfo
  {journal} {Phys. Rev. Lett.}\ }\textbf {\bibinfo {volume} {85}},\ \bibinfo
  {pages} {1883--1886} (\bibinfo {year} {2000})}\BibitemShut {NoStop}%
   \bibitem [{\citenamefont {Ahmadi}\ and\ \citenamefont
  {et~al.}(2016)}]{Ahmadi2016}%
  \BibitemOpen
  \bibfield  {author} {\bibinfo {author} {\bibfnamefont {M.}~\bibnamefont
  {Ahmadi}, \bibinfo {author} {\bibfnamefont {M.}\ \bibnamefont
  {Baquero-Ruiz}}, \bibinfo {author} {\bibfnamefont {W.}\ \bibnamefont
  {Bertsche}}, \bibinfo {author} {\bibfnamefont {E.}\ \bibnamefont
  {Butler}}, \bibinfo {author} {\bibfnamefont {A.}\ \bibnamefont
  {Capra}}, \bibinfo {author} {\bibfnamefont {C.}\ \bibnamefont
  {Carruth}}, \bibinfo {author} {\bibfnamefont {C.~L.}\ \bibnamefont
  {Cesar}}, \bibinfo {author} {\bibfnamefont {M.}\ \bibnamefont
  {Charlton}}, \bibinfo {author} {\bibfnamefont {A.~E.}\ \bibnamefont
  {Charman}}, \bibinfo {author} {\bibfnamefont {S.}\ \bibnamefont
  {Eriksson}}} \bibinfo {author} {\bibnamefont {et~al.,}}\ }\href {\doibase
  10.1038/nature16491} {\bibfield  {journal} {\bibinfo  {journal} {Nature}\
  }\textbf {\bibinfo {volume} {529}},\ \bibinfo {pages} {373--376} (\bibinfo
  {year} {2016})}\BibitemShut {NoStop}%
\bibitem [{\citenamefont {Ahmadi}\ \emph {et~al.}(2018)\citenamefont {Ahmadi},
  \citenamefont {Alves}, \citenamefont {Baker}, \citenamefont {Bertsche},
  \citenamefont {Capra}, \citenamefont {Carruth}, \citenamefont {Cesar},
  \citenamefont {Charlton}, \citenamefont {Cohen}, \citenamefont {Collister}}]{Ahmadi2018}%
  \BibitemOpen
  \bibfield  {author} {\bibinfo {author} {\bibfnamefont {M.}~\bibnamefont
  {Ahmadi}}, \bibinfo {author} {\bibfnamefont {B.~X.~R.}\ \bibnamefont
  {Alves}}, \bibinfo {author} {\bibfnamefont {C.~J.}\ \bibnamefont {Baker}},
  \bibinfo {author} {\bibfnamefont {W.}~\bibnamefont {Bertsche}}, \bibinfo
  {author} {\bibfnamefont {A.}~\bibnamefont {Capra}}, \bibinfo {author}
  {\bibfnamefont {C.}~\bibnamefont {Carruth}}, \bibinfo {author} {\bibfnamefont
  {C.~L.}\ \bibnamefont {Cesar}}, \bibinfo {author} {\bibfnamefont
  {M.}~\bibnamefont {Charlton}}, \bibinfo {author} {\bibfnamefont
  {S.}~\bibnamefont {Cohen}}, \bibinfo {author} {\bibfnamefont
  {R.}~\bibnamefont {Collister}} \bibinfo {author} {\bibnamefont
  {et~al.,}} (\bibinfo {collaboration} {ALPHA Collaboration}),\ }\bibfield
  {title} {\enquote {\bibinfo {title} {Enhanced control and reproducibility of
  non-neutral plasmas},}\ }\href {\doibase 10.1103/PhysRevLett.120.025001}
  {\bibfield  {journal} {\bibinfo  {journal} {Phys. Rev. Lett.}\ }\textbf
  {\bibinfo {volume} {120}},\ \bibinfo {pages} {025001} (\bibinfo {year}
  {2018})}\BibitemShut {NoStop}%
\bibitem [{\citenamefont {O’Neil}\ and\ \citenamefont
  {Driscoll}(1979)}]{ONeil1979}%
  \BibitemOpen
  \bibfield  {author} {\bibinfo {author} {\bibfnamefont {T.~M.}\ \bibnamefont
  {O’Neil}}\ and\ \bibinfo {author} {\bibfnamefont {C.~F.}\ \bibnamefont
  {Driscoll}},\ }\bibfield  {title} {\enquote {\bibinfo {title} {Transport to
  thermal equilibrium of a pure electron plasma},}\ }\href {\doibase
  10.1063/1.862577} {\bibfield  {journal} {\bibinfo  {journal} {The Physics of
  Fluids}\ }\textbf {\bibinfo {volume} {22}},\ \bibinfo {pages} {266--277}
  (\bibinfo {year} {1979})},\ \Eprint
  {http://arxiv.org/abs/https://aip.scitation.org/doi/pdf/10.1063/1.862577}
  {https://aip.scitation.org/doi/pdf/10.1063/1.862577} \BibitemShut {NoStop}%
  \bibitem [{\citenamefont {Malmberg}\ and\ \citenamefont
  {Driscoll}(1980)}]{Malmberg1980}%
  \BibitemOpen
  \bibfield  {author} {\bibinfo {author} {\bibfnamefont {J.~H.}\ \bibnamefont
  {Malmberg}}\ and\ \bibinfo {author} {\bibfnamefont {C.~F.}\ \bibnamefont
  {Driscoll}},\ }\bibfield  {title} {\enquote {\bibinfo {title} {Long-time
  containment of a pure electron plasma},}\ }\href {\doibase
  10.1103/PhysRevLett.44.654} {\bibfield  {journal} {\bibinfo  {journal} {Phys.
  Rev. Lett.}\ }\textbf {\bibinfo {volume} {44}},\ \bibinfo {pages} {654--657}
  (\bibinfo {year} {1980})}\BibitemShut {NoStop}%
   \bibitem [{\citenamefont {Driscoll}, \citenamefont {Malmberg},\ and\
  \citenamefont {Fine}(1988)}]{Driscoll1988}%
  \BibitemOpen
  \bibfield  {author} {\bibinfo {author} {\bibfnamefont {C.~F.}\ \bibnamefont
  {Driscoll}}, \bibinfo {author} {\bibfnamefont {J.~H.}\ \bibnamefont
  {Malmberg}}, \ and\ \bibinfo {author} {\bibfnamefont {K.~S.}\ \bibnamefont
  {Fine}},\ }\bibfield  {title} {\enquote {\bibinfo {title} {Observation of
  transport to thermal equilibrium in pure electron plasmas},}\ }\href
  {\doibase 10.1103/PhysRevLett.60.1290} {\bibfield  {journal} {\bibinfo
  {journal} {Phys. Rev. Lett.}\ }\textbf {\bibinfo {volume} {60}},\ \bibinfo
  {pages} {1290--1293} (\bibinfo {year} {1988})}\BibitemShut {NoStop}%
\bibitem [{\citenamefont {Dubin}\ and\ \citenamefont
  {O'Neil}(1999)}]{Dubin1999}%
  \BibitemOpen
  \bibfield  {author} {\bibinfo {author} {\bibfnamefont {D.~H.~E.}\
  \bibnamefont {Dubin}}\ and\ \bibinfo {author} {\bibfnamefont {T.~M.}\
  \bibnamefont {O'Neil}},\ }\href {\doibase 10.1103/RevModPhys.71.87}
  {\bibfield  {journal} {\bibinfo  {journal} {Rev. Mod. Phys.}\ }\textbf
  {\bibinfo {volume} {71}},\ \bibinfo {pages} {87--172} (\bibinfo {year}
  {1999})}\BibitemShut {NoStop}%
\bibitem [{\citenamefont {Fine}\ \emph {et~al.}(1995)\citenamefont {Fine},
  \citenamefont {Cass}, \citenamefont {Flynn},\ and\ \citenamefont
  {Driscoll}}]{Fine1995}%
  \BibitemOpen
  \bibfield  {author} {\bibinfo {author} {\bibfnamefont {K.~S.}\ \bibnamefont
  {Fine}}, \bibinfo {author} {\bibfnamefont {A.~C.}\ \bibnamefont {Cass}},
  \bibinfo {author} {\bibfnamefont {W.~G.}\ \bibnamefont {Flynn}}, \ and\
  \bibinfo {author} {\bibfnamefont {C.~F.}\ \bibnamefont {Driscoll}},\ }\href
  {\doibase 10.1103/PhysRevLett.75.3277} {\bibfield  {journal} {\bibinfo
  {journal} {Phys. Rev. Lett.}\ }\textbf {\bibinfo {volume} {75}},\ \bibinfo
  {pages} {3277--3280} (\bibinfo {year} {1995})}\BibitemShut {NoStop}%
  \bibitem [{\citenamefont {Briggs}, \citenamefont {Daugherty},\ and\
  \citenamefont {Levy}(1970)}]{Briggs1970}%
  \BibitemOpen
  \bibfield  {author} {\bibinfo {author} {\bibfnamefont {R.~J.}\ \bibnamefont
  {Briggs}}, \bibinfo {author} {\bibfnamefont {J.~D.}\ \bibnamefont
  {Daugherty}}, \ and\ \bibinfo {author} {\bibfnamefont {R.~H.}\ \bibnamefont
  {Levy}},\ }\bibfield  {title} {\enquote {\bibinfo {title} {Role of landau
  damping in crossed‐field electron beams and inviscid shear flow},}\ }\href
  {\doibase 10.1063/1.1692936} {\bibfield  {journal} {\bibinfo  {journal} {The
  Physics of Fluids}\ }\textbf {\bibinfo {volume} {13}},\ \bibinfo {pages}
  {421--432} (\bibinfo {year} {1970})},\ \Eprint
  {http://arxiv.org/abs/https://aip.scitation.org/doi/pdf/10.1063/1.1692936}
  {https://aip.scitation.org/doi/pdf/10.1063/1.1692936} \BibitemShut {NoStop}%
  \bibitem [{\citenamefont {Ganesh}\ and\ \citenamefont
  {Pahari}(2006)}]{Ganesh2006}%
  \BibitemOpen
  \bibfield  {author} {\bibinfo {author} {\bibfnamefont {R.}~\bibnamefont
  {Ganesh}}\ and\ \bibinfo {author} {\bibfnamefont {S.}~\bibnamefont
  {Pahari}},\ }in\ \href {\doibase 10.1142/9789812771025_fmatter} {\emph
  {\bibinfo {booktitle} {Frontiers in Turbulence and Coherent Structures}}},\
  Vol.~\bibinfo {volume} {6}\ (\bibinfo  {publisher} {World Scientific},\
  \bibinfo {year} {2006})\ p.\ \bibinfo {pages} {471}\BibitemShut {NoStop}%
\bibitem [{\citenamefont {Rome}\ \emph {et~al.}(2000)\citenamefont {Rome},
  \citenamefont {Brunetti}, \citenamefont {Califano}, \citenamefont
  {Pegoraro},\ and\ \citenamefont {Pozzoli}}]{Rome2000}%
  \BibitemOpen
  \bibfield  {author} {\bibinfo {author} {\bibfnamefont {M.}~\bibnamefont
  {Rome}}, \bibinfo {author} {\bibfnamefont {M.}~\bibnamefont {Brunetti}},
  \bibinfo {author} {\bibfnamefont {F.}~\bibnamefont {Califano}}, \bibinfo
  {author} {\bibfnamefont {F.}~\bibnamefont {Pegoraro}}, \ and\ \bibinfo
  {author} {\bibfnamefont {R.}~\bibnamefont {Pozzoli}},\ }\bibfield  {title}
  {\enquote {\bibinfo {title} {Motion of extended vortices in an inhomogeneous
  pure electron plasma},}\ }\href {\doibase 10.1063/1.874135} {\bibfield
  {journal} {\bibinfo  {journal} {Physics of Plasmas}\ }\textbf {\bibinfo
  {volume} {7}},\ \bibinfo {pages} {2856--2865} (\bibinfo {year} {2000})},\
  \Eprint {http://arxiv.org/abs/https://doi.org/10.1063/1.874135}
  {https://doi.org/10.1063/1.874135} \BibitemShut {NoStop}%
\bibitem [{\citenamefont {Sengupta}\ and\ \citenamefont
  {Ganesh}(2014)}]{Sengupta2014}%
  \BibitemOpen
  \bibfield  {author} {\bibinfo {author} {\bibfnamefont {M.}~\bibnamefont
  {Sengupta}}\ and\ \bibinfo {author} {\bibfnamefont {R.}~\bibnamefont
  {Ganesh}},\ }\bibfield  {title} {\enquote {\bibinfo {title} {Inertia driven
  radial breathing and nonlinear relaxation in cylindrically confined pure
  electron plasma},}\ }\href {\doibase 10.1063/1.4866022} {\bibfield  {journal}
  {\bibinfo  {journal} {Physics of Plasmas}\ }\textbf {\bibinfo {volume}
  {21}},\ \bibinfo {pages} {022116} (\bibinfo {year} {2014})},\ \Eprint
  {http://arxiv.org/abs/https://doi.org/10.1063/1.4866022}
  {https://doi.org/10.1063/1.4866022} \BibitemShut {NoStop}%
 \bibitem [{\citenamefont {Birdsall}\ and\ \citenamefont
  {Langdon}(2004)}]{Birdsall}%
  \BibitemOpen
  \bibfield  {author} {\bibinfo {author} {\bibfnamefont {C.}~\bibnamefont
  {Birdsall}}\ and\ \bibinfo {author} {\bibfnamefont {A.}~\bibnamefont
  {Langdon}},\ }\href {https://books.google.co.in/books?id=S2lqgDTm6a4C} {\emph
  {\bibinfo {title} {Plasma Physics via Computer Simulation}}},\ Series in
  Plasma Physics and Fluid Dynamics\ (\bibinfo  {publisher} {Taylor \&
  Francis},\ \bibinfo {year} {2004})\BibitemShut {NoStop}%
  \bibitem [{\citenamefont {Niemann}\ \emph {et~al.}(2019)\citenamefont
  {Niemann}, \citenamefont {Meiners}, \citenamefont {Mielke}, \citenamefont
  {Borchert}, \citenamefont {Cornejo}, \citenamefont {Ulmer},\ and\
  \citenamefont {Ospelkaus}}]{Niemann2019}%
  \BibitemOpen
  \bibfield  {author} {\bibinfo {author} {\bibfnamefont {M.}~\bibnamefont
  {Niemann}}, \bibinfo {author} {\bibfnamefont {T.}~\bibnamefont {Meiners}},
  \bibinfo {author} {\bibfnamefont {J.}~\bibnamefont {Mielke}}, \bibinfo
  {author} {\bibfnamefont {M.~J.}\ \bibnamefont {Borchert}}, \bibinfo {author}
  {\bibfnamefont {J.~M.}\ \bibnamefont {Cornejo}}, \bibinfo {author}
  {\bibfnamefont {S.}~\bibnamefont {Ulmer}}, \ and\ \bibinfo {author}
  {\bibfnamefont {C.}~\bibnamefont {Ospelkaus}},\ }\href {\doibase
  10.1088/1361-6501/ab5722} {\bibfield  {journal} {\bibinfo  {journal}
  {Measurement Science and Technology}\ }\textbf {\bibinfo {volume} {31}},\
  \bibinfo {pages} {035003} (\bibinfo {year} {2019})}\BibitemShut {NoStop}%
\bibitem [{\citenamefont {Schuh}\ \emph {et~al.}(2019)\citenamefont {Schuh},
  \citenamefont {Hei\ss{}e}, \citenamefont {Eronen}, \citenamefont {Ketter},
  \citenamefont {K\"ohler-Langes}, \citenamefont {Rau}, \citenamefont {Segal},
  \citenamefont {Quint}, \citenamefont {Sturm},\ and\ \citenamefont
  {Blaum}}]{Schuh2019}%
  \BibitemOpen
  \bibfield  {author} {\bibinfo {author} {\bibfnamefont {M.}~\bibnamefont
  {Schuh}}, \bibinfo {author} {\bibfnamefont {F.}~\bibnamefont {Hei\ss{}e}},
  \bibinfo {author} {\bibfnamefont {T.}~\bibnamefont {Eronen}}, \bibinfo
  {author} {\bibfnamefont {J.}~\bibnamefont {Ketter}}, \bibinfo {author}
  {\bibfnamefont {F.}~\bibnamefont {K\"ohler-Langes}}, \bibinfo {author}
  {\bibfnamefont {S.}~\bibnamefont {Rau}}, \bibinfo {author} {\bibfnamefont
  {T.}~\bibnamefont {Segal}}, \bibinfo {author} {\bibfnamefont
  {W.}~\bibnamefont {Quint}}, \bibinfo {author} {\bibfnamefont
  {S.}~\bibnamefont {Sturm}}, \ and\ \bibinfo {author} {\bibfnamefont
  {K.}~\bibnamefont {Blaum}},\ }\bibfield  {title} {\enquote {\bibinfo {title}
  {Image charge shift in high-precision penning traps},}\ }\href {\doibase
  10.1103/PhysRevA.100.023411} {\bibfield  {journal} {\bibinfo  {journal}
  {Phys. Rev. A}\ }\textbf {\bibinfo {volume} {100}},\ \bibinfo {pages}
  {023411} (\bibinfo {year} {2019})}\BibitemShut {NoStop}%
\bibitem [{\citenamefont {Guti\'errez}\ \emph {et~al.}(2019)\citenamefont
  {Guti\'errez}, \citenamefont {Berrocal}, \citenamefont {Dom\'{\i}nguez},
  \citenamefont {Arrazola}, \citenamefont {Block}, \citenamefont {Solano},\
  and\ \citenamefont {Rodr\'{\i}guez}}]{Guti2019}%
  \BibitemOpen
  \bibfield  {author} {\bibinfo {author} {\bibfnamefont {M.~J.}\ \bibnamefont
  {Guti\'errez}}, \bibinfo {author} {\bibfnamefont {J.}~\bibnamefont
  {Berrocal}}, \bibinfo {author} {\bibfnamefont {F.}~\bibnamefont
  {Dom\'{\i}nguez}}, \bibinfo {author} {\bibfnamefont {I.}~\bibnamefont
  {Arrazola}}, \bibinfo {author} {\bibfnamefont {M.}~\bibnamefont {Block}},
  \bibinfo {author} {\bibfnamefont {E.}~\bibnamefont {Solano}}, \ and\ \bibinfo
  {author} {\bibfnamefont {D.}~\bibnamefont {Rodr\'{\i}guez}},\ }\bibfield
  {title} {\enquote {\bibinfo {title} {Dynamics of an unbalanced two-ion
  crystal in a penning trap for application in optical mass spectrometry},}\
  }\href {\doibase 10.1103/PhysRevA.100.063415} {\bibfield  {journal} {\bibinfo
   {journal} {Phys. Rev. A}\ }\textbf {\bibinfo {volume} {100}},\ \bibinfo
  {pages} {063415} (\bibinfo {year} {2019})}\BibitemShut {NoStop}%
\bibitem [{\citenamefont {Peurrung}, \citenamefont {Kouzes},\ and\
  \citenamefont {Barlow}(1996)}]{Peurrung1996}%
  \BibitemOpen
  \bibfield  {author} {\bibinfo {author} {\bibfnamefont {A.}~\bibnamefont
  {Peurrung}}, \bibinfo {author} {\bibfnamefont {R.}~\bibnamefont {Kouzes}}, \
  and\ \bibinfo {author} {\bibfnamefont {S.}~\bibnamefont {Barlow}},\
  }\bibfield  {title} {\enquote {\bibinfo {title} {The non-neutral plasma: an
  introduction to physics with relevance to cyclotron resonance mass
  spectrometry},}\ }\href {\doibase
  https://doi.org/10.1016/S0168-1176(96)04395-9} {\bibfield  {journal}
  {\bibinfo  {journal} {International Journal of Mass Spectrometry and Ion
  Processes}\ }\textbf {\bibinfo {volume} {157-158}},\ \bibinfo {pages} {39 --
  83} (\bibinfo {year} {1996})}\BibitemShut {NoStop}%
\bibitem [{\citenamefont {Fajans}\ and\ \citenamefont
  {Surko}(2020)}]{Fajans2020}%
  \BibitemOpen
  \bibfield  {author} {\bibinfo {author} {\bibfnamefont {J.}~\bibnamefont
  {Fajans}}\ and\ \bibinfo {author} {\bibfnamefont {C.~M.}\ \bibnamefont
  {Surko}},\ }\bibfield  {title} {\enquote {\bibinfo {title} {Plasma and
  trap-based techniques for science with antimatter},}\ }\href {\doibase
  10.1063/1.5131273} {\bibfield  {journal} {\bibinfo  {journal} {Physics of
  Plasmas}\ }\textbf {\bibinfo {volume} {27}},\ \bibinfo {pages} {030601}
  (\bibinfo {year} {2020})},\ \Eprint
  {http://arxiv.org/abs/https://doi.org/10.1063/1.5131273}
  {https://doi.org/10.1063/1.5131273} \BibitemShut {NoStop}%
  \bibitem [{\citenamefont {Morigi}\ \emph {et~al.}(1999)\citenamefont {Morigi},
  \citenamefont {Eschner}, \citenamefont {Cirac},\ and\ \citenamefont
  {Zoller}}]{Morigi1999}%
  \BibitemOpen
  \bibfield  {author} {\bibinfo {author} {\bibfnamefont {G.}~\bibnamefont
  {Morigi}}, \bibinfo {author} {\bibfnamefont {J.}~\bibnamefont {Eschner}},
  \bibinfo {author} {\bibfnamefont {J.~I.}\ \bibnamefont {Cirac}}, \ and\
  \bibinfo {author} {\bibfnamefont {P.}~\bibnamefont {Zoller}},\ }\bibfield
  {title} {\enquote {\bibinfo {title} {Laser cooling of two trapped ions:
  Sideband cooling beyond the lamb-dicke limit},}\ }\href {\doibase
  10.1103/PhysRevA.59.3797} {\bibfield  {journal} {\bibinfo  {journal} {Phys.
  Rev. A}\ }\textbf {\bibinfo {volume} {59}},\ \bibinfo {pages} {3797--3808}
  (\bibinfo {year} {1999})}\BibitemShut {NoStop}%
  \bibitem [{\citenamefont {Morigi}\ and\ \citenamefont
  {Walther}(2001)}]{Morigi2001}%
  \BibitemOpen
  \bibfield  {author} {\bibinfo {author} {\bibfnamefont {G.}~\bibnamefont
  {Morigi}}\ and\ \bibinfo {author} {\bibfnamefont {H.}~\bibnamefont
  {Walther}},\ }\bibfield  {title} {\enquote {\bibinfo {title} {Two-species
  coulomb chains for quantum information},}\ }\href {\doibase
  10.1007/s100530170275} {\bibfield  {journal} {\bibinfo  {journal} {The
  European Physical Journal D - Atomic, Molecular, Optical and Plasma Physics}\
  }\textbf {\bibinfo {volume} {13}} (\bibinfo {year} {2001}),\
  10.1007/s100530170275}\BibitemShut {NoStop}%
\bibitem [{\citenamefont {Douglas}, \citenamefont {Frank},\ and\ \citenamefont
  {Mao}(2005)}]{Douglas2005}%
  \BibitemOpen
  \bibfield  {author} {\bibinfo {author} {\bibfnamefont {D.~J.}\ \bibnamefont
  {Douglas}}, \bibinfo {author} {\bibfnamefont {A.~J.}\ \bibnamefont {Frank}},
  \ and\ \bibinfo {author} {\bibfnamefont {D.}~\bibnamefont {Mao}},\ }\bibfield
   {title} {\enquote {\bibinfo {title} {Linear ion traps in mass
  spectrometry},}\ }\href {\doibase https://doi.org/10.1002/mas.20004}
  {\bibfield  {journal} {\bibinfo  {journal} {Mass Spectrometry Reviews}\
  }\textbf {\bibinfo {volume} {24}},\ \bibinfo {pages} {1--29} (\bibinfo {year}
  {2005})}
  \BibitemShut {NoStop}%
  \bibitem [{\citenamefont {Greaves}\ and\ \citenamefont
  {Surko}(2002)}]{Greaves2002}%
  \BibitemOpen
  \bibfield  {author} {\bibinfo {author} {\bibfnamefont {R.~G.}\ \bibnamefont
  {Greaves}}\ and\ \bibinfo {author} {\bibfnamefont {C.~M.}\ \bibnamefont
  {Surko}},\ }\bibfield  {title} {\enquote {\bibinfo {title} {Practical limits
  on positron accumulation and the creation of electron-positron plasmas},}\
  }\href {\doibase 10.1063/1.1454263} {\bibfield  {journal} {\bibinfo
  {journal} {AIP Conference Proceedings}\ }\textbf {\bibinfo {volume} {606}},\
  \bibinfo {pages} {10--23} (\bibinfo {year} {2002})},\ \Eprint
  {http://arxiv.org/abs/https://aip.scitation.org/doi/pdf/10.1063/1.1454263}
  {https://aip.scitation.org/doi/pdf/10.1063/1.1454263} \BibitemShut {NoStop}%
  \bibitem [{\citenamefont {Higaki}\ \emph {et~al.}(2017)\citenamefont {Higaki},
  \citenamefont {Kaga}, \citenamefont {Fukushima}, \citenamefont {Okamoto},
  \citenamefont {Nagata}, \citenamefont {Kanai},\ and\ \citenamefont
  {Yamazaki}}]{Higaki2017}%
  \BibitemOpen
  \bibfield  {author} {\bibinfo {author} {\bibfnamefont {H.}~\bibnamefont
  {Higaki}}, \bibinfo {author} {\bibfnamefont {C.}~\bibnamefont {Kaga}},
  \bibinfo {author} {\bibfnamefont {K.}~\bibnamefont {Fukushima}}, \bibinfo
  {author} {\bibfnamefont {H.}~\bibnamefont {Okamoto}}, \bibinfo {author}
  {\bibfnamefont {Y.}~\bibnamefont {Nagata}}, \bibinfo {author} {\bibfnamefont
  {Y.}~\bibnamefont {Kanai}}, \ and\ \bibinfo {author} {\bibfnamefont
  {Y.}~\bibnamefont {Yamazaki}},\ }\bibfield  {title} {\enquote {\bibinfo
  {title} {Simultaneous confinement of low-energy electrons and positrons in a
  compact magnetic mirror trap},}\ }\href {\doibase 10.1088/1367-2630/aa5a45}
  {\bibfield  {journal} {\bibinfo  {journal} {New Journal of Physics}\ }\textbf
  {\bibinfo {volume} {19}},\ \bibinfo {pages} {023016} (\bibinfo {year}
  {2017})}\BibitemShut {NoStop}%
  \bibitem [{\citenamefont {Hicks}, \citenamefont {Bowman},\ and\ \citenamefont
  {Godden}(2019)}]{Hicks2019}%
  \BibitemOpen
  \bibfield  {author} {\bibinfo {author} {\bibfnamefont {N.~K.}\ \bibnamefont
  {Hicks}}, \bibinfo {author} {\bibfnamefont {A.}~\bibnamefont {Bowman}}, \
  and\ \bibinfo {author} {\bibfnamefont {K.}~\bibnamefont {Godden}},\
  }\bibfield  {title} {\enquote {\bibinfo {title} {Particle-in-cell simulation
  of quasi-neutral plasma trapping by rf multipole electric fields},}\ }\href
  {\doibase 10.3390/physics1030028} {\bibfield  {journal} {\bibinfo  {journal}
  {Physics}\ }\textbf {\bibinfo {volume} {1}},\ \bibinfo {pages} {392--401}
  (\bibinfo {year} {2019})}\BibitemShut {NoStop}%
  \bibitem [{\citenamefont {Nakajima}, \citenamefont {Himura},\ and\
  \citenamefont {Sanpei}(2021)}]{nakajima2021}%
  \BibitemOpen
  \bibfield  {author} {\bibinfo {author} {\bibfnamefont {Y.}~\bibnamefont
  {Nakajima}}, \bibinfo {author} {\bibfnamefont {H.}~\bibnamefont {Himura}}, \
  and\ \bibinfo {author} {\bibfnamefont {A.}~\bibnamefont {Sanpei}},\
  }\bibfield  {title} {\enquote {\bibinfo {title} {Counter differential
  rigid-rotation equilibrium of electrically non-neutral two-fluid plasma with
  finite pressure},}\ }\href {\doibase 10.1017/S0022377821000854} {\bibfield
  {journal} {\bibinfo  {journal} {Journal of Plasma Physics}\ }\textbf
  {\bibinfo {volume} {87}},\ \bibinfo {pages} {905870415} (\bibinfo {year}
  {2021})}\BibitemShut {NoStop}%
  \bibitem [{\citenamefont {Janes}\ \emph {et~al.}(1966)\citenamefont {Janes},
  \citenamefont {Levy}, \citenamefont {Bethe},\ and\ \citenamefont
  {Feld}}]{Janes1966}%
  \BibitemOpen
  \bibfield  {author} {\bibinfo {author} {\bibfnamefont {G.~S.}\ \bibnamefont
  {Janes}}, \bibinfo {author} {\bibfnamefont {R.~H.}\ \bibnamefont {Levy}},
  \bibinfo {author} {\bibfnamefont {H.~A.}\ \bibnamefont {Bethe}}, \ and\
  \bibinfo {author} {\bibfnamefont {B.~T.}\ \bibnamefont {Feld}},\ }\href
  {\doibase 10.1103/PhysRev.145.925} {\bibfield  {journal} {\bibinfo  {journal}
  {Phys. Rev.}\ }\textbf {\bibinfo {volume} {145}},\ \bibinfo {pages}
  {925--952} (\bibinfo {year} {1966})}\BibitemShut {NoStop}%
\bibitem [{\citenamefont {Daugherty}, \citenamefont {Eninger},\ and\
  \citenamefont {Janes}(1969)}]{Daugherty1969}%
  \BibitemOpen
  \bibfield  {author} {\bibinfo {author} {\bibfnamefont {J.~D.}\ \bibnamefont
  {Daugherty}}, \bibinfo {author} {\bibfnamefont {J.~E.}\ \bibnamefont
  {Eninger}}, \ and\ \bibinfo {author} {\bibfnamefont {G.~S.}\ \bibnamefont
  {Janes}},\ }\bibfield  {title} {\enquote {\bibinfo {title} {Experiments on
  the injection and containment of electron plasmas in a toroidal apparatus},}\
  }\href {\doibase 10.1063/1.1692411} {\bibfield  {journal} {\bibinfo
  {journal} {The Physics of Fluids}\ }\textbf {\bibinfo {volume} {12}},\
  \bibinfo {pages} {2677--2693} (\bibinfo {year} {1969})},\ \Eprint
  {http://arxiv.org/abs/https://aip.scitation.org/doi/pdf/10.1063/1.1692411}
  {https://aip.scitation.org/doi/pdf/10.1063/1.1692411} \BibitemShut {NoStop}%
\bibitem [{\citenamefont {Clark}\ \emph {et~al.}(1976)\citenamefont {Clark},
  \citenamefont {Korn}, \citenamefont {Mondelli},\ and\ \citenamefont
  {Rostoker}}]{Clark1976}%
  \BibitemOpen
  \bibfield  {author} {\bibinfo {author} {\bibfnamefont {W.}~\bibnamefont
  {Clark}}, \bibinfo {author} {\bibfnamefont {P.}~\bibnamefont {Korn}},
  \bibinfo {author} {\bibfnamefont {A.}~\bibnamefont {Mondelli}}, \ and\
  \bibinfo {author} {\bibfnamefont {N.}~\bibnamefont {Rostoker}},\ }\href
  {\doibase 10.1103/PhysRevLett.37.592} {\bibfield  {journal} {\bibinfo
  {journal} {Phys. Rev. Lett.}\ }\textbf {\bibinfo {volume} {37}},\ \bibinfo
  {pages} {592--595} (\bibinfo {year} {1976})}\BibitemShut {NoStop}%
\bibitem [{\citenamefont {Zaveri}\ \emph {et~al.}(1992)\citenamefont {Zaveri},
  \citenamefont {John}, \citenamefont {Avinash},\ and\ \citenamefont
  {Kaw}}]{Zaveri1992}%
  \BibitemOpen
  \bibfield  {author} {\bibinfo {author} {\bibfnamefont {P.}~\bibnamefont
  {Zaveri}}, \bibinfo {author} {\bibfnamefont {P.~I.}\ \bibnamefont {John}},
  \bibinfo {author} {\bibfnamefont {K.}~\bibnamefont {Avinash}}, \ and\
  \bibinfo {author} {\bibfnamefont {P.~K.}\ \bibnamefont {Kaw}},\ }\bibfield
  {title} {\enquote {\bibinfo {title} {Low-aspect-ratio toroidal equilibria of
  electron plasmas},}\ }\href {\doibase 10.1103/PhysRevLett.68.3295} {\bibfield
  {journal} {\bibinfo  {journal} {Phys. Rev. Lett.}\ }\textbf {\bibinfo
  {volume} {68}},\ \bibinfo {pages} {3295--3298} (\bibinfo {year}
  {1992})}\BibitemShut {NoStop}%
\bibitem [{\citenamefont {Khirwadkar}\ \emph {et~al.}(1993)\citenamefont
  {Khirwadkar}, \citenamefont {John}, \citenamefont {Avinash}, \citenamefont
  {Agarwal},\ and\ \citenamefont {Kaw}}]{Khirwadkar1993}%
  \BibitemOpen
  \bibfield  {author} {\bibinfo {author} {\bibfnamefont {S.~S.}\ \bibnamefont
  {Khirwadkar}}, \bibinfo {author} {\bibfnamefont {P.~I.}\ \bibnamefont
  {John}}, \bibinfo {author} {\bibfnamefont {K.}~\bibnamefont {Avinash}},
  \bibinfo {author} {\bibfnamefont {A.~K.}\ \bibnamefont {Agarwal}}, \ and\
  \bibinfo {author} {\bibfnamefont {P.~K.}\ \bibnamefont {Kaw}},\ }\bibfield
  {title} {\enquote {\bibinfo {title} {Steady state formation of a toroidal
  electron plasma},}\ }\href {\doibase 10.1103/PhysRevLett.71.4334} {\bibfield
  {journal} {\bibinfo  {journal} {Phys. Rev. Lett.}\ }\textbf {\bibinfo
  {volume} {71}},\ \bibinfo {pages} {4334--4337} (\bibinfo {year}
  {1993})}\BibitemShut {NoStop}%
\bibitem [{\citenamefont {Lachhvani}\ \emph {et~al.}(2017)\citenamefont
  {Lachhvani}, \citenamefont {Pahari}, \citenamefont {Sengupta}, \citenamefont
  {Yeole}, \citenamefont {Bajpai},\ and\ \citenamefont
  {Chattopadhyay}}]{Lachhvani2017}%
  \BibitemOpen
  \bibfield  {author} {\bibinfo {author} {\bibfnamefont {L.}~\bibnamefont
  {Lachhvani}}, \bibinfo {author} {\bibfnamefont {S.}~\bibnamefont {Pahari}},
  \bibinfo {author} {\bibfnamefont {S.}~\bibnamefont {Sengupta}}, \bibinfo
  {author} {\bibfnamefont {Y.~G.}\ \bibnamefont {Yeole}}, \bibinfo {author}
  {\bibfnamefont {M.}~\bibnamefont {Bajpai}}, \ and\ \bibinfo {author}
  {\bibfnamefont {P.~K.}\ \bibnamefont {Chattopadhyay}},\ }\href {\doibase
  10.1063/1.5009013} {\bibfield  {journal} {\bibinfo  {journal} {Physics of
  Plasmas}\ }\textbf {\bibinfo {volume} {24}},\ \bibinfo {pages} {102132}
  (\bibinfo {year} {2017})},\ \Eprint
  {http://arxiv.org/abs/https://doi.org/10.1063/1.5009013}
  {https://doi.org/10.1063/1.5009013} \BibitemShut {NoStop}%
\bibitem [{\citenamefont {Stoneking}\ \emph {et~al.}(2004)\citenamefont
  {Stoneking}, \citenamefont {Growdon}, \citenamefont {Milne},\ and\
  \citenamefont {Peterson}}]{Stoneking2004}%
  \BibitemOpen
  \bibfield  {author} {\bibinfo {author} {\bibfnamefont {M.~R.}\ \bibnamefont
  {Stoneking}}, \bibinfo {author} {\bibfnamefont {M.~A.}\ \bibnamefont
  {Growdon}}, \bibinfo {author} {\bibfnamefont {M.~L.}\ \bibnamefont {Milne}},
  \ and\ \bibinfo {author} {\bibfnamefont {R.~T.}\ \bibnamefont {Peterson}},\
  }\href {\doibase 10.1103/PhysRevLett.92.095003} {\bibfield  {journal}
  {\bibinfo  {journal} {Phys. Rev. Lett.}\ }\textbf {\bibinfo {volume} {92}},\
  \bibinfo {pages} {095003} (\bibinfo {year} {2004})}\BibitemShut {NoStop}%
  \bibitem [{\citenamefont {{Davidson}}(1990)}]{Davidson}%
  \BibitemOpen
  \bibfield  {author} {\bibinfo {author} {\bibfnamefont {R.~C.}\ \bibnamefont
  {{Davidson}}},\ }\href@noop {} {\emph {\bibinfo {title} {An introduction to
  the physics of nonneutral plasmas.}}}\ (\bibinfo  {publisher} {IEEE},\
  \bibinfo {year} {1990})\BibitemShut {NoStop}%
\bibitem [{\citenamefont {Marksteiner}\ \emph {et~al.}(2008)\citenamefont
  {Marksteiner}, \citenamefont {Pedersen}, \citenamefont {Berkery},
  \citenamefont {Hahn}, \citenamefont {Mendez}, \citenamefont {Durand~de
  Gevigney},\ and\ \citenamefont {Himura}}]{Marksteiner2008}%
  \BibitemOpen
  \bibfield  {author} {\bibinfo {author} {\bibfnamefont {Q.~R.}\ \bibnamefont
  {Marksteiner}}, \bibinfo {author} {\bibfnamefont {T.~S.}\ \bibnamefont
  {Pedersen}}, \bibinfo {author} {\bibfnamefont {J.~W.}\ \bibnamefont
  {Berkery}}, \bibinfo {author} {\bibfnamefont {M.~S.}\ \bibnamefont {Hahn}},
  \bibinfo {author} {\bibfnamefont {J.~M.}\ \bibnamefont {Mendez}}, \bibinfo
  {author} {\bibfnamefont {B.}~\bibnamefont {Durand~de Gevigney}}, \ and\
  \bibinfo {author} {\bibfnamefont {H.}~\bibnamefont {Himura}},\ }\href
  {\doibase 10.1103/PhysRevLett.100.065002} {\bibfield  {journal} {\bibinfo
  {journal} {Phys. Rev. Lett.}\ }\textbf {\bibinfo {volume} {100}},\ \bibinfo
  {pages} {065002} (\bibinfo {year} {2008})}\BibitemShut {NoStop}%
\bibitem [{\citenamefont {Saitoh}\ \emph {et~al.}(2010)\citenamefont {Saitoh},
  \citenamefont {Yoshida}, \citenamefont {Morikawa}, \citenamefont {Yano},
  \citenamefont {Hayashi}, \citenamefont {Mizushima}, \citenamefont {Kawai},
  \citenamefont {Kobayashi},\ and\ \citenamefont {Mikami}}]{Saitoh2010}%
  \BibitemOpen
  \bibfield  {author} {\bibinfo {author} {\bibfnamefont {H.}~\bibnamefont
  {Saitoh}}, \bibinfo {author} {\bibfnamefont {Z.}~\bibnamefont {Yoshida}},
  \bibinfo {author} {\bibfnamefont {J.}~\bibnamefont {Morikawa}}, \bibinfo
  {author} {\bibfnamefont {Y.}~\bibnamefont {Yano}}, \bibinfo {author}
  {\bibfnamefont {H.}~\bibnamefont {Hayashi}}, \bibinfo {author} {\bibfnamefont
  {T.}~\bibnamefont {Mizushima}}, \bibinfo {author} {\bibfnamefont
  {Y.}~\bibnamefont {Kawai}}, \bibinfo {author} {\bibfnamefont
  {M.}~\bibnamefont {Kobayashi}}, \ and\ \bibinfo {author} {\bibfnamefont
  {H.}~\bibnamefont {Mikami}},\ }\bibfield  {title} {\enquote {\bibinfo {title}
  {Confinement of electron plasma by levitating dipole magnet},}\ }\href
  {\doibase 10.1063/1.3514207} {\bibfield  {journal} {\bibinfo  {journal}
  {Physics of Plasmas}\ }\textbf {\bibinfo {volume} {17}},\ \bibinfo {pages}
  {112111} (\bibinfo {year} {2010})},\ \Eprint
  {http://arxiv.org/abs/https://doi.org/10.1063/1.3514207}
  {https://doi.org/10.1063/1.3514207} \BibitemShut {NoStop}%
  \bibitem [{\citenamefont {Khamaru}, \citenamefont {Sengupta},\ and\
  \citenamefont {Ganesh}(2019)}]{Khamaru2019}%
  \BibitemOpen
  \bibfield  {author} {\bibinfo {author} {\bibfnamefont {S.}~\bibnamefont
  {Khamaru}}, \bibinfo {author} {\bibfnamefont {M.}~\bibnamefont {Sengupta}}, \
  and\ \bibinfo {author} {\bibfnamefont {R.}~\bibnamefont {Ganesh}},\
  }\bibfield  {title} {\enquote {\bibinfo {title} {Dynamics of a toroidal pure
  electron plasma using 3d pic simulations},}\ }\href {\doibase
  10.1063/1.5111747} {\bibfield  {journal} {\bibinfo  {journal} {Physics of
  Plasmas}\ }\textbf {\bibinfo {volume} {26}},\ \bibinfo {pages} {112106}
  (\bibinfo {year} {2019})},\ \Eprint
  {http://arxiv.org/abs/https://doi.org/10.1063/1.5111747}
  {https://doi.org/10.1063/1.5111747} \BibitemShut {NoStop}%
  \bibitem [{\citenamefont {Khamaru}, \citenamefont {Ganesh},\ and\ \citenamefont
  {Sengupta}(2021{\natexlab{a}})}]{Khamaru2021}%
  \BibitemOpen
  \bibfield  {author} {\bibinfo {author} {\bibfnamefont {S.}~\bibnamefont
  {Khamaru}}, \bibinfo {author} {\bibfnamefont {R.}~\bibnamefont {Ganesh}}, \
  and\ \bibinfo {author} {\bibfnamefont {M.}~\bibnamefont {Sengupta}},\
  }\bibfield  {title} {\enquote {\bibinfo {title} {A novel quiescent
  quasi-steady state of a toroidal electron plasma},}\ }\href {\doibase
  10.1063/5.0032880} {\bibfield  {journal} {\bibinfo  {journal} {Physics of
  Plasmas}\ }\textbf {\bibinfo {volume} {28}},\ \bibinfo {pages} {042101}
  (\bibinfo {year} {2021}{\natexlab{a}})},\ \Eprint
  {http://arxiv.org/abs/https://doi.org/10.1063/5.0032880}
  {https://doi.org/10.1063/5.0032880} \BibitemShut {NoStop}%
\bibitem [{\citenamefont {Khamaru}, \citenamefont {Ganesh},\ and\ \citenamefont
  {Sengupta}(2021{\natexlab{b}})}]{Khamaru2021Er}%
  \BibitemOpen
  \bibfield  {author} {\bibinfo {author} {\bibfnamefont {S.}~\bibnamefont
  {Khamaru}}, \bibinfo {author} {\bibfnamefont {R.}~\bibnamefont {Ganesh}}, \
  and\ \bibinfo {author} {\bibfnamefont {M.}~\bibnamefont {Sengupta}},\
  }\bibfield  {title} {\enquote {\bibinfo {title} {Erratum: “a novel
  quiescent quasi-steady state of a toroidal electron plasma” [phys. plasmas
  28, 042101 (2021)]},}\ }\href {\doibase 10.1063/5.0074384} {\bibfield
  {journal} {\bibinfo  {journal} {Physics of Plasmas}\ }\textbf {\bibinfo
  {volume} {28}},\ \bibinfo {pages} {119901} (\bibinfo {year}
  {2021}{\natexlab{b}})},\ \Eprint
  {http://arxiv.org/abs/https://doi.org/10.1063/5.0074384}
  {https://doi.org/10.1063/5.0074384} \BibitemShut {NoStop}%
  \bibitem [{\citenamefont {Crooks}\ and\ \citenamefont
  {O’Neil}(1996)}]{Crooks1996}%
  \BibitemOpen
  \bibfield  {author} {\bibinfo {author} {\bibfnamefont {S.~M.}\ \bibnamefont
  {Crooks}}\ and\ \bibinfo {author} {\bibfnamefont {T.~M.}\ \bibnamefont
  {O’Neil}},\ }\bibfield  {title} {\enquote {\bibinfo {title} {Transport in a
  toroidally confined pure electron plasma},}\ }\href {\doibase
  10.1063/1.871971} {\bibfield  {journal} {\bibinfo  {journal} {Physics of
  Plasmas}\ }\textbf {\bibinfo {volume} {3}},\ \bibinfo {pages} {2533--2537}
  (\bibinfo {year} {1996})},\ \Eprint
  {http://arxiv.org/abs/https://doi.org/10.1063/1.871971}
  {https://doi.org/10.1063/1.871971} \BibitemShut {NoStop}%
  \bibitem [{\citenamefont {Brambilla}\ \emph {et~al.}(2018)\citenamefont
  {Brambilla}, \citenamefont {Kalapotharakos}, \citenamefont {Timokhin},
  \citenamefont {Harding},\ and\ \citenamefont {Kazanas}}]{Brambilla2018}%
  \BibitemOpen
  \bibfield  {author} {\bibinfo {author} {\bibfnamefont {G.}~\bibnamefont
  {Brambilla}}, \bibinfo {author} {\bibfnamefont {C.}~\bibnamefont
  {Kalapotharakos}}, \bibinfo {author} {\bibfnamefont {A.~N.}\ \bibnamefont
  {Timokhin}}, \bibinfo {author} {\bibfnamefont {A.~K.}\ \bibnamefont
  {Harding}}, \ and\ \bibinfo {author} {\bibfnamefont {D.}~\bibnamefont
  {Kazanas}},\ }\bibfield  {title} {\enquote {\bibinfo {title}
  {Electron{\textendash}positron pair flow and current composition in the
  pulsar magnetosphere},}\ }\href {\doibase 10.3847/1538-4357/aab3e1}
  {\bibfield  {journal} {\bibinfo  {journal} {The Astrophysical Journal}\
  }\textbf {\bibinfo {volume} {858}},\ \bibinfo {pages} {81} (\bibinfo {year}
  {2018})}\BibitemShut {NoStop}%
\bibitem [{\citenamefont {Stoneking}\ \emph {et~al.}(2020)\citenamefont
  {Stoneking}, \citenamefont {Pedersen}, \citenamefont {Helander},
  \citenamefont {Chen}, \citenamefont {Hergenhahn}, \citenamefont {Stenson},
  \citenamefont {Fiksel}, \citenamefont {von~der Linden}, \citenamefont
  {Saitoh}, \citenamefont {Surko},\ and\ \citenamefont
  {et~al.}}]{Stoneking2020}%
  \BibitemOpen
  \bibfield  {author} {\bibinfo {author} {\bibfnamefont {M.~R.}\ \bibnamefont
  {Stoneking}}, \bibinfo {author} {\bibfnamefont {T.~S.}\ \bibnamefont
  {Pedersen}}, \bibinfo {author} {\bibfnamefont {P.}~\bibnamefont {Helander}},
  \bibinfo {author} {\bibfnamefont {H.}~\bibnamefont {Chen}}, \bibinfo {author}
  {\bibfnamefont {U.}~\bibnamefont {Hergenhahn}}, \bibinfo {author}
  {\bibfnamefont {E.~V.}\ \bibnamefont {Stenson}}, \bibinfo {author}
  {\bibfnamefont {G.}~\bibnamefont {Fiksel}}, \bibinfo {author} {\bibfnamefont
  {J.}~\bibnamefont {von~der Linden}}, \bibinfo {author} {\bibfnamefont
  {H.}~\bibnamefont {Saitoh}}, \bibinfo {author} {\bibfnamefont {C.~M.}\
  \bibnamefont {Surko}}, \ and\ \bibinfo {author} {\bibnamefont {et~al.}},\
  }\bibfield  {title} {\enquote {\bibinfo {title} {A new frontier in laboratory
  physics: magnetized electron–positron plasmas},}\ }\href {\doibase
  10.1017/S0022377820001385} {\bibfield  {journal} {\bibinfo  {journal}
  {Journal of Plasma Physics}\ }\textbf {\bibinfo {volume} {86}},\ \bibinfo
  {pages} {155860601} (\bibinfo {year} {2020})}\BibitemShut {NoStop}%
\bibitem [{\citenamefont {Singer}\ \emph {et~al.}(2021)\citenamefont {Singer},
  \citenamefont {Stoneking}, \citenamefont {Stenson}, \citenamefont {Nißl},
  \citenamefont {Deller}, \citenamefont {Card}, \citenamefont {Horn-Stanja},
  \citenamefont {Sunn~Pedersen}, \citenamefont {Saitoh},\ and\ \citenamefont
  {Hugenschmidt}}]{Singer2021}%
  \BibitemOpen
  \bibfield  {author} {\bibinfo {author} {\bibfnamefont {M.}~\bibnamefont
  {Singer}}, \bibinfo {author} {\bibfnamefont {M.~R.}\ \bibnamefont
  {Stoneking}}, \bibinfo {author} {\bibfnamefont {E.~V.}\ \bibnamefont
  {Stenson}}, \bibinfo {author} {\bibfnamefont {S.}~\bibnamefont {Nißl}},
  \bibinfo {author} {\bibfnamefont {A.}~\bibnamefont {Deller}}, \bibinfo
  {author} {\bibfnamefont {A.}~\bibnamefont {Card}}, \bibinfo {author}
  {\bibfnamefont {J.}~\bibnamefont {Horn-Stanja}}, \bibinfo {author}
  {\bibfnamefont {T.}~\bibnamefont {Sunn~Pedersen}}, \bibinfo {author}
  {\bibfnamefont {H.}~\bibnamefont {Saitoh}}, \ and\ \bibinfo {author}
  {\bibfnamefont {C.}~\bibnamefont {Hugenschmidt}},\ }\bibfield  {title}
  {\enquote {\bibinfo {title} {Injection of positrons into a dense electron
  plasma in a magnetic dipole trap},}\ }\href {\doibase 10.1063/5.0050881}
  {\bibfield  {journal} {\bibinfo  {journal} {Physics of Plasmas}\ }\textbf
  {\bibinfo {volume} {28}},\ \bibinfo {pages} {062506} (\bibinfo {year}
  {2021})},\ \Eprint {http://arxiv.org/abs/https://doi.org/10.1063/5.0050881}
  {https://doi.org/10.1063/5.0050881} \BibitemShut {NoStop}%
  \bibitem [{\citenamefont {White}, \citenamefont {Malmberg},\ and\ \citenamefont
  {Driscoll}(1982)}]{White1982}%
  \BibitemOpen
  \bibfield  {author} {\bibinfo {author} {\bibfnamefont {W.~D.}\ \bibnamefont
  {White}}, \bibinfo {author} {\bibfnamefont {J.~H.}\ \bibnamefont {Malmberg}},
  \ and\ \bibinfo {author} {\bibfnamefont {C.~F.}\ \bibnamefont {Driscoll}},\
  }\bibfield  {title} {\enquote {\bibinfo {title} {Resistive-wall
  destabilization of Diocotron waves},}\ }\href {\doibase
  10.1103/PhysRevLett.49.1822} {\bibfield  {journal} {\bibinfo  {journal}
  {Phys. Rev. Lett.}\ }\textbf {\bibinfo {volume} {49}},\ \bibinfo {pages}
  {1822--1826} (\bibinfo {year} {1982})}\BibitemShut {NoStop}%
\bibitem [{\citenamefont {Davidson}\ and\ \citenamefont
  {Chao}(1996)}]{Davidson1996}%
  \BibitemOpen
  \bibfield  {author} {\bibinfo {author} {\bibfnamefont {R.~C.}\ \bibnamefont
  {Davidson}}\ and\ \bibinfo {author} {\bibfnamefont {E.~H.}\ \bibnamefont
  {Chao}},\ }\bibfield  {title} {\enquote {\bibinfo {title} {l=1 electrostatic
  instability induced by electron‐neutral collisions in a nonneutral electron
  plasma interacting with background neutral gas},}\ }\href {\doibase
  10.1063/1.871610} {\bibfield  {journal} {\bibinfo  {journal} {Physics of
  Plasmas}\ }\textbf {\bibinfo {volume} {3}},\ \bibinfo {pages} {3279--3287}
  (\bibinfo {year} {1996})},\ \Eprint
  {http://arxiv.org/abs/https://doi.org/10.1063/1.871610}
  {https://doi.org/10.1063/1.871610} \BibitemShut {NoStop}%
\bibitem [{\citenamefont {Levy}, \citenamefont {Daugherty},\ and\ \citenamefont
  {Buneman}(1969)}]{Levy1969}%
  \BibitemOpen
  \bibfield  {author} {\bibinfo {author} {\bibfnamefont {R.~H.}\ \bibnamefont
  {Levy}}, \bibinfo {author} {\bibfnamefont {J.~D.}\ \bibnamefont {Daugherty}},
  \ and\ \bibinfo {author} {\bibfnamefont {O.}~\bibnamefont {Buneman}},\
  }\bibfield  {title} {\enquote {\bibinfo {title} {Ion resonance instability in
  grossly nonneutral plasmas},}\ }\href {\doibase 10.1063/1.1692404} {\bibfield
   {journal} {\bibinfo  {journal} {The Physics of Fluids}\ }\textbf {\bibinfo
  {volume} {12}},\ \bibinfo {pages} {2616--2629} (\bibinfo {year} {1969})},\
  \Eprint
  {http://arxiv.org/abs/https://aip.scitation.org/doi/pdf/10.1063/1.1692404}
  {https://aip.scitation.org/doi/pdf/10.1063/1.1692404} \BibitemShut {NoStop}%
  \bibitem [{\citenamefont {Davidson}\ and\ \citenamefont
  {Uhm}(1977)}]{Davidson1977}%
  \BibitemOpen
  \bibfield  {author} {\bibinfo {author} {\bibfnamefont {R.~C.}\ \bibnamefont
  {Davidson}}\ and\ \bibinfo {author} {\bibfnamefont {H.}~\bibnamefont {Uhm}},\
  }\bibfield  {title} {\enquote {\bibinfo {title} {Influence of strong
  self‐electric fields on the ion resonance instability in a nonneutral
  plasma column},}\ }\href {\doibase 10.1063/1.861813} {\bibfield  {journal}
  {\bibinfo  {journal} {The Physics of Fluids}\ }\textbf {\bibinfo {volume}
  {20}},\ \bibinfo {pages} {1938--1946} (\bibinfo {year} {1977})},\ \Eprint
  {http://arxiv.org/abs/https://aip.scitation.org/doi/pdf/10.1063/1.861813}
  {https://aip.scitation.org/doi/pdf/10.1063/1.861813} \BibitemShut {NoStop}%
\bibitem [{\citenamefont {Fajans}(1993)}]{Fajans1993}%
  \BibitemOpen
  \bibfield  {author} {\bibinfo {author} {\bibfnamefont {J.}~\bibnamefont
  {Fajans}},\ }\bibfield  {title} {\enquote {\bibinfo {title} {Transient ion
  resonance instability},}\ }\href {\doibase 10.1063/1.860649} {\bibfield
  {journal} {\bibinfo  {journal} {Physics of Fluids B: Plasma Physics}\
  }\textbf {\bibinfo {volume} {5}},\ \bibinfo {pages} {3127--3135} (\bibinfo
  {year} {1993})},\ \Eprint
  {http://arxiv.org/abs/https://doi.org/10.1063/1.860649}
  {https://doi.org/10.1063/1.860649} \BibitemShut {NoStop}%
  \bibitem [{\citenamefont {Peurrung}, \citenamefont {Notte},\ and\ \citenamefont
  {Fajans}(1993)}]{Peurrung1993}%
  \BibitemOpen
  \bibfield  {author} {\bibinfo {author} {\bibfnamefont {A.~J.}\ \bibnamefont
  {Peurrung}}, \bibinfo {author} {\bibfnamefont {J.}~\bibnamefont {Notte}}, \
  and\ \bibinfo {author} {\bibfnamefont {J.}~\bibnamefont {Fajans}},\
  }\bibfield  {title} {\enquote {\bibinfo {title} {Observation of the ion
  resonance instability},}\ }\href {\doibase 10.1103/PhysRevLett.70.295}
  {\bibfield  {journal} {\bibinfo  {journal} {Phys. Rev. Lett.}\ }\textbf
  {\bibinfo {volume} {70}},\ \bibinfo {pages} {295--298} (\bibinfo {year}
  {1993})}\BibitemShut {NoStop}%
  \bibitem [{\citenamefont {Lachhvani}\ \emph {et~al.}(2016)\citenamefont
  {Lachhvani}, \citenamefont {Pahari}, \citenamefont {Goswami}, \citenamefont
  {Bajpai}, \citenamefont {Yeole},\ and\ \citenamefont
  {Chattopadhyay}}]{Lachhvani2016}%
  \BibitemOpen
  \bibfield  {author} {\bibinfo {author} {\bibfnamefont {L.}~\bibnamefont
  {Lachhvani}}, \bibinfo {author} {\bibfnamefont {S.}~\bibnamefont {Pahari}},
  \bibinfo {author} {\bibfnamefont {R.}~\bibnamefont {Goswami}}, \bibinfo
  {author} {\bibfnamefont {M.}~\bibnamefont {Bajpai}}, \bibinfo {author}
  {\bibfnamefont {Y.}~\bibnamefont {Yeole}}, \ and\ \bibinfo {author}
  {\bibfnamefont {P.~K.}\ \bibnamefont {Chattopadhyay}},\ }\href {\doibase
  10.1063/1.4953440} {\bibfield  {journal} {\bibinfo  {journal} {Physics of
  Plasmas}\ }\textbf {\bibinfo {volume} {23}},\ \bibinfo {pages} {062109}
  (\bibinfo {year} {2016})},\ \Eprint
  {http://arxiv.org/abs/https://doi.org/10.1063/1.4953440}
  {https://doi.org/10.1063/1.4953440} \BibitemShut {NoStop}%
  \bibitem [{\citenamefont {Eckhouse}, \citenamefont {Fisher},\ and\
  \citenamefont {Rostoker}(1981)}]{Eckhouse1981}%
  \BibitemOpen
  \bibfield  {author} {\bibinfo {author} {\bibfnamefont {S.}~\bibnamefont
  {Eckhouse}}, \bibinfo {author} {\bibfnamefont {A.}~\bibnamefont {Fisher}}, \
  and\ \bibinfo {author} {\bibfnamefont {N.}~\bibnamefont {Rostoker}},\
  }\bibfield  {title} {\enquote {\bibinfo {title} {Possible observation of the
  ion‐resonance instability in a non‐neutral electron plasma},}\ }\href
  {\doibase 10.1063/1.92354} {\bibfield  {journal} {\bibinfo  {journal}
  {Applied Physics Letters}\ }\textbf {\bibinfo {volume} {38}},\ \bibinfo
  {pages} {318--320} (\bibinfo {year} {1981})},\ \Eprint
  {http://arxiv.org/abs/https://doi.org/10.1063/1.92354}
  {https://doi.org/10.1063/1.92354} \BibitemShut {NoStop}%
\bibitem [{\citenamefont {Bettega}\ \emph {et~al.}(2006)\citenamefont
  {Bettega}, \citenamefont {Cavaliere}, \citenamefont {Cavenago}, \citenamefont
  {De~Luca}, \citenamefont {Illiberi}, \citenamefont {Pozzoli},\ and\
  \citenamefont {Romé}}]{Bettega2006}%
  \BibitemOpen
  \bibfield  {author} {\bibinfo {author} {\bibfnamefont {G.}~\bibnamefont
  {Bettega}}, \bibinfo {author} {\bibfnamefont {F.}~\bibnamefont {Cavaliere}},
  \bibinfo {author} {\bibfnamefont {M.}~\bibnamefont {Cavenago}}, \bibinfo
  {author} {\bibfnamefont {F.}~\bibnamefont {De~Luca}}, \bibinfo {author}
  {\bibfnamefont {A.}~\bibnamefont {Illiberi}}, \bibinfo {author}
  {\bibfnamefont {R.}~\bibnamefont {Pozzoli}}, \ and\ \bibinfo {author}
  {\bibfnamefont {M.}~\bibnamefont {Romé}},\ }\bibfield  {title} {\enquote
  {\bibinfo {title} {Active control of the ion resonance instability by ion
  removing fields},}\ }\href {\doibase 10.1063/1.2363175} {\bibfield  {journal}
  {\bibinfo  {journal} {Physics of Plasmas}\ }\textbf {\bibinfo {volume}
  {13}},\ \bibinfo {pages} {112102} (\bibinfo {year} {2006})},\ \Eprint
  {http://arxiv.org/abs/https://doi.org/10.1063/1.2363175}
  {https://doi.org/10.1063/1.2363175} \BibitemShut {NoStop}%
\bibitem [{\citenamefont {Kabantsev}\ and\ \citenamefont
  {Driscoll}(2007)}]{Kabantsev2007}%
  \BibitemOpen
  \bibfield  {author} {\bibinfo {author} {\bibfnamefont {A.}~\bibnamefont
  {Kabantsev}}\ and\ \bibinfo {author} {\bibfnamefont {C.}~\bibnamefont
  {Driscoll}},\ }\bibfield  {title} {\enquote {\bibinfo {title} {Ion-induced
  instability of Diocotron modes in electron plasmas modelling curvature-driven
  flute modes},}\ }\href {\doibase 10.13182/FST07-A1324} {\bibfield  {journal}
  {\bibinfo  {journal} {Fusion Science and Technology}\ }\textbf {\bibinfo
  {volume} {51}},\ \bibinfo {pages} {96--99} (\bibinfo {year} {2007})},\
  \Eprint {http://arxiv.org/abs/https://doi.org/10.13182/FST07-A1324}
  {https://doi.org/10.13182/FST07-A1324} \BibitemShut {NoStop}%
  \bibitem [{\citenamefont {Stoneking}\ \emph {et~al.}(2002)\citenamefont
  {Stoneking}, \citenamefont {Fontana}, \citenamefont {Sampson},\ and\
  \citenamefont {Thuecks}}]{Stoneking2002}%
  \BibitemOpen
  \bibfield  {author} {\bibinfo {author} {\bibfnamefont {M.~R.}\ \bibnamefont
  {Stoneking}}, \bibinfo {author} {\bibfnamefont {P.~W.}\ \bibnamefont
  {Fontana}}, \bibinfo {author} {\bibfnamefont {R.~L.}\ \bibnamefont
  {Sampson}}, \ and\ \bibinfo {author} {\bibfnamefont {D.~J.}\ \bibnamefont
  {Thuecks}},\ }\bibfield  {title} {\enquote {\bibinfo {title} {Electron
  plasmas in a “partial” torus},}\ }\href {\doibase 10.1063/1.1445425}
  {\bibfield  {journal} {\bibinfo  {journal} {Physics of Plasmas}\ }\textbf
  {\bibinfo {volume} {9}},\ \bibinfo {pages} {766--771} (\bibinfo {year}
  {2002})},\ \Eprint {http://arxiv.org/abs/https://doi.org/10.1063/1.1445425}
  {https://doi.org/10.1063/1.1445425} \BibitemShut {NoStop}%
\bibitem [{\citenamefont {Marksteiner}\ \emph {et~al.}(2008)\citenamefont
  {Marksteiner}, \citenamefont {Pedersen}, \citenamefont {Berkery},
  \citenamefont {Hahn}, \citenamefont {Mendez}, \citenamefont {Durand~de
  Gevigney},\ and\ \citenamefont {Himura}}]{Marksteiner2008}%
  \BibitemOpen
  \bibfield  {author} {\bibinfo {author} {\bibfnamefont {Q.~R.}\ \bibnamefont
  {Marksteiner}}, \bibinfo {author} {\bibfnamefont {T.~S.}\ \bibnamefont
  {Pedersen}}, \bibinfo {author} {\bibfnamefont {J.~W.}\ \bibnamefont
  {Berkery}}, \bibinfo {author} {\bibfnamefont {M.~S.}\ \bibnamefont {Hahn}},
  \bibinfo {author} {\bibfnamefont {J.~M.}\ \bibnamefont {Mendez}}, \bibinfo
  {author} {\bibfnamefont {B.}~\bibnamefont {Durand~de Gevigney}}, \ and\
  \bibinfo {author} {\bibfnamefont {H.}~\bibnamefont {Himura}},\ }\bibfield
  {title} {\enquote {\bibinfo {title} {Observations of an ion-driven
  instability in non-neutral plasmas confined on magnetic surfaces},}\ }\href
  {\doibase 10.1103/PhysRevLett.100.065002} {\bibfield  {journal} {\bibinfo
  {journal} {Phys. Rev. Lett.}\ }\textbf {\bibinfo {volume} {100}},\ \bibinfo
  {pages} {065002} (\bibinfo {year} {2008})}\BibitemShut {NoStop}%
  \bibitem [{\citenamefont {Sengupta}\ and\ \citenamefont
  {Ganesh}(2015)}]{Sengupta2015}%
  \BibitemOpen
  \bibfield  {author} {\bibinfo {author} {\bibfnamefont {M.}~\bibnamefont
  {Sengupta}}\ and\ \bibinfo {author} {\bibfnamefont {R.}~\bibnamefont
  {Ganesh}},\ }\bibfield  {title} {\enquote {\bibinfo {title} {Linear and
  nonlinear evolution of the ion resonance instability in cylindrical traps: A
  numerical study},}\ }\href {\doibase 10.1063/1.4927126} {\bibfield  {journal}
  {\bibinfo  {journal} {Physics of Plasmas}\ }\textbf {\bibinfo {volume}
  {22}},\ \bibinfo {pages} {072112} (\bibinfo {year} {2015})},\ \Eprint
  {http://arxiv.org/abs/https://doi.org/10.1063/1.4927126}
  {https://doi.org/10.1063/1.4927126} \BibitemShut {NoStop}%
\bibitem [{\citenamefont {Sengupta}\ and\ \citenamefont
  {Ganesh}(2016)}]{Sengupta2016}%
  \BibitemOpen
  \bibfield  {author} {\bibinfo {author} {\bibfnamefont {M.}~\bibnamefont
  {Sengupta}}\ and\ \bibinfo {author} {\bibfnamefont {R.}~\bibnamefont
  {Ganesh}},\ }\bibfield  {title} {\enquote {\bibinfo {title} {Influence of
  electron-neutral elastic collisions on the instability of an ion-contaminated
  cylindrical electron plasma: 2d3v pic-with-mcc simulations},}\ }\href
  {\doibase 10.1063/1.4964913} {\bibfield  {journal} {\bibinfo  {journal}
  {Physics of Plasmas}\ }\textbf {\bibinfo {volume} {23}},\ \bibinfo {pages}
  {102111} (\bibinfo {year} {2016})},\ \Eprint
  {http://arxiv.org/abs/https://doi.org/10.1063/1.4964913}
  {https://doi.org/10.1063/1.4964913} \BibitemShut {NoStop}%
\bibitem [{\citenamefont {Sengupta}\ and\ \citenamefont
  {Ganesh}(2017)}]{Sengupta2017}%
  \BibitemOpen
  \bibfield  {author} {\bibinfo {author} {\bibfnamefont {M.}~\bibnamefont
  {Sengupta}}\ and\ \bibinfo {author} {\bibfnamefont {R.}~\bibnamefont
  {Ganesh}},\ }\bibfield  {title} {\enquote {\bibinfo {title} {Destabilization
  of a cylindrically confined electron plasma by impact ionization of background
  neutrals: 2d3v pic simulation with monte-carlo-collisions},}\ }\href
  {\doibase 10.1063/1.4978473} {\bibfield  {journal} {\bibinfo  {journal}
  {Physics of Plasmas}\ }\textbf {\bibinfo {volume} {24}},\ \bibinfo {pages}
  {032105} (\bibinfo {year} {2017})},\ \Eprint
  {http://arxiv.org/abs/https://doi.org/10.1063/1.4978473}
  {https://doi.org/10.1063/1.4978473} \BibitemShut {NoStop}%
\bibitem [{\citenamefont {Khamaru}, \citenamefont {Sengupta},\ and\
  \citenamefont {Ganesh}(2019)}]{Khamaru2019}%
  \BibitemOpen
  \bibfield  {author} {\bibinfo {author} {\bibfnamefont {S.}~\bibnamefont
  {Khamaru}}, \bibinfo {author} {\bibfnamefont {M.}~\bibnamefont {Sengupta}}, \
  and\ \bibinfo {author} {\bibfnamefont {R.}~\bibnamefont {Ganesh}},\
  }\bibfield  {title} {\enquote {\bibinfo {title} {Dynamics of a toroidal pure
  electron plasma using 3d pic simulations},}\ }\href {\doibase
  10.1063/1.5111747} {\bibfield  {journal} {\bibinfo  {journal} {Physics of
  Plasmas}\ }\textbf {\bibinfo {volume} {26}},\ \bibinfo {pages} {112106}
  (\bibinfo {year} {2019})},\ \Eprint
  {http://arxiv.org/abs/https://doi.org/10.1063/1.5111747}
  {https://doi.org/10.1063/1.5111747} \BibitemShut {NoStop}%
\bibitem [{\citenamefont {Sengupta}, \citenamefont {Khamaru},\ and\
  \citenamefont {Ganesh}(2021)}]{Sengupta2021}%
  \BibitemOpen
  \bibfield  {author} {\bibinfo {author} {\bibfnamefont {M.}~\bibnamefont
  {Sengupta}}, \bibinfo {author} {\bibfnamefont {S.}~\bibnamefont {Khamaru}}, \
  and\ \bibinfo {author} {\bibfnamefont {R.}~\bibnamefont {Ganesh}},\
  }\bibfield  {title} {\enquote {\bibinfo {title} {Self-organization of pure
  electron plasma in a partially toroidal magnetic-electrostatic trap: A 3d
  particle-in-cell simulation},}\ }\href {\doibase 10.1063/5.0055828}
  {\bibfield  {journal} {\bibinfo  {journal} {Journal of Applied Physics}\
  }\textbf {\bibinfo {volume} {130}},\ \bibinfo {pages} {133305} (\bibinfo
  {year} {2021})},\ \Eprint
  {http://arxiv.org/abs/https://doi.org/10.1063/5.0055828}
  {https://doi.org/10.1063/5.0055828} \BibitemShut {NoStop}%
  \bibitem [{\citenamefont {Huba}(2013)}]{Huba2013}%
  \BibitemOpen
  \bibfield  {author} {\bibinfo {author} {\bibfnamefont {J.~D.}\ \bibnamefont
  {Huba}},\ }\href {http://wwwppd.nrl.navy.mil/nrlformulary/} {\emph {\bibinfo
  {title} {Plasma Physics}}}\ (\bibinfo  {publisher} {Naval Research
  Laboratory},\ \bibinfo {address} {Washington, DC},\ \bibinfo {year} {2013})\
  p.~\bibinfo {pages} {28}\BibitemShut {NoStop}%
\end{thebibliography}

\providecommand{\noopsort}[1]{}\providecommand{\singleletter}[1]{#1}%
\end{document}